\documentclass[
    aps,
    prx,
    reprint,
    superscriptaddress,
    footinbib,
    amsmath,
    amssymb,
    longbibliography
]{revtex4-2}
 
\usepackage{graphicx}
\usepackage{bm}
\usepackage{mathtools}
\usepackage{physics}
\usepackage{braket}
\usepackage{bbold}
\usepackage{hyperref}

\begin{document}
 
\title{Quantum control of an electrically levitated nanoparticle with a trapped ion}
 
\author{Bernard Faulend}
\affiliation{
 Institute for Theoretical Physics and Vienna Center for Quantum Science and Technology, Vienna University of Technology, Wiedner Hauptstra{\ss}e 8-10, 1040 Vienna, Austria
}
\author{Dmitry S. Bykov}
\affiliation{Institut f\"ur Experimentalphysik, Universit\"at Innsbruck, Technikerstra{\ss}e 25, 6020 Innsbruck, Austria}
\author{Tracy E. Northup}
\affiliation{Institut f\"ur Experimentalphysik, Universit\"at Innsbruck, Technikerstra{\ss}e 25, 6020 Innsbruck, Austria}
\author{Carlos Gonzalez-Ballestero}
\email{carlos.gonzalez-ballestero@tuwien.ac.at}
\affiliation{
 Institute for Theoretical Physics and Vienna Center for Quantum Science and Technology, Vienna University of Technology, Wiedner Hauptstra{\ss}e 8-10, 1040 Vienna, Austria
}
 
\begin{abstract}
Quantum control of levitated nanoparticles offers a route towards macroscopic quantum states. Here we propose a novel way to achieve it for a nanoparticle levitated electrically in a linear Paul trap together with a single ion, a system recently realized experimentally. By suitably engineering the trap's AC potential, one can parametrically modulate the ion-nanoparticle distance and thus the Coulomb coupling to enable a resonant interaction. For realistic parameters, we calculate that this enables sympathetic cooling of the nanoparticle to its motional ground state by a single, continuously laser-cooled ion. Additional AC drives enable the preparation of squeezed nanoparticle states and of ion-nanoparticle entanglement. Moreover, non-Gaussian states such as Fock and cat states, prepared in the ion with standard trapped-ion techniques, can be transferred to the nanoparticle with high fidelities. We provide complete protocols for each of these states, including all known decoherence sources. Our results establish electrically trapped ion-nanoparticle systems as a platform for quantum control of levitated matter.
\end{abstract}
 
\maketitle

\begin{figure}[t]
	\centering
    \includegraphics[width=86mm]{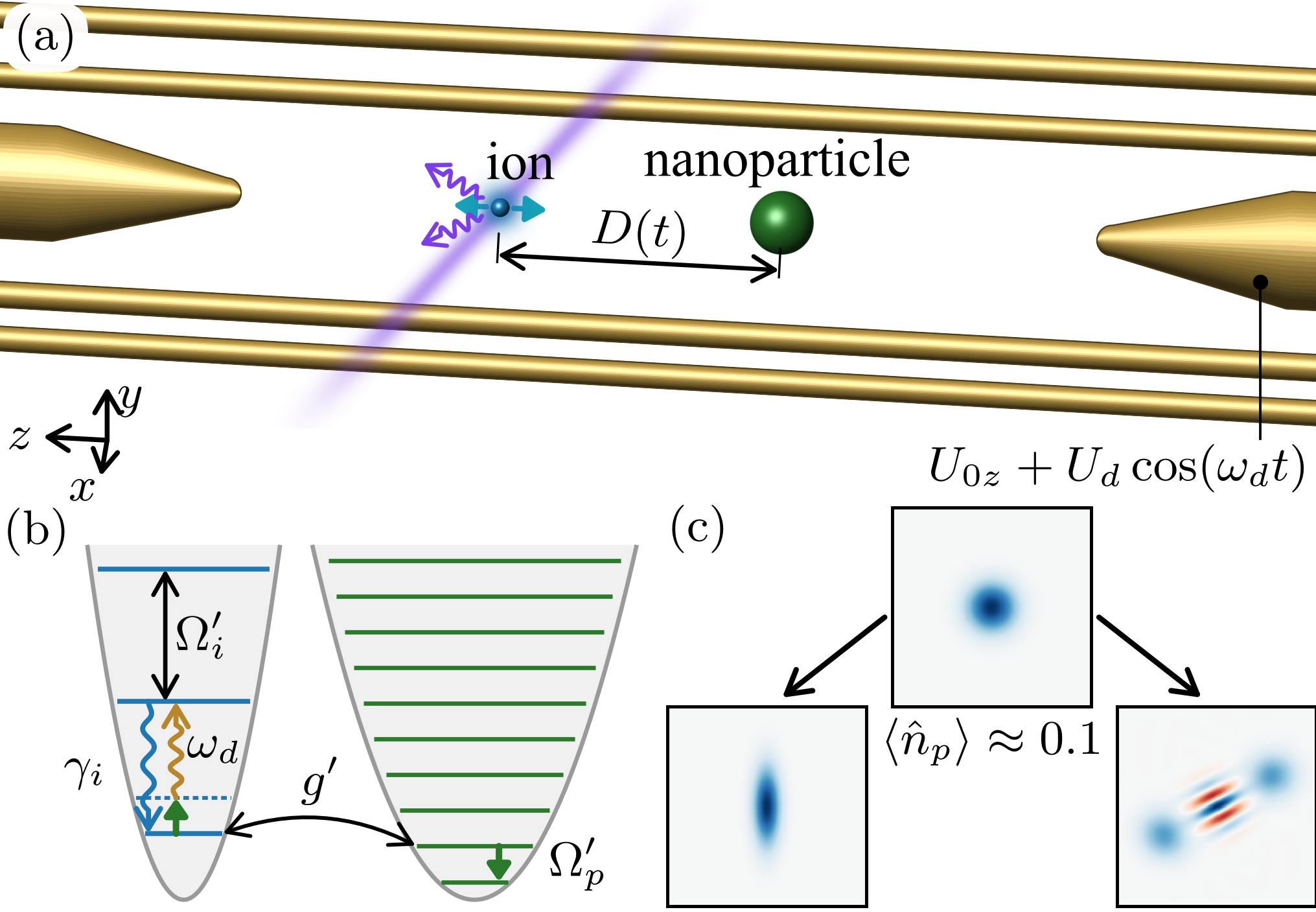}
    \caption{
    \textbf{The system.} (a) An ion and a nanoparticle are trapped in the same linear Paul trap and interact via the Coulomb force. An AC voltage $U_d\cos(\omega_dt)$ applied to the endcap electrodes modulates the ion-nanoparticle distance $D(t)$ by driving the ion motion along $z$. The ion motion is manipulated by trapped-ion techniques such as sideband cooling (rate $\gamma_i$). (b) Motional energy levels of the ion (i) and the nanoparticle (p) along $z$ (not to scale; frequency ratio is $\Omega_i'/\Omega_p'\sim10^3$). The drive at $\omega_d\approx\Omega_i'-\Omega_p'$ bridges the frequency gap, allowing resonant exchange of nanoparticle and ion phonons via the beam-splitter coupling (rate $g'$). (c) Our scheme allows for quantum control of the nanoparticle motion: from ground-state cooling ($\langle\hat{n}_p\rangle\approx0.1$, Sec.~\ref{sec3}) to squeezing (Sec.~\ref{sec4}) and non-Gaussian states (Sec.~\ref{sec5}).
    }
    \label{fig1}
\end{figure}
 
\section{Introduction}
 
Can the predictions of quantum mechanics be extended to macroscopic objects? How would gravity affect quantum superpositions of macroscopic objects? These questions, central to our understanding of the physical world, have been debated since the early days of quantum mechanics~\cite{Schrodinger1935, DeWitt2011, Feynman1995, Penrose1996, Bassi2013}, but remained in limbo between physics and philosophy until experimental advances in recent decades made it possible to study them directly~\cite{Arndt1999, OConnell2010, Chu2018, Fein2019, Bild2023, Pedalino2026}. In this context, experiments with levitated nanoparticles are one of the most promising routes towards preparing center-of-mass quantum superpositions in a regime that would allow tests of gravitational decoherence models or the preparation of gravitationally entangled states~\cite{RomeroIsart2010,RomeroIsart2011,RomeroIsart2011b,Bose2017,MarlettoVedral2017,RomeroIsart2017,Pino2018,Pedernales2020,GonzalezBallestero2021,Weiss2021,Pedernales2022,Neumeier2024,RodaLlordes2024,Bose2025,CunillSubiranas2026}. The main advantages of the platform are tunable trapping potentials, exquisite isolation from the environment, and quality factors exceeding $10^{10}$~\cite{Millen2020, GonzalezBallestero2021,Dania2024}.
 
Recent demonstrations of ground-state motional cooling~\cite{Delic2020, Tebbenjohanns2021, Magrini2021} and squeezing~\cite{Rossi2025, Kamba2025,SchmittKremer2026} for optically levitated nanoparticles constitute first steps en route to these goals. It will however be challenging to go beyond Gaussian state preparation with fully optical trapping because of the fast decoherence due to photon recoil heating~\cite{Jain2016}, a fundamental consequence of laser shot noise. 
Electrical trapping of  nanoparticles~\cite{Millen2015, Fonseca2016, Conangla2018, Conagla2019, Bykov2019, Goldwater2019, Dania2024, Bykov2025} is a promising solution as it would avoid this noise. Moreover, ions and a nanoparticle have been co-trapped electrically in the same linear Paul trap~\cite{Bykov2025}, opening the way to extend the toolbox used to control both electronic and motional degrees of freedom~\cite{Leibfried2003} of trapped ions to control nanoparticle motion~\cite{Pflanzer2013, Martinetz2020, Toros2021, Leontica2021, Schut2025}. However, removing the laser also removes its advantages: mechanical frequencies are reduced to a few kilohertz, making ground-state cooling and coherent quantum control more challenging~\cite{Walker2019, Streltsov2021, GonzalezBallestero2021}. Sympathetic cooling enabled by coupling to a continuously cooled ion provides a path to cool the nanoparticle, but it will not allow us to reach the ground state: the ion has a much higher mechanical frequency than the nanoparticle, leading to far-off-resonant coupling~\cite{Gupta2025}. Therefore, the questions of whether it is possible to resonantly couple ion and nanoparticle motion, cool the nanoparticle to the ground state, or prepare nonclassical states of the nanoparticle using this setup have remained open.
 
In this work we answer all three questions in the affirmative. 
Specifically, we propose to overcome the ion-nanoparticle frequency mismatch in current experiments in linear Paul traps by applying an AC voltage to the endcap electrodes in addition to the standard DC voltage [Fig.~\ref{fig1}(a)]. This AC voltage drives the classical motion of the ion and the nanoparticle. This results in a parametric modulation of their position-dependent Coulomb interaction, which can be tuned to induce resonant beam-splitter and two-mode squeezing interactions that would be absent otherwise. We show that, with feasible experimental parameters, the beam-splitter interaction allows the nanoparticle to be cooled to its motional ground state. Additionally, adding a second tone to the AC voltage drive applied to the endcap electrodes, both beam-splitter and two-mode squeezing terms can be enhanced simultaneously, leading to either a squeezed nanoparticle state or a Gaussian entangled state of the ion and nanoparticle motion, depending on the chosen voltage frequencies. Finally, we show how, using the ion's electronic qubit as an additional degree of freedom, it is possible to prepare non-Gaussian motional states of the ion and transfer them to the nanoparticle with high fidelity. Our work thus demonstrates all required capabilities for macroscopic quantum state preparation, establishing hybrid ion-nanoparticle platforms as a fully functional alternative route to optical trapping in levitated optomechanics.

The paper is organized as follows: in Sec.~\ref{sec2} we describe the theory of classical and quantum dynamics of the ion and the nanoparticle in the driven trap. In Sec.~\ref{sec3} we analyze the cooling of nanoparticle motion, followed by the analysis of two-tone driving and preparation of squeezed and entangled Gaussian states in Sec.~\ref{sec4}. Results on non-Gaussian state preparation are presented in Sec.~\ref{sec5}, experimental considerations are discussed in Sec.~\ref{sec_exp}, and we give our concluding remarks in Sec.~\ref{sec6}.
 
\section{Ion-nanoparticle dynamics in a driven trap}\label{sec2}
 
In this section we outline the derivation of the master equation governing the dynamics of an ion and a nanoparticle co-trapped in the same linear Paul trap, with an additional driving voltage applied to the endcap electrodes (Fig.~\ref{fig1}). 
The electric potential at a position $\mathbf{R}$ near the geometric center of a linear dual-frequency Paul trap~\cite{Dehmelt1995, Trypogeorgos2016, Leefer2017, Bykov2025}~\footnote{Note that a two-tone Paul trap with slow and fast AC frequencies is needed in order to co-trap both the ion and the nanoparticle~\cite{Bykov2025}.} with an added endcap drive is
\begin{equation}\label{fullpotential}
    \phi\left(\bm{R},t \right)=\phi_{\rm trap}\left(\bm{R},t \right)+\phi_{\rm drive}\left(\bm{R},t \right).
\end{equation}
The first term is the potential for a linear Paul trap, given by~\cite{Leibfried2003, Gupta2025}
\begin{multline}\label{potential0}
    \phi_{\rm trap} \left(\bm{R},t \right) = \frac{\alpha_{\rm DC}U_{0z}}{2d_z^2}\left(2R_z^2-R_x^2-R_y^2\right)\\
    +\frac{\alpha_{\rm AC}}{2 d_r^2}
    \left(R_x^2-R_y^2\right)\Big(U_{\rm comp}+U_s\cos(\omega_s t)+U_f\cos(\omega_f t)\Big).
\end{multline}
The endcap electrodes provide confinement along the longitudinal axis of a linear Paul trap with a DC voltage, while the RF electrodes parallel to the $z$ axis provide radial confinement. The second term is the added AC endcap potential, given by $\phi_{\rm drive}(\bm{R},t)=-U_{\rm drive}(t)R_z/d_z$ and resulting in a time-dependent homogeneous electric field along the $z$ axis. Here, $\alpha_{\rm AC, DC}$ denote trap geometric factors, $U_{0z, s, f, \rm comp, drive}$ denote voltage amplitudes, $d_{r, z}$ denote distances from the trap center to the radial and longitudinal electrodes, and $\omega_{s, f, d}$ denote voltage frequencies (see Appendix~\ref{A-EP} for typical parameter values).
 
\subsection{Classical dynamics}
 
Let us consider first the classical dynamics of an ion and a nanoparticle, with masses $M_i$, $M_p$ and charges $Q_i$, $Q_p$, respectively, trapped in the potential of Eq.~\eqref{fullpotential}. Without the time-dependent drive, the coupled dynamics were analyzed in detail in Ref.~\cite{Gupta2025}, where it was shown that the motion along the $x$, $y$ and $z$ axes is uncoupled~\footnote{This follows from the separability of $\phi_{\rm trap}$. When both the nanoparticle and the ion are present in the trap, their interaction couples the motion along different axes, but only as a third-order effect in the fluctuations around equilibrium, as discussed in Appendix~\ref{A-HO}.} and that the only stable equilibrium has the ion and the nanoparticle lying on the $z$ axis, at the positions $z_{i,0}$ and $z_{p,0}$ given by~\cite{Gupta2025}
\begin{equation}
    z_{p,0}=-z_{i,0}M_i\Omega_{i}^2/(M_p\Omega_{p}^2)=-D_0(1+M_p\Omega_{p}^2/(M_i\Omega_{i}^2))^{-1},
\end{equation} 
with the equilibrium distance
\begin{equation}
    D_0=z_{i,0}-z_{p,0}=\sqrt[3]{\frac{Q_i Q_p}{4\pi\varepsilon_0}\left(\frac{1}{M_p\Omega_{p}^2}+\frac{1}{M_i\Omega_{i}^2}\right)},
\end{equation}
where we have taken $z_i>z_p$ without loss of generality. Here, $\Omega_{i,p}=\sqrt{2Q_{i,p}\alpha_{\rm DC}U_{0z}/(d_z^2M_{i,p})}$ are the bare mechanical frequencies along $z$ generated by $\phi_{\rm trap}$. Expanding the Coulomb interaction to second order in the displacements from equilibrium, we can write the ion and the nanoparticle motion as two coupled harmonic oscillators with the renormalized frequencies~\cite{Gupta2025}
\begin{equation}
   \Omega_{j}' = \sqrt{\Omega_{j}^2 + \frac{Q_i Q_p}{2 \pi \varepsilon_0 M_j D_{\rm 0}^3}}\qquad(j \in \{i,p\}),
   \label{Freq_renormalized}
\end{equation}
and a motional coupling rate $\mathcal{K}_j$, where
\begin{equation}
    \mathcal{K}_j^2=\frac{Q_iQ_p}{2\pi\varepsilon_0M_jD_0^3}.
    \label{K0}
\end{equation}
The above motional coupling is inefficient as it cannot bridge the large gap between ion and nanoparticle motional frequencies ($\Omega_i'\sim 10^3\Omega_p'$). This prevents the use of the ion for quantum control of the nanoparticle.

Our proposed solution is to use the endcap drive $U_{\rm drive}$ to modulate the ion-nanoparticle distance, and with it the Coulomb coupling. Parametric modulation is a versatile tool for engineering interactions that are absent from the static Hamiltonian: in superconducting circuits, for example, it is used to convert single photons between modes of different frequencies~\cite{ZakkaBajjani2011}, to realize entangling gates between strongly detuned qubits~\cite{McKay2016,Reagor2018}, to amplify and squeeze quantum noise~\cite{CastellanosBeltran2008}, and to engineer dissipation that stabilizes nonclassical states~\cite{Leghtas2015}. A related parametric coupling approach was proposed for trapped electrons~\cite{Osada2022}, while the trapped-ion approaches, based on electrode modulation~\cite{Gorman2014,Hou2024} or on a driven mode combined with the Coulomb nonlinearity~\cite{Nikolova2023}, have so far been limited to modes of comparable frequencies. Applying the idea to our strongly off-resonant system requires a drive $U_{\rm drive}(t)$ specifically engineered to avoid parametric effects, as we will see below.  The drive acts along the $z$ axis and leaves the motion in the $xy$ plane uncoupled. Both objects thus remain on the $z$ axis, and their motion along it is described by the equations
\begin{equation}\label{di_cl}
    \ddot{z}_i=-\Omega_i^2z_i+\frac{Q_iQ_p}{4\pi\varepsilon_0M_i(z_i-z_p)^2}-\gamma_i\dot{z}_i+\frac{Q_i}{M_id_z}U_{\rm drive}(t),
\end{equation}
\begin{equation}\label{dp_cl}
    \ddot{z}_p=-\Omega_p^2z_p-\frac{Q_iQ_p}{4\pi\varepsilon_0M_p(z_i-z_p)^2}-\gamma_p\dot{z}_p+\frac{Q_p}{M_pd_z}U_{\rm drive}(t).
\end{equation}
Here, $\gamma_i$ and $\gamma_p\equiv\gamma_{\rm gas}$ denote the damping rates of the ion and the nanoparticle, respectively. The damping of the nanoparticle is dominated by collisions with residual gas molecules, and that of the ion is provided by laser cooling~\footnote{Eqs.~\eqref{di_cl} and~\eqref{dp_cl} remain valid in the presence of the laser feed-forward used in our protocols, as we discuss in Sec.~\ref{sec_exp}.}.
 
The drive makes the ion and the nanoparticle oscillate around their equilibrium positions. To illustrate the physics of the system, we look first into the regime of small oscillations. As for the undriven system we expand the Coulomb interaction to second order around the equilibrium, the small deviations $\delta z_j(t)=z_j(t)-z_{j,0}$ obey the linear equations
\begin{equation}\label{lin_cl}
    \delta\ddot{z}_j=-\Omega_j'(t)^2\,\delta z_j+\mathcal{K}_j^2(t)\,\delta z_{\bar{j}}-\gamma_j\,\delta\dot{z}_j,
\end{equation}
where $\bar{j}=p$ for $j=i$ and vice versa, and
\begin{equation}\label{Freq_t}
    \mathcal{K}_j^2(t)=\frac{Q_iQ_p}{2\pi\varepsilon_0M_jD(t)^3},\qquad \Omega_j'(t)^2=\Omega_j^2+\mathcal{K}_j^2(t),
\end{equation}
with the classical distance $D(t)=z_{i}(t)-z_{p}(t)$. Both the frequencies $\Omega_j'(t)$ and the coupling $\mathcal{K}_j(t)$ depend on time through $D(t)$; for $D=D_0$ they reduce to Eq.~\eqref{Freq_renormalized} and Eq.~\eqref{K0}. 

To parametrically enhance the weak coupling between ion and nanoparticle motion, a first approach is to apply a monochromatic endcap drive $U_{\rm drive}(t) \propto \cos(\omega_d t)$, where $\omega_d$ is the drive frequency tuned to $\omega_d\approx\Omega_i'-\Omega_p'$ to bridge the frequency gap. In the harmonic (weak-drive) regime this leads to ion and nanoparticle oscillating at different amplitudes due to their charge-to-mass ratio differing by 6 orders of magnitude (see Appendix~\ref{A-EP}), and results in a distance modulation $D(t)\approx D_0 + A_{\rm cl}\cos(\omega_d t)$ (with $A_{\rm cl}$ the oscillation amplitude) and gives a coupling $\mathcal{K}_j^2(t)\approx\mathcal{K}_j^2|_{D_0}[1-3a\cos(\omega_d t)]$ for $a=A_{\rm cl}/D_0\ll1$. Having a large coupling requires a non-negligible modulation $a$ at a frequency very close to the ion's, $\omega_d\sim\Omega_i'$, which has two consequences. First, the ion behaves as a driven nonlinear (Duffing) oscillator whose resonance frequency increases with its amplitude~\cite{Landau1976}. The resonance condition is therefore shifted to $\omega_d=\Omega_i'-\Omega_p'+\Delta_d$, with a drive-induced shift $\Delta_d$ that grows quadratically with $A_{\rm cl}$ (see Appendix~\ref{A-PR} for details). Second, the ion frequency $\Omega_i'(t)$ in Eq.~\eqref{Freq_t} has a component modulated at $2\omega_d\approx2\Omega_i'$, which drives the ion's parametric resonance and heats its motion. This component is mainly a consequence of the second harmonic of $D(t)$, which exists because of the Coulomb nonlinearity, as detailed in Appendix~\ref{A-PR}. We can suppress this effect by adding a second tone at $2\omega_d$, which drives the ion in antiphase with this harmonic. Hereafter we therefore use the drive voltage
\begin{equation}\label{Udrive}
    U_{\rm drive}(t)=U_d\cos(\omega_d t)+U_d^{(2)}\cos(2\omega_d t+\varphi_2),
\end{equation}
where the amplitude $U_d^{(2)}$ and the phase $\varphi_2$ of the compensation tone can be chosen to suppress the second harmonic of $D(t)$.

\begin{figure}[t]
	\centering
    \includegraphics[width=86mm]{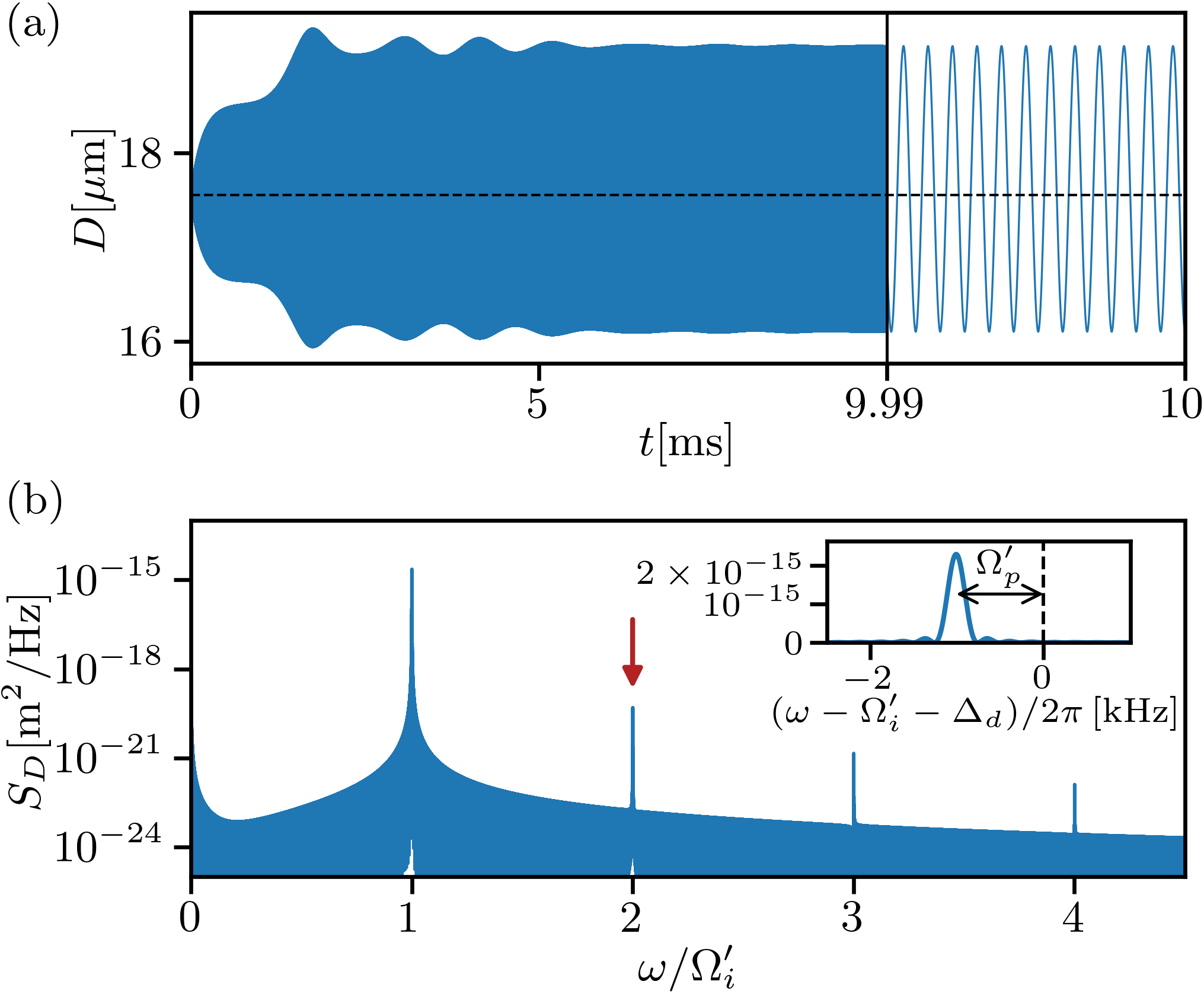}
    \caption{\textbf{Classical dynamics.} (a) Time dependence of the classical ion-nanoparticle distance $D(t)$ calculated from Eqs.~\eqref{di_cl}-\eqref{dp_cl}, with experimental parameters from Table~\ref{Table1} in Appendix~\ref{A-EP}, $U_d=60\,\rm \mu V$, $\omega_d=\Omega_i'-\Omega_p'+\Delta_d$, $\gamma_i=2\pi\times350\,\rm Hz$ and the second-harmonic compensation tone $U_d^{(2)}=2.7\,\rm mV$ [Eq.~\eqref{Udrive}]. After the voltage modulation is turned on, there are transient nonlinear oscillations, which are stabilized by the ion damping on a timescale $\gamma_i^{-1}$. The dashed horizontal line shows the equilibrium distance without driving, $D_0$. (b) Spectrum $S_D$ of the distance obtained from the time trace of $D(t)$ between $t=6\,\rm ms$ and $t=10\,\rm ms$, showing almost pure harmonic oscillations at $\omega_d$. The red arrow indicates the second harmonic, which is suppressed by the compensation tone. Inset: magnified view around the drive-shifted ion frequency $\Omega_i'+\Delta_d$ (Appendix~\ref{A-PR}); the peak at $\omega_d$ is red-detuned from it by $\Omega_p'$, which makes the beam-splitter interaction resonant.}
    \label{fig2}
\end{figure}

To show that the above intuition from the linearized dynamics holds in the presence of strong driving, Fig.~\ref{fig2} shows the full nonlinear dynamics obtained by numerically solving Eqs.~\eqref{di_cl} and~\eqref{dp_cl} with the drive~\eqref{Udrive}. 
For our system of interest, which consists of a $^{40}\mathrm{Ca}^{+}$ ion and a silica nanoparticle with $R_p=50\rm \,nm$ radius (corresponding to $M_p=1\times10^{-18}$~kg) and charge $Q_p=30\rm\, e$, the bare frequencies are $\Omega_i/2\pi=7.1\times10^5\rm\, Hz$ and $\Omega_p/2\pi=1.0\times10^3\rm\, Hz$, the renormalized frequencies are $\Omega_i'/2\pi=1.2\times10^6\rm\, Hz$ and $\Omega_p'/2\pi=1.0\times10^3\rm\, Hz$, and the static equilibrium distance is $D_0=17.6\,\rm\mu m$.  As shown in Fig.~\ref{fig2}(a), after the drive is turned on, there are transient nonlinear oscillations, which are stabilized by the ion damping on a timescale $\sim\gamma_i^{-1}$. Afterwards, the oscillations become almost purely harmonic at the drive frequency $\omega_d$, with a distance modulation amplitude $A_{\rm cl}\approx1.5\,\rm\mu m$ for $U_d=60\,\rm\mu V$. This is confirmed by Fig.~\ref{fig2}(b), which shows that the spectrum of $D(t)$ is dominated by the first harmonic, with higher harmonics suppressed by several orders of magnitude.
In this case the compensation tone suppresses the second harmonic of $D(t)$ by a further factor of $\approx3$.
 
\subsection{Quantum dynamics}\label{sec2b}
 
Let us now focus on the quantum dynamics of the ion-nanoparticle system. The Hamiltonian is given by
\begin{multline}
    \hat{H}=\frac{1}{2M_i}\hat{P}_i^2+\frac{1}{2M_p}\hat{P}_p^2+\frac{1}{2}M_i\Omega_i^2\hat{z}_i^2+\frac{1}{2}M_p\Omega_p^2\hat{z}_p^2\\+\frac{Q_iQ_p}{4\pi\varepsilon_0}\frac{1}{|\hat{z}_i-\hat{z}_p|}-\frac{U_{\rm drive}(t)Q_i}{d_z}\hat{z}_i-\frac{U_{\rm drive}(t)Q_p}{d_z}\hat{z}_p,
    \label{H_full}
\end{multline}
with $\hat{z}_j$ the position operator and $\hat{P}_j$ its associated canonical momentum.
To eliminate the time-dependent classical components of the position, we move to a displaced frame $\hat{z}_j\to\hat{z}_j-z_{j,\rm cl}(t)\mathbb{1}$ with the unitary $\hat{D}(t)=\hat{D}_i(t)\hat{D}_p(t)$ given by
\begin{equation}\label{disptrafo}
    \hat{D}_j(t)=\exp\Big(-\frac{\rm i}{\hbar}(p_{j,\rm cl}(t)\hat{z}_j-z_{j,\rm cl}(t)\hat{P}_j)\Big),
\end{equation}
where $p_{j,\rm cl}=M_j\dot{z}_{j,\rm cl}$, such that $\hat{D}_j\hat{z}_j\hat{D}_j^\dagger=\hat{z}_j+z_{j,\rm cl}$ and $\hat{D}_j\hat{P}_j\hat{D}_j^\dagger=\hat{P}_j+p_{j,\rm cl}$. To cancel the driving terms linear in system operators, $z_{j, \rm cl}(t)$ should satisfy the classical equations of motion~\eqref{di_cl} and~\eqref{dp_cl} without damping (damping terms are introduced later by the dissipation; see below). The transformation Eq.~\eqref{disptrafo} thus represents translation to a frame following the classical trajectories~\cite{RodaLlordes2024,RodaLlordes2024b}.
The Hamiltonian transforms to
\begin{equation}
    \hat{H'}=\hat{D}\hat{H}\hat{D}^{\dag}-\mathrm{i}\hbar \hat{D}\dot{\hat{D}}^\dag.
\end{equation}

After expanding the Coulomb interaction to second order in quantum fluctuations, the Hamiltonian in the displaced frame is
\begin{multline}
    \hat{H'}=\frac{1}{2M_i}\hat{P}_i^2+\frac{1}{2M_p}\hat{P}_p^2+\frac{1}{2}M_i\Omega_i'(t)^2\hat{z}_i^2+\frac{1}{2}M_p\Omega_p'(t)^2\hat{z}_p^2\\-M_i\mathcal{K}_i^2(t)\hat{z}_i\hat{z}_p,
    \label{Hdisplaced}
\end{multline}
with the time-dependent frequencies $\Omega_j'(t)$ and coupling $\mathcal{K}_i(t)$ of Eq.~\eqref{Freq_t}, as in the classical equations~\eqref{lin_cl}. The validity of the expansion above is confirmed in Appendix~\ref{A-HO}, where we show that, for the parameters chosen throughout this work, higher-order terms contribute negligibly to the dynamics.
The distance $D_{\rm cl}=z_{i,\rm cl}-z_{p, \rm cl}$ is time-dependent and, in light of the results in the previous section (Fig.~\ref{fig2}), can be assumed  to oscillate harmonically, $D_{\rm cl}(t)=D_0(1+a\cos(\omega_d t))$. The mechanical frequencies can be expanded as $\Omega_j'(t)\approx\Omega_j'+\delta\Omega_j'\cos(\omega_d t)$, with $\delta\Omega_j'=-3a(\Omega_j'^2-\Omega_j^2)/(2\Omega_j')$ and $|\delta\Omega_j'|\ll \Omega_j'$, which allows us to define the ladder operators in a standard way through $\hat{z}_j=x_{\mathrm{zpm},j}(\hat{a}_j+\hat{a}_j^\dagger)$, with $x_{\mathrm{zpm},j}=\sqrt{\hbar/(2M_j\Omega_j')}$. The Hamiltonian can be written in terms of ladder operators as
\begin{multline}
    \hat{H}'/\hbar=\Omega_i'\hat{a}_i^{\dag}\hat{a}_i+\Omega_p'\hat{a}_p^{\dag}\hat{a}_p+g(t)(\hat{a}_i+\hat{a}_i^{\dag})(\hat{a}_p+\hat{a}_p^{\dag})\\+\frac{\delta\Omega_i'}{2}\cos(\omega_d t)(\hat{a}_i^2+(\hat{a}_i^{\dag})^2+2\hat{a}_i^{\dag}\hat{a}_i).
    \label{H_ab}
\end{multline}
The contribution proportional to $\delta\Omega_p'$ is negligible for realistic nanoparticle parameters. The coupling strength is given by
\begin{equation}
    g(t)=-\frac{Q_iQ_px_{\mathrm{zpm},i}x_{\mathrm{zpm},p}}{2\pi\varepsilon_0\hbar D_{\rm cl}(t)^3}\approx g_0+\tilde{g}\cos(\omega_d t),
\end{equation}
where $g_0=-Q_iQ_px_{\mathrm{zpm},i}x_{\mathrm{zpm},p}/(2\pi\varepsilon_0\hbar D_0^3)$ is the static coupling and $\tilde{g}=-3ag_0$ is the modulation of the coupling.

In the next step, we move the ion operators to the frame rotating with frequency $\omega_d=\Omega_i'-\Omega_p'+\Delta_d$, where $\Delta_d$ is the drive-induced frequency shift introduced in Sec.~\ref{sec2} and derived in Appendix~\ref{A-PR}, and neglect all fast-rotating terms under a rotating-wave approximation (RWA). Although $|g_0|, |\tilde{g}|, |\delta\Omega_i'|\ll \Omega_i'$, this step requires care: the terms proportional to $g_0$ and to $\delta\Omega_i'$, which both oscillate at $\approx\omega_d$ in this frame, have much larger rates than $\tilde{g}$ (for our parameters, $|\delta\Omega_i'|/|\tilde{g}|\approx 100$), and their combined second-order contribution is resonant and of the same order as $\tilde{g}$. Taking it into account (see Appendix~\ref{A-PR}), we arrive at the Hamiltonian
\begin{equation}
    \hat{H}_{\rm RWA}/\hbar=\Omega_p'(\hat{a}_i^{\dag}\hat{a}_i+\hat{a}_p^{\dag}\hat{a}_p)+\frac{g'}{2}(\hat{a}_i\hat{a}_p+\hat{a}_i^{\dag}\hat{a}_p+\mathrm{H.c.}).
    \label{H_RWA}
\end{equation}
with the effective coupling
\begin{equation}
    g'=\tilde{g}\,\frac{\Omega_i^2-\Omega_p'^2}{\Omega_i'^2-\Omega_p'^2}\approx\tilde{g}\,\frac{\Omega_i^2}{\Omega_i'^2},
    \label{gprime}
\end{equation}
where $\Omega_i^2/\Omega_i'^2\approx0.34$ for our parameters. Physically, the drive modulates the curvature of the Coulomb potential, which enters both the coupling ($\tilde g$) and the ion frequency ($\delta\Omega_i'$). Thus, besides the direct exchange through $\tilde g$, phonons are exchanged by the static coupling $g_0$, off-resonant by $\omega_d$, with the frequency gap bridged by modulation $\delta\Omega_i'\cos(\omega_d t)$ of the ion frequency. Conceptually this is analogous to a Raman transition, with amplitude $\propto g_0\delta\Omega_i'/\omega_d$, and it interferes destructively with direct exchange.
 
Note that one could also move to a different frame, this time rotating with frequencies $\Omega_i'$ for the ion and $\Omega_p'$ for the nanoparticle operators, and assume $|g'|\ll\Omega_p'$ in order to also drop terms rotating with $2\Omega_p'$. This would lead to a pure beam-splitter Hamiltonian $\frac{g'}{2}(\hat{a}_i\hat{a}^\dag_p+\hat{a}_i^{\dag}\hat{a}_p)$. However, in Eq.~\eqref{H_RWA} we do not use this approximation and drop only the terms rotating at frequencies of order $\Omega_i'$. This provides a more accurate picture while still retaining an exactly solvable quadratic Hamiltonian.
 
Apart from the coherent dynamics, there are several sources of dissipation. The ion dissipation arises from continuous Doppler or sideband cooling, and heating due to photon recoil, electrical noise and off-resonant excitation. 
In the original frame of the Hamiltonian \eqref{H_full}, it can be described by the dissipator~\cite{Gupta2025}
\begin{equation}\label{dissipatorI}
    \mathcal{D}_i \left( \hat{\rho} \right)=\left( \Gamma_i + \gamma_i \right) \mathcal{L}_{\hat{a}_i,\hat{a}_i^\dagger} \left(\hat{\rho}\right) + \Gamma_i \mathcal{L}_{\hat{a}_i^\dagger,\hat{a}_i} \left(\hat{\rho}\right),
\end{equation}
where $\mathcal{L}_{\hat{A},\hat{B}} \left( \hat{\rho} \right) \equiv \hat{A} \hat{\rho} \hat{B} - \frac{1}{2} \{\hat{B} \hat{A},\hat{\rho}\}$ and $\Gamma_i$ is the ion heating rate. The sources of dissipation for the nanoparticle are collisions with residual gas molecules (with associated heating rate $\Gamma_{\rm gas}$) and effective trap displacements. Causes of the trap displacement noise (with rate $\Gamma_{\rm td}$) include voltage noise on the trap electrodes and mechanical vibrations~\cite{Gupta2025}. The dissipator for the nanoparticle is given by
\begin{multline}
    \mathcal{D}_p \left( \hat{\rho} \right)=\left( \Gamma_{\rm gas} + \gamma_{\rm gas} \right) \mathcal{L}_{\hat{a}_p,\hat{a}_p^\dagger} \left(\hat{\rho}\right) + \Gamma_{\rm gas} \mathcal{L}_{\hat{a}_p^\dagger,\hat{a}_p}\left(\hat{\rho}\right) \\- \Gamma_{\rm td} \comm{\hat{Q}_{p}}{\comm{\hat{Q}_{p}}{\hat{\rho}}},
    \label{Dp}
\end{multline}
where $\hat{Q}_p=(\hat{a}_p+\hat{a}^\dag_p)/\sqrt{2}$ is the dimensionless nanoparticle position quadrature. The gas term is the long-wavelength limit of the collisional master equation~\cite{Joos1985, Gallis1990, Hornberger2003, Vacchini2009}. It describes the first and second moments exactly, and is therefore sufficient for the Gaussian states discussed in Secs.~\ref{sec3} (cooling) and~\ref{sec4} (squeezing and entanglement). For the non-Gaussian states in Sec.~\ref{sec5}, we use the full collisional dissipator, derived in Appendix~\ref{A-GAS}, where we also show that nanoparticle decoherence due to blackbody radiation is negligible. In the frame displaced with $\hat{D}(t)$, the transformation of the dissipators yields additional effective driving terms that are canceled by choosing $z_{j,\rm cl}(t)$ that satisfies Eqs.~\eqref{di_cl} and~\eqref{dp_cl} with damping,  while moving to the rotating frame of the ion operators leaves the dissipator invariant. Finally, the full system dynamics are described by the master equation
\begin{equation}
    \dot{\hat{\rho}}=-\frac{\rm i}{\hbar}[\hat{H}_{\rm RWA}, \hat{\rho}]+\mathcal{D}_p(\hat{\rho})+\mathcal{D}_i(\hat{\rho}).
    \label{master-eq}
\end{equation}
 
The decoherence rates are estimated in Sec.~\ref{sec_exp}, and the complete list of system parameters is provided in Table~\ref{Table1} in Appendix~\ref{A-EP}. For the steady-state oscillation amplitude of $A_{\rm cl}\approx 1.5\,\mathrm{\mu m}$ [Fig.~\ref{fig2}], we obtain a strong coupling $|g'|/2\pi\approx 320\,\rm Hz\approx 0.3\,(\Omega_p'/2\pi)$, with $\Delta_d/2\pi\approx1.5\,\rm kHz$. The second-order drive effects also generate a parametric term $\chi_i(\hat{a}_i^2+(\hat{a}_i^\dagger)^2)$ in Eq.~\eqref{H_RWA}, with rate $|\chi_i|/2\pi\approx 0.1\,\rm kHz$ for the parameters of Fig.~\ref{fig2} and using the compensation tone at $2\omega_d$. Second-order drive effects including the parametric resonance are treated in more detail in Appendix~\ref{A-PR}. Note that the results of Secs.~\ref{sec3} and~\ref{sec4}, based on Eq.~\eqref{H_RWA} and its two-tone generalization, have been verified against an exact numerical solution of the full model, as described in Appendix~\ref{A-FL}.

\section{Cooling}\label{sec3}
 
Let us assess the impact of the resonant ion-nanoparticle coupling engineered above on the sympathetic cooling of a nanoparticle by a continuously cooled ion.
Since the system is described by a quadratic master equation, Gaussianity is preserved during state evolution and the state can be completely characterized through the expectation values of ladder operators (which are zero in the displaced frame) and their covariance matrix. The covariance matrix is defined as $\bar{\sigma}_{\lambda\mu}=\frac{1}{2}\langle \hat{B}_\lambda \hat{B}_\mu+\hat{B}_\mu \hat{B}_\lambda \rangle-\langle \hat{B}_\lambda\rangle\langle \hat{B}_\mu\rangle$, where $\hat{\bm{B}}=[\hat{a}_i, \hat{a}^\dag_i, \hat{a}_p, \hat{a}^\dag_p]^T$ is the vector of ladder operators. The system's dynamics are obtained by solving the linear system of differential equations
\begin{equation}\label{covmatstatic}
    \dv{}{t}\bar\sigma = \bar{A}\bar\sigma+\bar\sigma\bar{A}^T+\bar{C},
\end{equation}
with explicit forms of matrices $\bar{A}$ and $\bar{C}$ given in Appendix~\ref{A-GD}. The steady state is obtained from the condition $\dv{}{t}\bar\sigma_{\rm ss}=0$. To assess the cooling performance, we compute the nanoparticle phonon occupation from the covariance matrix as $\langle\hat{n}_p\rangle=\langle \hat{a}^\dag_p\hat{a}_p\rangle=\bar{\sigma}_{34}-1/2$, and the nanoparticle state purity as $\det(2\bar{\Sigma}_p)^{-1/2}$~\footnote{With the full collisional dissipator, this expression is not exact because the state is not exactly Gaussian. The purity estimate from the covariance matrix is conservative in our case. For details, see Appendix~\ref{A-GAS}.}, where  $\bar{\Sigma}_p$ is the $2\times2$ covariance matrix of the nanoparticle quadratures $\hat{Q}_p$ and $\hat{\Pi}_p=(\hat{a}_p-\hat{a}^\dag_p)/(\mathrm{i}\sqrt{2})$, normalized such that $\bar{\Sigma}_p=\mathbb{1}/2$ in the vacuum state. The exact results can be well approximated by
\begin{equation}
    \langle \hat{n}_p\rangle_{\rm ss}\approx \frac{(\Gamma_i+\Gamma_{\rm gas}+\Gamma_{\rm td})g'^2+(\Gamma_{\rm gas}+\Gamma_{\rm td})\gamma_i^2}{\gamma_ig'^2},
    \label{n_ss}
\end{equation}
which is derived from the analytical solution under the assumption of negligible squeezing and two-mode correlations in the steady state, i.e., $\langle\hat{a}_i^2\rangle\approx\langle\hat{a}_p^2\rangle\approx\langle\hat{a}_i\hat{a}_p\rangle\approx0$. This expression reaches a minimum of
\begin{equation}
    \langle \hat{n}_{p}\rangle_{\rm ss, opt}=\frac{2\sqrt{(\Gamma_i+\Gamma_{\rm gas}+\Gamma_{\rm td})(\Gamma_{\rm gas}+\Gamma_{\rm td})}}{|g'|}
    \label{n_ss_min}
\end{equation}
at the optimal ion cooling rate
\begin{equation}
    \gamma_{i, \rm opt}=|g'|\sqrt{(\Gamma_i+\Gamma_{\rm gas}+\Gamma_{\rm td})/(\Gamma_{\rm gas}+\Gamma_{\rm td})}.
    \label{gamma_i_opt}
\end{equation}
The timescale at which the steady-state occupation is reached is given by $(2|\max_k(\mathrm{Re}\lambda_k)|)^{-1}$, where $\lambda_k$ $(k=1,...4)$ are the eigenvalues of $\bar{A}$~\footnote{System's stability requires all eigenvalues of $\bar{A}$ to have negative real parts, and the one with the smallest $|\mathrm{Re}\lambda_j|$ dictates the decay rate.}. The associated rate can be approximated as
\begin{equation}
    \max_k(\mathrm{Re}\lambda_k)\approx-\frac{\gamma_i}{4}+\mathrm{Re}\sqrt{\frac{\gamma_i^2}{16}-\Omega_p'^2+\Omega_p'\sqrt{g'^2-\frac{\gamma_i^2}{4}}}.
    \label{cooling_rate}
\end{equation}
This expression assumes $\gamma_{\rm gas}\ll \Omega_j',\gamma_i$ and becomes exact for $\gamma_{\rm gas}\to0$.

\begin{figure}[t]
	\centering
    \includegraphics[width=86mm]{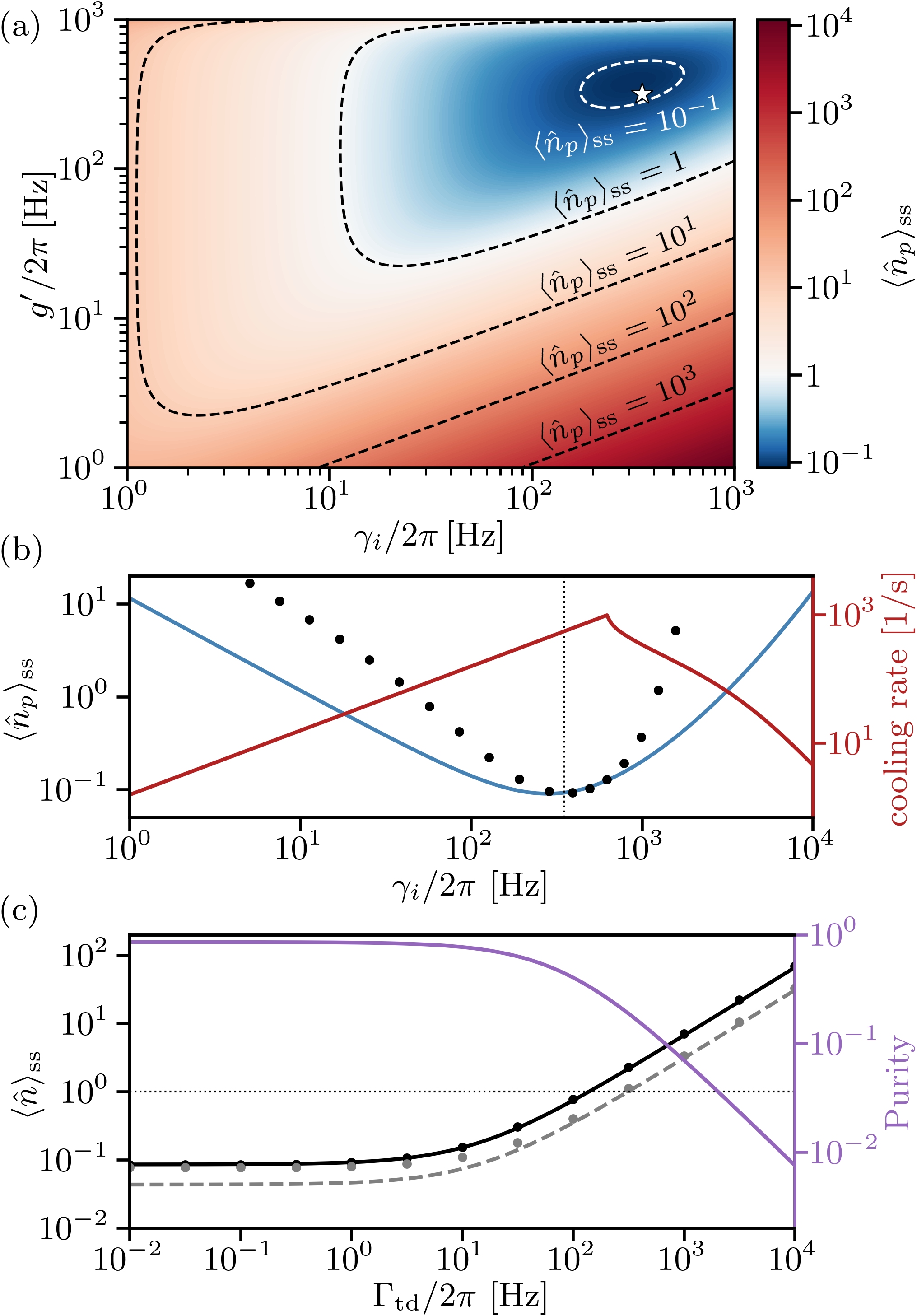}
    \caption{\textbf{Cooling.} (a) Steady-state nanoparticle phonon occupation  $\langle\hat{n}_p\rangle_{\rm ss}$ as a function of ion sideband cooling rate $\gamma_i$ and coupling strength $g'$. The star marks the parameters used in Fig.~\ref{fig2}. (b) Steady-state nanoparticle phonon occupation $\langle\hat{n}_p\rangle_{\rm ss}$ (blue) and nanoparticle cooling rate (Eq.~\eqref{cooling_rate}, red) as a function of $\gamma_i$ for a fixed $g'/2\pi=320\,\rm Hz$. The black dots indicate the time-averaged exact numerical solution for $\langle\hat{n}_p\rangle_{\rm ss}$ using the drive parameters of Fig.~\ref{fig2} (see Appendix~\ref{A-FL} for details). Away from the optimum, the dots deviate from the blue line because $\gamma_i$ also modifies the classical motion, and thereby also $g'$ and the resonance condition; optimizing the AC drive parameters for each $\gamma_i$ should suppress this deviation. The dotted vertical line marks $\gamma_i/2\pi=350\,\mathrm{Hz}\approx|g'|/2\pi$. (c) Steady-state phonon occupation number of nanoparticle (black line) and ion (dashed grey line) as a function of $\Gamma_{\rm td}$ for $g'/2\pi=320\,\rm Hz$ and the optimal $\gamma_i/2\pi=350\,\rm Hz$; dots indicate the corresponding time-averaged exact solution. Motional purity of the nanoparticle is shown in purple. All results are obtained for parameters in Table~\ref{Table1}.}
    \label{fig3}
\end{figure}
 
The results obtained for steady-state phonon occupations, purities and cooling rates are shown in Fig.~\ref{fig3}. Entering the ground-state cooling regime $\langle \hat{n}_p\rangle_{\rm ss}<1$ is possible with continuous sideband cooling of the ion and $\Gamma_{\rm td}/2\pi\lesssim100\,\rm Hz$, well within reach of experiments according to our estimates (see Sec.~\ref{sec_exp}). We observe that there is an optimal cooling rate $\gamma_i\approx |g'|$ as predicted above. This can be understood intuitively: for $\gamma_i\gg |g'|$ the ion motion is overdamped, which suppresses its exchange of energy with the nanoparticle, and the effective cooling rate scales as $g'^2/\gamma_i$. Conceptually, the cooling mechanism is identical to resolved-sideband cooling in optomechanics, with the ion motion acting as the energy sink instead of the optical mode. The optimal cooling rate is analogous to the known result in optomechanics~\cite{Wilson-Rae2008}. The relaxation time is minimized for $\gamma_i\approx 2|g'|$, and typically the system approaches its steady state on a timescale of 1--10~ms.
 
As seen in Fig.~\ref{fig3}(a), there is also an optimal coupling rate $g'$. Generally, a value of $|g'|$ that is too large is suboptimal for nanoparticle cooling because it enhances the two-mode squeezing term that induces heating. With our parameters, $\langle\hat{n}_p\rangle_{\rm ss}$ is minimized for $|g'|\approx 2\pi\times 0.38\,\rm kHz$, which requires $A_{\rm cl}/D_0\approx0.1$ due to the renormalization factor in Eq.~\eqref{gprime}. The parameters of Fig.~\ref{fig2} give $|g'|/2\pi\approx 0.32\,\rm kHz$ (the value used in Fig.~\ref{fig3}(b)), close to this optimum; these parameters also minimize $\langle\hat{n}_p\rangle_{\rm ss}$ in the exact solution of the full model, where larger amplitudes are limited by the nonlinear response of the ion (Appendix~\ref{A-FL}). For $\Gamma_{\rm td}/2\pi\lesssim1$~Hz, which we consider reachable in near-term experiments based on considerations in Sec.~\ref{sec_exp}, it is possible to achieve cooling deep into the ground-state regime, $\langle\hat{n}_p\rangle_{\rm ss}\approx 0.09$, in agreement with the exact solution of the full model [dots in Fig.~\ref{fig3}(b),(c)]. Note that the exact and approximate solutions for the ion differ slightly due to the small residual parametric drive that the compensation voltage is not able to fully cancel (see  Appendix~\ref{A-PR} for details).
Without the compensation tone, this parametric term would become more relevant and limit the cooling to $\langle\hat{n}_p\rangle_{\rm ss}\approx 0.16$, reached for $U_d\approx50\,\rm\mu V$ at a smaller drive frequency, where $A_{\rm cl}\approx1.0\,\rm\mu m$ and $|g'|/2\pi\approx0.21\,\rm kHz$ (Appendix~\ref{A-FL}). Increasing $|g'|$ further does not help in this case, as it amplifies the ion parametric drive; in addition, the Hamiltonian~\eqref{H_RWA} becomes unstable for $|g'|>\Omega_p'$ due to the two-mode squeezing term.
 
\begin{figure*}
	\centering
    \includegraphics[width=178mm]{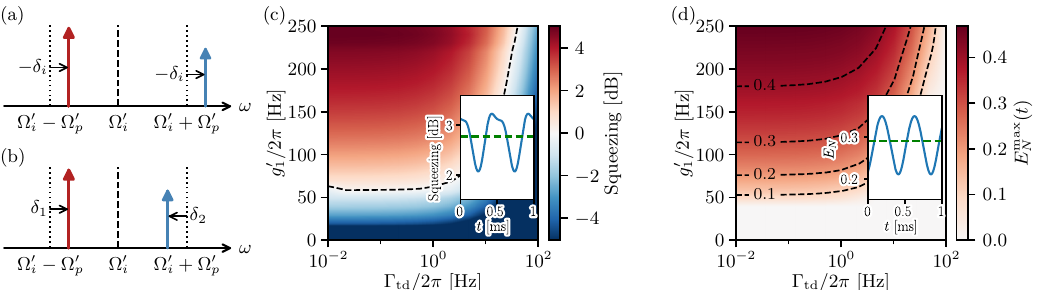}
    \caption{\textbf{Squeezing and entanglement.} 
    (a-b) Illustration of the two-tone driving schemes that generates squeezing (a) and entanglement (b). (c) Maximum nanoparticle squeezing (in dB) as a function of $\Gamma_{\rm td}$ and $g_1'$, obtained by optimizing over $\gamma_i$, $g_2'$, $\delta_i$ and the residual parametric term, for the worst case of realistic parameter errors (see text for details). Inset: time dependence of the squeezing for $\Gamma_{\rm td}/2\pi=1$~Hz and $g_1'/2\pi=150$~Hz; the dashed green line is the prediction obtained with the time-independent Hamiltonian~\eqref{H_st} only, and the trap-displacement dissipator averaged over the rotation, i.e., the first line in Eq.~\eqref{D_td_rot}. (d) Maximum ion-nanoparticle logarithmic negativity $E_N$ as a function of $\Gamma_{\rm td}$ and $g_1'$, obtained by optimizing over $\gamma_i$, $\delta$, $\delta_i$ and $g_2'$, for the worst case of realistic parameter errors. Inset: time dependence of $E_N$ for $\Gamma_{\rm td}/2\pi=1$~Hz and $g_1'/2\pi=150$~Hz; the dashed green line is the time-independent prediction. Other parameters used to obtain the results are taken from Table~\ref{Table1}.}
    \label{fig4}
\end{figure*}
 
\section{Gaussian state preparation}\label{sec4}

The ability to reach a high-purity motional state for both ion and nanoparticle opens the door for non-classical state preparation. In this section we show how, extending the above idea to more general AC drives, one can prepare non-classical Gaussian motional states of the ion-nanoparticle system.

\subsection{Two-tone driving}
 
The Hamiltonian~\eqref{H_RWA} contains a two-mode squeezing term $\hat{a}_i\hat{a}_p+\mathrm{H.c.}$ which can be used to prepare squeezed motional states of the nanoparticle and entanglement between the ion and the nanoparticle motion. This term can be resonantly enhanced by driving the classical motion at frequency $\omega_d=\Omega_i'+\Omega_p'+\Delta_d$, in the same way as the beam-splitter term is enhanced at $\omega_d=\Omega_i'-\Omega_p'+\Delta_d$. Driving only the two-mode squeezing term leads to unstable dynamics, but the instability can be avoided if both two-mode squeezing and beam-splitter terms are driven simultaneously, by applying a two-tone AC voltage with frequencies $\omega_{d1, d2}$ to the endcap electrodes:
\begin{multline}
    \phi_{\rm drive}\to -\frac{R_z}{d_z}\Big(U_{d1}\cos(\omega_{d1}t)+U_{d2}\cos(\omega_{d2}t)\Big)+\\+U_d^{(2)}\cos((\omega_{d1}+\omega_{d2})t+\varphi_2),
    \label{phi2tone}
\end{multline}
where the tone at $\omega_{d1}+\omega_{d2}$ cancels the parametric resonance terms. Compensation tones at $2\omega_{d1}$ and $2\omega_{d2}$ are not required in this case: as in the single-tone case, the corresponding parametric terms are detuned by $\approx2\Omega_p'$, and their amplitudes are smaller since each tone drives a smaller distance modulation; we verified numerically that canceling them does not improve the squeezing and only slightly increases the entanglement (Appendix~\ref{A-FL}). Following similar steps as above, this leads to the Hamiltonian
\begin{multline}
    \hat{H}_2'/\hbar=\Omega_i'\hat{a}_i^\dagger\hat{a}_i+\Omega_p'\hat{a}_p^\dagger\hat{a}_p\\+\Big(g_0+\tilde{g}_1\cos(\omega_{d1}t)+\tilde{g}_2\cos(\omega_{d2}t)\Big)\Big(\hat{a}_i+\hat{a}_i^\dagger\Big)\Big(\hat{a}_p+\hat{a}_p^\dagger\Big),
\end{multline}
where $\tilde{g}_{1,2}=-3(A_{{\rm cl},1,2}/D_0)g_0$, with $A_{{\rm cl},1,2}$ the amplitudes of the classical distance oscillations at $\omega_{d1,2}$ (we omit the frequency modulation terms of Eq.~\eqref{H_ab} for brevity). The drive frequencies are chosen as $\omega_{d1}=\Omega_i'-\Omega_p'+\delta_1$ and $\omega_{d2}=\Omega_i'+\Omega_p'-\delta_2$ (see Fig.~\ref{fig4}(a),(b)), where $\Omega_{i,p}'$ include the drive-induced frequency shifts; the importance of the detunings $\delta_{1,2}$ will be explained shortly. We transform to the frame rotating with $(\omega_{d1}+\omega_{d2})/2=\Omega_i'-\delta_i$ for ion operators and with $(\omega_{d2}-\omega_{d1})/2=\Omega_p'-\delta$ for particle operators, where $\delta=(\delta_1+\delta_2)/2$ and $\delta_i=(\delta_2-\delta_1)/2$, and neglect terms rotating at $\sim \Omega_i'$, taking into account their second-order contributions as in Sec.~\ref{sec2b}. This renormalizes the couplings to $g_{1,2}'$, given by Eq.~\eqref{gprime} with $\tilde{g}\to\tilde{g}_{1,2}$, and results in a Hamiltonian
\begin{equation}
    \hat{H}_{\rm RWA, 2}=\hat{H}_{\rm st}+\hat{H}_{\rm slow}(t).
    \label{H_RWA2}
\end{equation}
Here,
\begin{equation}
    \hat{H}_{\rm st}/\hbar=\delta\hat{a}_p^\dag \hat{a}_p+\delta_i\hat{a}_i^\dag \hat{a}_i+\frac{g_1'}{2}(\hat{a}_i\hat{a}_p^\dag+\hat{a}_i^\dag\hat{a}_p)+\frac{g_2'}{2}(\hat{a}_i\hat{a}_p+\hat{a}_i^\dag\hat{a}_p^\dag).
    \label{H_st}
\end{equation}
is the time-independent part of the Hamiltonian, while
\begin{multline}
    \hat{H}_{\rm slow}/\hbar=\frac{g_1'}{2}\Big(\hat a_i\hat a_p  e^{-\mathrm{i}2(\Omega_p'-\delta)t}+\mathrm{H.c.}\Big)\\+\frac{g_2'}{2}\Big(\hat a_i^\dagger\hat a_p e^{-\mathrm{i}2(\Omega_p'-\delta)t}+\mathrm{H.c.}\Big).
    \label{H_slow}
\end{multline}
is the time-dependent part that contains terms slowly rotating at $\sim2\Omega_p'$. Unlike the single-tone driving case, here we cannot eliminate all time dependence by moving to an appropriately rotating frame~\footnote{For weak coupling $|g_{1,2}'|\ll\Omega_p'$, we could make an additional RWA and neglect $\hat{H}_{\rm slow}$, but we are interested in the strong-coupling regime where more squeezing and entanglement can be achieved.}. We restrict ourselves to the regime $|g_2'|<|g_1'|$, where the static Hamiltonian~\eqref{H_st} is dynamically stable for $\delta=0$. Other system parameters are the same as in previous sections.
 
Two further points are important in this frame. First, the trap-displacement dissipator in Eq.~\eqref{master-eq} becomes time-dependent, since the lab-frame position quadrature transforms as $\hat{Q}_p\to \hat{Q}_p\cos\theta+\hat{\Pi}_p\sin\theta$ with $\theta=(\Omega_p'-\delta)t$ and $\hat{\Pi}_p=\left(\hat{a}_p-\hat{a}_p^\dagger\right)/(\rm i\sqrt{2})$ the dimensionless nanoparticle momentum. It becomes
\begin{multline}
    -\frac{\Gamma_{\rm td}}{2}\Big(\comm{\hat{Q}_p}{\comm{\hat{Q}_p}{\hat{\rho}}}+\comm{\hat{\Pi}_p}{\comm{\hat{\Pi}_p}{\hat{\rho}}}\Big)\\
    -\frac{\Gamma_{\rm td}}{2}\cos2\theta\Big(\comm{\hat{Q}_p}{\comm{\hat{Q}_p}{\hat{\rho}}}-\comm{\hat{\Pi}_p}{\comm{\hat{\Pi}_p}{\hat{\rho}}}\Big)\\
    -\frac{\Gamma_{\rm td}}{2}\sin2\theta\Big(\comm{\hat{Q}_p}{\comm{\hat{\Pi}_p}{\hat{\rho}}}+\comm{\hat{\Pi}_p}{\comm{\hat{Q}_p}{\hat{\rho}}}\Big).
    \label{D_td_rot}
\end{multline}
The first, time-averaged term equals $\Gamma_{\rm td}\big(\mathcal{L}_{\hat{a}_p,\hat{a}_p^\dagger}+\mathcal{L}_{\hat{a}_p^\dagger,\hat{a}_p}\big)(\hat{\rho})$, i.e., it heats both quadratures equally with the same total rate $\Gamma_{\rm td}$, while the other terms oscillate at $2(\Omega_p'-\delta)$. We keep all terms in the numerical calculations. Second, the second-order drive effects also produce an ion parametric term $\chi_{12}(\hat{a}_i^2+(\hat{a}_i^\dagger)^2)$ driven at $\omega_{d1}+\omega_{d2}$, which is exactly resonant in this frame and makes the ion motion unstable unless $\gamma_i\gtrsim4|\chi_{12}|$, with $|\chi_{12}|\approx|g_1'g_2'|/(2\pi\times 125\,{\rm Hz})$ (Appendix~\ref{A-PR}). In analogy to the single-tone case discussed in previous sections, the compensation tone at $\omega_{d1}+\omega_{d2}$ in Eq.~\eqref{phi2tone} is chosen to nullify the corresponding Fourier component of $D_{\rm cl}(t)$, which cancels this parametric term. The cancellation can also be made incomplete on purpose: by changing the amplitude and phase of the compensation tone, we leave a residual ion parametric term $\chi(e^{\mathrm{i}\varphi}\hat{a}_i^2+\mathrm{H.c.})$, with the rate $\chi$ and phase $\varphi$ set by the compensation tone. This term squeezes the ion motion, and we use $\chi$ and $\varphi$ as additional parameters to optimize the nanoparticle squeezing in the next subsection.

\subsection{Squeezing}
 
We first focus on the case $\delta=\delta_i=0$, where the static Hamiltonian can be written 
as
\begin{equation}
    \hat H_{\rm st}(\delta=0)/\hbar=\mathcal{G}\left(\hat a_i \hat b_p^\dagger+\hat a_i^\dagger \hat b_p\right),
    \label{H_st_sq}
\end{equation}
in terms of a squeezed nanoparticle mode
\begin{equation}
    \hat b_p=\cosh r\,\hat a_p+\sinh r\,\hat a_p^\dagger ,
    \quad 
    \tanh r=\frac{g_2'}{g_1'} ,
    \label{b_mode}
\end{equation}
with $\mathcal{G}=(1/2)\sqrt{(g_1')^2-(g_2')^2}$.
The coherent dynamics thus describe a beam-splitter coupling between the ion and the squeezed mode $\hat{b}_p$, whose vacuum is the position-squeezed state $\ket{S(r)_p}=\hat{S}_p(r)\ket{0_p}$ of the nanoparticle, where $\hat{S}_p(r)=\exp\Big(\frac{r}{2}(\hat{a}_p^2-(\hat{a}_p^\dagger)^2)\Big)$. It is straightforward to show that, in the simplified case where $\Gamma_i=0$ and in the absence of nanoparticle decoherence, the steady state  satisfying $\dot{\hat{\rho}}_{\rm ss}=-(\mathrm{i}/\hbar)[\hat{H}_{\rm st},\hat{\rho}_{\rm ss}]+\mathcal{D}_i( \hat{\rho}_{\rm ss})=0$ is the pure state $\ket{0_i}\ket{S(r)_p}$, i.e., a vacuum-squeezed nanoparticle motional state. The amount of squeezing, i.e., the ratio of minimum phase-space variance to its vacuum value, is given by $2 \lambda_{\rm min}(\bar{\Sigma}_p)=\exp(-2 r)$, with $\lambda_{\rm min}(\bar{\Sigma}_p)$ the minimum eigenvalue of the nanoparticle covariance matrix defined in the previous section. The squeezing parameter $r$, defined in Eq.~\eqref{b_mode}, diverges as the system approaches instability for $g_2'/g_1'\to 1$, but in practice experimental constraints would prevent operating too close to instability. Assuming $g_2'/g_1'=0.9$, we get $r=1.47$, or equivalently 12.8 dB of squeezing, where squeezing in dB is defined as $-10\log_{10}[2\lambda_{\rm min}(\bar{\Sigma}_p)]$.
 
The maximum achievable squeezing in the presence of all (time-dependent) dissipators and the full time-dependent Hamiltonian $\hat{H}_{\rm RWA, 2}$ is shown in Fig.~\ref{fig4}(c), as a function of $g_1'$ and $\Gamma_{\rm td}$. The dissipation neglected in the estimation above considerably reduces the state purity and the achievable squeezing to $\approx4.7$~dB for $g_1'/2\pi=250$~Hz and $\Gamma_{\rm td}/2\pi=1$~Hz. Position squeezing below the zero-point fluctuations is possible for $\Gamma_{\rm td}/2\pi\lesssim40$~Hz. Since the dissipator rotates at frequency $\Omega_p'-\delta$, the trap noise heats both nanoparticle quadratures on average (first line in Eq.~\eqref{D_td_rot}), and the squeezed quadrature is only protected by the cooling of the mode $\hat{b}_p$ at the rate $\sim4\mathcal{G}^2/\gamma_i$, which is reduced for $g_2'\to g_1'$. 

To obtain the results, we numerically calculated the time-periodic steady state from the monodromy matrix (Floquet theory) by solving a discrete Lyapunov equation (see Appendix~\ref{A-FL}), and maximized $\max_t r(t)$ over $\gamma_i$, $g_2'$ (with the constraint $|g_2'|<0.9|g_1'|$), $\delta_i$, and the residual parametric term $\chi$, $\varphi$, keeping $\delta=0$. Since the optimum lies close to the boundary of dynamical stability, we maximized the worst case over realistic fluctuations of the parameters: $\pm1$~Hz in the nanoparticle frequency, which is stable to a few mHz in Paul traps~\cite{Dania2024}, $\pm5$~Hz in the ion frequency, i.e., $4\times10^{-6}\,\Omega_i'$, comparable to the secular-frequency stability achieved in ion traps~\cite{Johnson2016,Zhang2022}, and $\pm0.5\%$ and $\pm5\%$ in $g_2'/g_1'$ and $\chi$, respectively, each varied separately. The resulting optima lie at the edge of this tolerance range: for example, at the squeezing optimum for $g_1'/2\pi=150$~Hz, the dynamics becomes unstable for a nanoparticle-frequency error slightly beyond $1$~Hz, and some combinations of simultaneous errors within the tolerances are unstable. Requiring stability for all combinations of errors and maximizing their worst case reduces the results only slightly, e.g., the squeezing from $3.2$ to $3.0$~dB at $g_1'/2\pi=150$~Hz and from $4.7$ to $4.6$~dB at $g_1'/2\pi=250$~Hz ($\Gamma_{\rm td}/2\pi=1$~Hz).
 
The time-dependent part of the Hamiltonian~\eqref{H_slow} leads to oscillations in the amount of squeezing (inset of Fig.~\ref{fig4}(c)) which grow in magnitude as $|g_{1,2}'|$ increase. It also tilts the squeezed quadrature relative to the position quadrature. Typical phase-space tilt angles are a few degrees and grow with the coupling strengths. We verified these results against the exact numerical solution of the two-tone problem (Appendix~\ref{A-FL}), which gives $3.0$~dB compared to $3.2$~dB at $g_1'/2\pi=150\,\rm Hz$. For stronger two-tone driving, the disagreement becomes larger due to neglected higher-order terms in $A_{\rm cl}/D_0$, with smaller gains in squeezing than predicted by the effective model. Eventually the full driven system becomes classically unstable for $g_1'/2\pi\gtrsim250\,\rm Hz$, which sets the upper limit of the maps in Fig.~\ref{fig4}(c),(d).
 
\begin{figure*}
	\centering
    \includegraphics[width=178mm]{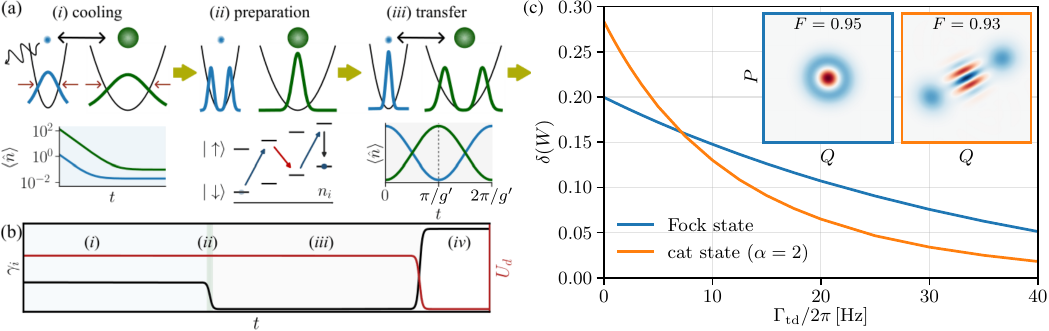}
    \caption{\textbf{State transfer.} (a) Illustration of the state transfer protocol: step (\textit{i}) is ground-state cooling, step (\textit{ii}) is ion motional state preparation via the qubit, step (\textit{iii}) is state transfer from the ion to the nanoparticle. (b) Schematic time dependence of the ion cooling rate $\gamma_i$ (sideband cooling in step (\textit{i}), Doppler cooling in step (\textit{iv})) and driving voltage $U_d$ used for implementing the protocol. After we decouple the ion and the nanoparticle in step (\textit{iv}), the nanoparticle is left in a nonclassical state. (c) Wigner negativity volumes and Wigner functions (insets, for $\Gamma_{\rm td}/2\pi=1$~Hz), with the state fidelities $F=\bra{\psi}\hat{\rho}_p\ket{\psi}$) after the transfer, i.e., right after step (\textit{iii}), for the single-phonon Fock state and a Schr\"{o}dinger cat state with $\alpha=2$. Parameters used to obtain the results are taken from Table~\ref{Table1}.}
    \label{fig5}
\end{figure*}
 
\subsection{Entanglement}
 
When $\delta\ne 0$, the term $\delta\hat{a}^\dagger_p\hat{a}_p$ breaks the symmetric form of the Hamiltonian~\eqref{H_st} and causes its steady state to become non-separable (i.e., entangled). In the squeezed-mode basis, the detuning term becomes
\begin{equation}
    \delta \hat a_p^\dagger \hat a_p=\delta \cosh(2r)\,\hat b_p^\dagger \hat b_p-\frac{\delta}{2}\sinh(2r)
    \left(\hat b_p^2+\hat b_p^{\dagger 2}\right)
\end{equation}
up to an irrelevant constant.
The second term is a parametric squeezing term for $\hat b_p$; together with the beam-splitter interaction with the ion in Eq.~\eqref{H_st_sq}, it correlates ion and nanoparticle motion, making the steady-state non-separable. Entanglement for a two-mode Gaussian state can be quantified through the logarithmic negativity, defined as $E_N = \max(0, -\ln (2\tilde{\nu}_-))$, where $\tilde{\nu}_-$ is the smallest symplectic eigenvalue of the partially transposed covariance matrix $\tilde{\sigma}$ of the quadratures of both modes (normalized as $\bar{\Sigma}_p$), obtained by applying the time-reversal operation to the quadratures of one subsystem. The Peres--Horodecki criterion~\cite{Peres1996, Horodecki1996} for Gaussian states establishes that the system is entangled if and only if $\tilde{\nu}_- < 1/2$. 
 
Results for the maximum entanglement over time, $\max_t E_N(t)$, optimized over $\delta$, $\delta_i$, $\gamma_i$ and $|g_2'|<0.9|g_1'|$ (with $\chi=0$) are shown in Fig.~\ref{fig4}(d) as a function of $g_1'$ and $\Gamma_{\rm td}$. The optimization was done analogously to the optimization in the previous subsection, including the parameter fluctuations. The optimum keeps the beam-splitter tone close to resonance, $|\delta_1|\lesssim0.2|g_1'|$, while the two-mode squeezing tone is detuned by $\delta_2\approx-(1.4$--$1.5)g_1'$. There are significant oscillations in $E_N$ around the value predicted with the time-independent Hamiltonian $\hat{H}_{\rm st}$ already for $|g_1'|/\Omega_p'\sim 0.2$, and again the amplitude of oscillations increases further with the coupling strengths. The frequency of these oscillations also increases with coupling strengths, approaching $2\Omega_p'$ as $|g_1'|$ becomes comparable to $\Omega_p'$. For realistic values of $\Gamma_{\rm td}$, it is possible to get time maxima of $E_N\approx0.46$ at $g_1'/2\pi=250\,\rm Hz$ (a slight overestimate compared to the exact solution, see Appendix~\ref{A-FL}). These values are conservative, since the approximate Gaussian dissipator used to describe gas collisions, Eq.~\eqref{Dp}, reduces the entanglement more than the full collisional model (as shown in Appendix~\ref{A-GAS}, the full model predicts $E_N\approx 0.56$ instead of 0.46 at $g_1'/2\pi=250\,\rm Hz$). We note that, although the entanglement is not a steady-state property, driving makes the oscillations in $E_N$ persistent. For smaller coupling strengths, steady-state entanglement can be achieved to a good approximation at the expense of smaller $E_N$.

The above results for both squeezing and entanglement are a proof of principle. Higher values of squeezing (one- and two-mode) could be reached in several ways, such as with direct modulation of the trapping potentials or even driving the ion-nanoparticle system into a dynamically unstable regime in a controlled way~\cite{Kustura2022}. Our approach could also be extended to prepare entangled states of two particles entangled by a mediating ion.

\section{State Transfer}\label{sec5}
 
So far we have only considered the dynamics of the coupled bosonic modes of the ion and the nanoparticle, but one of this platform's core advantages is that the electronic transition of the ion provides an intrinsic two-level degree of freedom. This can be exploited to prepare arbitrary states of the ion motion and transfer them to the nanoparticle, as done in other spin-mechanical systems~\cite{OConnell2010, Kienzler2016, Johnson2017, Fluhmann2019, Bild2023}. 
Specifically, in the single-tone drive configuration of Secs.~\ref{sec2}-\ref{sec3}, and in the regime $\vert g'\vert \ll \Omega_p'$, the Hamiltonian~\eqref{H_RWA} is dominated by the beam-splitter interaction, which swaps the motional states of the ion and the nanoparticle after $t_{\rm swap}=\pi/|g'|$. 
The ion can be prepared in a non-Gaussian motional state using standard trapped-ion techniques~\cite{Meekhof1996, Monroe1996}, e.g., a Fock state $\vert 1_i\rangle$ can be prepared from the ground state $\vert 0_i\rangle$ by driving the blue sideband of the qubit to map the state $\ket{\downarrow, 0_i}$ to $\ket{\uparrow, 1_i}$, and then resetting the qubit to the ground state. In this section we propose a protocol to realize this state transfer in the ion-nanoparticle system.
 
The protocol is illustrated in Fig.~\ref{fig5}(a-b) and consists of the following steps:
 
(\textit{i}) Cooling the ion and the nanoparticle to the ground state, $\langle \hat{n}_{i,p}\rangle \ll 1$, as outlined in Sec.~\ref{sec3}.
 
(\textit{ii}) Ion state preparation using the internal ion levels, assumed instantaneous on the timescale of the nanoparticle motion. This is justified because the ion's motional states can be prepared in a few tens of microseconds~\cite{Monroe1996}, which is much shorter than the swap time $t_{\rm swap}=\pi/|g'|\sim\,$ms. During this step ion cooling is turned off to avoid motional decoherence.
 
(\textit{iii}) State transfer, achieved after the interaction time $t_{\rm swap}\approx\pi/|g'|$. During this step ion cooling remains off so that the motional dynamics are dominated by coherent interaction, and the coherence of the non-Gaussian state is preserved. To achieve this we need $\Omega_p'\gg|g'|\gg \Gamma_{i},\Gamma_{\rm td},\gamma_c$ with $\gamma_c$ the gas-molecule collision rate (see Appendix~\ref{A-GAS}), i.e., a coupling rate $g'$ larger than the decoherence, but not so large that the beam-splitter term stops being dominant over the two-mode squeezing term. Since the cooling force vanishes on the classical trajectory, which is mainly controlled by the drive (Sec.~\ref{sec_exp}), switching off the cooling leaves the classical motion and thus $g'$ unchanged from the values in Sec.~\ref{sec2}.
 
(\textit{iv}) Turning off the ion-nanoparticle interaction to prevent the motional state from being transferred back to the ion. Note that turning off the voltage modulation $\phi_{\rm drive}$ is not enough, because the interaction is created by the classical motion and thus remains active during the time it takes for ion and nanoparticle to ring down to their static equilibrium positions. To shut down the interaction quickly, we propose to simultaneously switch off the drive and damp the classical motion of the ion with a short Doppler-cooling pulse without the feed-forward (see Sec.~\ref{sec_exp} for details), which takes tens of microseconds, much less than $t_{\rm swap}$.
 
In Fig.~\ref{fig5}(c) we show the predictions of this protocol for the preparation of the single-phonon Fock state and a Schr\"{o}dinger cat state of the nanoparticle motion. We plot the Wigner negativity volume, defined from the Wigner function as $\delta(W)=\int_{W(x,p)<0}|W(x,p)|dxdp$, right after step (\textit{iii}) as a function of the nanoparticle trap noise $\Gamma_{\rm td}$. The insets show the Wigner functions of the nanoparticle state after step (\textit{iii}) for $\Gamma_{\rm td}/2\pi=1$~Hz. Our results were obtained by numerically solving the master equation~\eqref{master-eq} with the Hamiltonian~\eqref{H_RWA} supplemented by the cubic term~\eqref{H3RWA} of Appendix~\ref{A-HO}, and without the gas collision term. Gas collisions are then included as described in Appendix~\ref{A-GAS}: a collision occurs during the transfer only with probability $1-e^{-\gamma_ct_{\rm swap}}\approx3\times10^{-3}$, and it kicks the nanoparticle far from the target state, so we multiply the fidelity and the Wigner negativity volume of the simulated state by $e^{-\gamma_ct_{\rm swap}}$. We used the initial states $\ket{1_i}\bra{1_i}\otimes \hat{\rho}_{p, \rm ss}$ and $\mathcal{N}^{-1}(\ket{\alpha_i}+\ket{-\alpha_i})(\bra{\alpha_i}+\bra{-\alpha_i})\otimes \hat{\rho}_{p, \rm ss}$, with $\mathcal{N}=2(1+e^{-2|\alpha|^2})$. The state $\hat{\rho}_{p, \rm ss}$ is the nanoparticle Gaussian state with $\langle \hat{n}_p\rangle_{\rm ss}$ obtained from the cooling method in Sec.~\ref{sec3} for the chosen system parameters. For $\Gamma_{\rm td}/2\pi=1$~Hz, fidelities $F=\bra{\psi}\hat{\rho}_p\ket{\psi}$ with the target state $\ket{\psi}$ (up to a phase rotation) of 0.95 and 0.93 can be obtained for generation of Fock and cat states, respectively. After the state transfer, the nanoparticle will remain in the non-Gaussian state for as long as the decoherence allows. For $\Gamma_{\rm td}/2\pi=1$~Hz, Wigner negativity persists for tens of mechanical periods, limited mainly by the trap noise: for the cat state with $\alpha=2$, the negativity volume $\delta(W)$ decreases from $0.26$ to $0.1$ in $\approx10$~ms, and for the Fock state from $0.19$ to $0.1$ in $\approx17$~ms.
 
The two-mode squeezing term present in the Hamiltonian~\eqref{H_RWA} induces fast oscillations with frequency $O(\Omega_p')$ and relative amplitude $O(g'^2/\Omega_p'^2)$ on top of the coherent population exchange induced by the beam-splitter interaction. The only first-order effect of these oscillations is a relative shift of the optimal $t_{\rm swap}$ of order $g'/\Omega_p'$.\footnote{The shift can be estimated as the quotient of the slope (time derivative, $O(g'^2/\Omega_p')$) of the rapidly oscillating correction and the curvature (second time derivative, $O(g'^2)$) of the slow RWA trajectory at the unperturbed $t_{\rm swap}$.} For the parameters we used (the same parameters as in Secs.~\ref{sec2} and~\ref{sec3}, and $g'/2\pi=320$~Hz), the optimal transfer time becomes $t_{\rm swap}\approx 1.06\,\pi/|g'|$. The state fidelity is resilient to slight differences in experimental protocol. For example, if the swap time $t_{\rm swap}=\pi/|g'|$ is used instead of the optimal one, the cat-state fidelity is reduced from 0.93 to 0.87. We assume the interaction is switched off instantly in step (\textit{iv}), which is justified since the damping occurs on a timescale much shorter than $t_{\rm swap}$, as argued in Sec.~\ref{sec_exp}. 
Our results thus show that non-Gaussian state preparation of massive levitated nanoparticles, a core milestone in opto- and nanomechanics~\cite{GonzalezBallestero2021}, is within reach of current ion-trap setups.
 
Apart from preparation of non-Gaussian states of the nanoparticle, the state transfer protocol can also be used to prepare entangled non-Gaussian states of ion-nanoparticle motion. A simple way to achieve this is to use an interaction time of $t_{\rm ent}\approx0.5\pi/|g'|$, which allows the preparation of single-phonon Bell states $(\ket{1_i}\ket{0_p}-\mathrm{i}\ket{0_i}\ket{1_p})/\sqrt{2}$ or, starting from an ion cat state with amplitude $\alpha$, two-mode cat states $\propto \ket{(\alpha/\sqrt{2})_i}\ket{(-\mathrm{i}\alpha/\sqrt{2})_p}+\ket{(-\alpha/\sqrt{2})_i}\ket{(\mathrm{i}\alpha/\sqrt{2})_p}$. We note that, unlike the Gaussian entangled states of Sec.~\ref{sec4} or the non-Gaussian nanoparticle states discussed above, these states can only be prepared transiently: the beam-splitter interaction will disentangle the modes, and the only way to turn it off is by damping the ion's motion, causing the state to decohere.
 
\section{Experimental considerations}\label{sec_exp}
 
\subsection{Decoherence rates}
 
To assess whether the protocols outlined in previous sections are realistic, it is crucial to estimate the decoherence rates. The sideband cooling rate $\gamma_i$ can be tuned by laser power up to $\gamma_i\sim 2\pi\times1\,\rm kHz$~\cite{Roos1999}, while an ion heating rate of $\Gamma_i=2\pi\times 0.1\,\rm Hz$ is realistic in macroscopic linear Paul traps at room temperature~\cite{Poulsen2012}. At an ultrahigh-vacuum pressure of $P=5\times 10^{-12}\,\rm mbar$, the nanoparticle gas damping rate is $\gamma_{\rm gas}=2\pi\times1.7\,\rm nHz$ (details in Appendix~\ref{A-EP}), and the corresponding heating rate is $\Gamma_{\rm gas}=\gamma_{\rm gas}k_BT/\hbar\Omega_p'=2\pi\times 10\,\rm Hz$. This is the rate of energy diffusion; the collision rate is only $\gamma_c\approx1.7\,\rm s^{-1}$, but each collision transfers $\approx38$ phonons, so that collisions limit the coherence of non-Gaussian states at the rate $\gamma_c$ rather than $\Gamma_{\rm gas}$ (Appendix~\ref{A-GAS}). 

It is not known with certainty what dominates the trap noise rate $\Gamma_{\rm td}$ in current experiments, which makes it harder to estimate. In a recent experiment~\cite{Dania2024}, a heating rate of $3.3\times10^4$ phonons/s, i.e., $\Gamma_{\rm td}\approx2\pi\times 5\,\rm kHz$, was measured at $7\times10^{-11}$~mbar for a nanoparticle with $M_p\approx 4\times 10^{-17}\,\rm kg$, $Q_p\approx 300\,e$ and $\Omega_p/2\pi=1.28$~kHz, and attributed to either mechanical vibrations or voltage noise on the electrodes. The rate $\Gamma_{\rm td}$ can be obtained from the force-noise power spectral density as $\Gamma_{\rm td}=S_F(\Omega_p')/(2\hbar M_p\Omega_p')$, where $S_F(\omega)=\int d\tau\,e^{\mathrm{i}\omega\tau}\langle F(\tau)F(0)\rangle$ is the two-sided spectral density. If vibrations dominate the noise, assuming a fixed (one-sided) acceleration spectrum $S_{\ddot{x}}=2S_F/M_p^2$, the trap noise rate scales with $M_p$. Analogously, if voltage noise is dominant, assuming a fixed voltage spectrum $S_V$, the trap noise rate scales with $Q_p^2/M_p$. For the acceleration noise measured on the vacuum chamber in Ref.~\cite{Dania2024}, $S_{\ddot{x}}\approx8\times 10^{-10}\,\rm (m/s^2)^2/Hz$ at $1.28$~kHz, we obtain $\Gamma_{\rm td}/2\pi\approx50\,\rm Hz$ for the parameters of this work. 

Vibrational noise can be reduced by orders of magnitude by placing the setup in an acoustic enclosure, which attenuates airborne sound by $\approx40$~dB at 1~kHz~\cite{HerzanAEK2002}, and by mounting the trap on a passive vibration isolation stage, which demonstrated attenuation of 250~dB at 1~kHz frequency~\cite{Chan1999}. Regarding the electrical noise, only the endcap electrodes generate fields along $z$: the radial electrodes generate fields only in the $xy$ plane so that their noise does not couple to motion along $z$. Voltage fluctuations between the endcap electrodes generate a homogeneous field along $z$. For the thermal (Johnson) noise of a resistance $R$ in an endcap circuit, this gives $\Gamma_{\rm td}\approx Q_p^2k_BTR/(\hbar M_p\Omega_p'd_z^2)$~\cite{Turchette2000, Brownnutt2015}, so that assumed rate $\Gamma_{\rm td}=2\pi\times1\,\rm Hz$ requires $R\lesssim130\,\Omega$ at $\Omega_p'$, corresponding to a voltage noise of $\lesssim1.5\,\rm nV/\sqrt{Hz}$. This can be reached by low-pass filtering the DC voltages close to the trap, e.g., with an $RC$ filter, whose real impedance at $\Omega_p'$ is only $\approx1/(\Omega_p'^2RC^2)$, and by coupling the drive tones at MHz frequencies through high-pass filters that suppress the noise of the drive source at $\Omega_p'$~\cite{Brownnutt2015}. Based on these considerations, the noise rate of $\Gamma_{\rm td}=2\pi\times1\,\rm Hz$ used throughout the paper should be realistically achievable.

The discussion above assumed the experiment operated at room temperature and at achievable ultrahigh vacuum levels. Operating the setup at a cryogenic temperature 4~K would allow reaching $10^{-15}$~mbar level vacuum~\cite{Micke2019}, reducing $\Gamma_{\rm gas}$, as well as reducing Johnson noise in electrical circuits, which affects $\Gamma_{\rm td}$ and $\Gamma_{i}$. Therefore, all noise rates can be decreased by 2-3 more orders of magnitude, at the price of increased experimental complexity.
 
\subsection{Cooling protocol details}

To achieve the nanoparticle cooling described in Sec.~\ref{sec3}, one can follow a similar multi-step protocol as in conventional ion cooling: (\textit{i}) trapping the ion and the nanoparticle, (\textit{ii}) Doppler cooling the ion to an occupation of $\sim 10$ phonons, (\textit{iii}) turning on the voltage modulation on the endcap electrodes (including the compensation tone at $2\omega_d$) and waiting until steady-state phonon occupations are achieved while the ion is continuously Doppler-cooled, (\textit{iv}) switching to sideband cooling of the ion, which becomes efficient at these low occupations, and waiting for the new steady state, in which the ground-state cooling regime is reached for both the ion and the nanoparticle. It is important that the switch to sideband cooling is not made until after step (\textit{iii}) because the nanoparticle occupation number after step (\textit{ii}) is $\sim 10^6$~\cite{Gupta2025}, and transferring these excitations to the ion via the beam-splitter Hamiltonian would render sideband cooling ineffective. 
In step (\textit{iii}), because of the nonlinear classical response, the drive is best switched on at a slightly larger frequency than optimal $\omega_d$ and then swept down to the operating point (Appendix~\ref{A-FL}). Keeping $\langle\hat{n}_p\rangle_{\rm ss}\lesssim0.1$ then requires $\omega_d-\Omega_i'$ to be stable within $\approx2\pi\times30\,\rm Hz$, i.e., a relative stability of $\Omega_i'\propto U_{0z}^{1/2}$ of $\approx2.5\times10^{-5}$, or $\approx5\times10^{-5}$ of the DC voltage $U_{0z}$.
 
Sideband cooling of the ion also needs some care. As opposed to usual experiments in which the ion sits at an equilibrium position, our method relies on the ion following an oscillating classical trajectory $z_{i,\rm cl}(t)$. In the displaced frame given by Eq.~\eqref{disptrafo}, this appears as an additional phase factor in the laser-induced interaction Hamiltonian responsible for cooling,
\begin{equation}
    \hat{H}_L\propto\hat{\sigma}^+\exp\Big(\mathrm{i}k_zz_{i,\rm cl}(t)\Big)\exp\Big(\mathrm{i}(k_z\hat{z}_i-\omega_Lt+\phi)\Big)+\mathrm{H.c.},
\end{equation}
where $\hat{\sigma}^+$ is the qubit raising operator, $k_z=2\pi/\lambda_L$ is the wavenumber and $\omega_L$ is the frequency of the laser. The classical oscillation amplitude above the equilibrium position, $A_{\rm cl}$, is comparable to the $729\,\rm nm$ wavelength of the $^{40}\mathrm{Ca}^{+}$ optical qubit transition (or optical transitions in general), so the phase factor $\exp\Big(\mathrm{i}k_zz_{i,\rm cl}(t)\Big)$ remains explicitly in the Hamiltonian after performing the Lamb--Dicke expansion,
\begin{equation}
    \exp(\mathrm{i}k_z\hat{z}_i)= 1+\mathrm{i}\eta(\hat{a}_i+\hat{a}_i^\dag)+O(\eta^2),
\end{equation}
where $\eta=k_zx_{\mathrm{zpm},i}\ll 1$ is the Lamb-Dicke parameter. The time-dependent phase factor makes it difficult to reach and maintain the sideband resonance condition needed for cooling. Since this factor
is deterministic, however, it can be canceled by applying phase or frequency feed-forward on the laser field. A possible implementation uses a programmed modulator phase-locked to the endcap drive, that applies a frequency shift $\Delta f(t)=-k_z\dot{z}_{i, \rm cl}/2\pi$. In this way, the optical coupling becomes independent of deterministic classical oscillations. For typical parameters ($\Omega_i'\sim2\pi\times1\,\rm MHz$, $A_{\rm cl}\sim 1\,\rm \mu m$), the required amplitude of the laser frequency modulation is $\Delta f_{\rm max}=A_{\rm cl}\Omega_i'/\lambda_L\sim 10\,\rm MHz$, which is achievable with a standard acousto-optic modulator~\cite{Aldous2017}. We note that this phase cancellation is necessary for all protocols outlined in the paper, including the Doppler cooling in step (\textit{iii}), where the peak Doppler shift of $\approx30\,\rm MHz$ at $397\,\rm nm$ exceeds the transition linewidth. 

With the feed-forward, the cooling force acting on the ion becomes $-M_i\gamma_i(\dot{z}_i-\dot{z}_{i,\rm cl})$: it damps deviations from the programmed classical trajectory $z_{i,\rm cl}(t)$ but vanishes on the trajectory. The damping force $-M_i\gamma_i\dot{z}_{i,\rm cl}$ assumed in Eq.~\eqref{di_cl} -- needed for the classical trajectories to be approximately harmonic --  can then be supplied by the drive itself, by adding to $U_{\rm drive}(t)$ the component $-\gamma_iM_id_z\dot{z}_{i,\rm cl}(t)/Q_i$, which oscillates at $\omega_d$ in antiphase with the ion velocity. The total force on the ion is then the same as in Eq.~\eqref{di_cl} for any trajectory, so that Eqs.~\eqref{di_cl} and~\eqref{dp_cl} and all results of this work remain unchanged. For the parameters of Fig.~\ref{fig2}, this adds an $18\,\rm\mu V$ component to the $60\,\rm\mu V$ drive, which results in a single tone at $\omega_d$ with amplitude $49\,\rm\mu V$, shifted by $0.27$~rad with respect to the ion motion, while the tone at $2\omega_d$ is unchanged. Since the cooling force vanishes on the classical trajectory, the classical motion does not change when switching from Doppler to sideband cooling, or when the cooling is switched off during the state transfer (Sec.~\ref{sec5}).
 
In the last step of the state transfer protocol, it is important that the ion's classical motion is damped on a timescale much shorter than the state transfer time $t_{\rm swap}\approx \pi/g'\sim\,\rm ms$. Sideband cooling is not suitable for this, since the classical motion corresponds to $\sim10^4$ phonons and $k_zA_{\rm cl}\approx13$ is far outside the Lamb--Dicke regime. Instead, the drive (including the tone at $2\omega_d$) and the feed-forward are switched off and the ion classical motion is Doppler-cooled. With a damping rate of $\gamma_i/2\pi\approx10\,\rm kHz$~\cite{Bykov2025}, the classical motion is damped within a few tens of microseconds. During this time $t_{\rm damp}$, only $\sim(g't_{\rm damp}/2)^2\langle\hat{n}_i\rangle\lesssim10^{-2}$ excitations of the Doppler-cooled ion leak to the nanoparticle.

Finally, the radial modes of the nanoparticle, which are not cooled via the ion, have to be precooled: thermal radial motion with amplitude $A_r$ shifts the ion frequency along z by $\approx-(A_r^2/D_0^2)\,\Omega_i'$, which stays within the required stability only for radial temperatures $\lesssim30$~mK. Temperatures of a few mK have been reached by feedback cooling of a nanoparticle in a Paul trap~\cite{Dania2021}. Since the radial modes are heated only at a rate $\approx\gamma_{\rm gas}T\approx3\,\mu$K/s by the residual gas, the feedback and the detection light it requires can be switched off after a short cooling period ($\sim1$~s). All protocols described in the paper can then be repeated for minutes to hours before the radial motion has to be cooled again, so that photon recoil does not affect the protocols.
 
\subsection{Detection}
 
Any state preparation protocol has to be followed by detection that verifies it. The detection is more challenging when the prepared state is not a steady state, as is the case in Secs.~\ref{sec4} and~\ref{sec5}, because continuous detection schemes are not applicable. The most challenging case is verification of non-Gaussian states such as Schr\"odinger cat states (Sec.~\ref{sec5}) because it requires resolving interference features significantly smaller than the zero-point fluctuations. A potential detection scheme is to implement pulsed homodyne measurement of the position quadrature at a certain time $t=t_{\rm det}$; varying $t_{\rm det}$ would give information about arbitrary quadrature at initial time $t_0$. This scheme in principle allows for full state tomography, but sweeping through a few angles would already be enough to verify the existence of quantum interference and Wigner negativity~\cite{Fiurasek2013, Chabaud2021, Zaw2022, Chen2024}. The key requirement for the measurement is resolving interference fringes with a homodyne pulse length $\tau\ll \Omega_p'^{-1}$ so that the quadrature is effectively frozen during readout. For a cat state with $\alpha=2$, this implies a resolution $\delta x\lesssim x_{\mathrm{zpm}, p}/4\alpha\sim 10^{-11}\,$m with a pulse duration of $\tau\lesssim 10^{-2}\times2\pi/\Omega_p'\sim 10\,\rm \mu s$. A recent experiment achieved an imprecision of $\sqrt{S_{\rm imp}}\sim 10^{-14}\,\rm m/\sqrt{Hz}$ in a continuous free-space measurement~\cite{Dinter2026}, which corresponds to $\delta x=\sqrt{S_{\rm imp}/\tau}\sim 3\times10^{-12}\,\rm m$ assuming the same noise floor holds for a $\tau=10\,\rm \mu s$ pulse window. This argument suggests that detection requirements are feasible in state-of-the-art setups, even without placing the nanoparticle in an optical cavity.
 
\section{Conclusion}\label{sec6}
 
We have shown that it is possible to achieve the long-standing goals of quantum levitodynamics with levitated nanoparticles co-trapped with ions in electrical traps. Specifically, we have proposed protocols for motional ground-state cooling of a levitated nanoparticle and for the preparation of squeezed, entangled, and non-Gaussian states. Furthermore, we predict that these protocols are feasible in state-of-the-art setups. The advantages of this platform for macroscopic quantum science are the absence of laser recoil heating, tunability of the potential and availability of established trapped-ion techniques to engineer the quantum state of the nanoparticle.
 
Our work is only the first stepping stone, as the system's versatility opens many further pathways to quantum control of mechanical motion. For example, the motional nonlinearities discussed in Appendix~\ref{A-HO}, which can be enhanced by stronger or additional drives, could be exploited to prepare non-Gaussian states without using the qubit transition. A second route for non-Gaussian state preparation is to engineer nonlinear dissipation using a two-photon drive of the ion transition, as we outline in Appendix~\ref{A-ND}. Moreover, it is possible to exploit more mechanical modes: the coupling between the motion along different axes can be tailored, and several particles and ions can be co-trapped in the same trap~\cite{Bykov2025, Gupta2025}. This range of options, with the possible addition of optical cavities providing control and coupling mechanisms, paves the way to more refined protocols for quantum control using collective motional states~\cite{Leibfried2003, Casabone2015}.

The potential of this platform for tests of fundamental quantum physics and gravity is illustrated by the estimate $\mu\approx17.5$ of the macroscopicity parameter defined in Ref.~\cite{Nimmrichter2013}, which can be increased to $\mu\approx18$--$19.5$ by optimizing the nanoparticle size and charge (details are in Appendix~\ref{A-NH}). This value exceeds the current experimental record~\cite{Pedalino2026} by 2-4 orders of magnitude and the record for mechanical resonators~\cite{Schrinski2023} by 6-8 orders of magnitude. Reducing the decoherence rates, e.g., by operating at cryogenic temperatures, could increase $\mu$ by several orders of magnitude more. In the long run, co-trapping several nanoparticles and ions, with the ions mediating the coupling and serving as an interface for preparing and verifying the joint quantum states, could extend such tests from superpositions of a single nanoparticle to entangled states of several massive objects.
 
\begin{acknowledgments}
 
We are grateful to Nils J. Engelsen, Piotr T. Grochowski, Andrea Navoni and Lorenzo Papa for useful and stimulating discussions. This research was funded in part by the Austrian Science Fund (FWF) [10.55776/COE1], [10.55776/PIN3404324].
 
\end{acknowledgments}
 
\appendix
 
\section{Experimental parameters}\label{A-EP}
 
\begin{table*}[t]
\caption{Experimental parameters}
\begin{ruledtabular}
\begin{tabular}{lll}
\textbf{Parameter} & \textbf{Description} & \textbf{Value} \\
\hline
$R_p$ & Nanoparticle radius & $50 \times 10^{-9}~\mathrm{m}$ \\
$M_p$ & Nanoparticle mass & $1.0 \times 10^{-18}~\mathrm{kg}$ \\
$Q_p$ & Nanoparticle charge & $30 \times 1.6 \times 10^{-19}~\mathrm{C}$ \\
$M_i$ & Ion mass & $6.63 \times 10^{-26}~\mathrm{kg}$ \\
$Q_i$ & Ion charge & $1.6 \times 10^{-19}~\mathrm{C}$ \\
$[d_r,\,d_z]$ & Distances between electrode and trap center & $[0.9,\,1.7] ~\mathrm{mm}$\\
$[\alpha_{\rm AC},\,\alpha_{\rm DC}]$ & Geometric factors  & $[0.93,\,0.19]$\\
$[U_{0z},U_s,U_f,U_{\rm comp}]$ & Voltage amplitudes & $[63,120,1350,0.68] ~\mathrm{V}$
\\
$[U_d,\,U_d^{(2)}]$ & Drive and compensation-tone amplitudes & $[60~\mathrm{\mu V},\,2.7~\mathrm{mV}]$
\\
$\omega_s$ & Slow frequency & $2\pi \times 7  ~\mathrm{kHz}$
\\
$\omega_f$ & Fast frequency & $2\pi \times 17.5  ~\mathrm{MHz}$\\
$T$ & Temperature of surrounding gas & $300~\mathrm{K}$ \\
$P$ & Pressure of surrounding gas & $5 \times 10^{-12}~\mathrm{mbar}$ \\
$\Gamma_i$ & Ion heating rate & $2\pi \times 0.1 ~\mathrm{Hz}$ \\
$\Gamma_{\rm td}$ & Nanoparticle trap displacement noise rate & $2\pi \times 1~\mathrm{Hz}$
\\
$\gamma_{\rm gas}$ & Damping rate of the nanoparticle due to surrounding gas & $2\pi \times 1.7 ~\mathrm{nHz}$ 
\\
$\Gamma_{\rm gas}$ & Nanoparticle heating rate due to gas collisions & $2\pi\times 10\,\rm Hz$
\\
$[\Omega_{xp},\, \Omega_{yp},\, \Omega_{zp}]$ & Mechanical frequencies of nanoparticle ($\Omega_{zp}\equiv\Omega_p$) & $2\pi \times [1.75,\, 1.67,\, 1.00] ~\mathrm{kHz}$ \\
$[\Omega_{xi},\, \Omega_{yi},\, \Omega_{zi}]$ & Mechanical frequencies of ion ($\Omega_{zi}\equiv\Omega_i$) & $2\pi \times [3.72,\, 3.70,\, 0.71] ~\mathrm{MHz}$ \\
$[\Omega_i',\, \Omega_p']$ & Renormalized $z$ frequencies of ion and nanoparticle & $2\pi \times [1.22~\mathrm{MHz},\, 1.03~\mathrm{kHz}]$ \\
$D_0$ &  Ion-nanoparticle equilibrium separation (along $z$) & $17.6 ~\mathrm{\mu m}$
\end{tabular}
\end{ruledtabular}
\label{Table1}
\end{table*}
 
The experimental parameters used throughout the paper are shown in Table~\ref{Table1}. The trap parameters are the same as in Ref.~\cite{Bykov2025}, while the mechanical frequencies and equilibrium separation were calculated according to the procedure outlined in Ref.~\cite{Gupta2025}. The nanoparticle parameters and voltage amplitudes are chosen to maximize the cooling and state preparation performance, primarily by increasing the zero-point motion and reducing the equilibrium separation, while keeping the mechanical frequency at $\sim 1$~kHz to avoid various technical noise sources that become relevant at low frequencies. The gas damping rate is calculated as~\cite{Li2011, Beresnev1990}
\begin{equation}
    \gamma_{\rm gas} = 0.619 \frac{6 \pi R_p^2}{M_p} P \sqrt{\frac{2 m_0}{\pi k_B T}},
\end{equation}
where $m_0=3.35\times10^{-27}$~kg is the mass of a hydrogen molecule and $P$ the gas pressure.
 
\section{Gaussian dynamics}\label{A-GD}
 
As mentioned in the main text, state evolution with a quadratic master equation preserves Gaussianity, so it is possible to completely characterize the system with equations of motion for expectation values and covariance matrix of ladder operators. We define the vector of ladder operators $\hat{\bm{B}}=[\hat{a}_i, \hat{a}^\dagger_i, \hat{a}_p, \hat{a}^\dagger_p]^T$, its vector of expectation values $\langle\hat{\bm{B}}\rangle$, and the covariance matrix as $\bar{\sigma}_{\lambda\mu}=\frac{1}{2}\langle \hat{B}_\lambda \hat{B}_\mu+\hat{B}_\mu \hat{B}_\lambda \rangle-\langle \hat{B}_\lambda\rangle\langle \hat{B}_\mu\rangle$. The equation of motion for the expectation value of a general operator $\hat{O}$ is obtained from the master equation
\begin{equation}
    \dot{\hat{\rho}}=-\frac{\mathrm{i}}{\hbar}[\hat{H}_{\mathrm{ RWA}},\hat{\rho}]+\mathcal{D}_p(\hat{\rho})+\mathcal{D}_i(\hat{\rho})
\end{equation}
as
\begin{multline}\label{eomOgeneral}
    \dv{\langle\hat O\rangle}{t}= -\frac{\rm i}{\hbar} \Tr[\comm{\hat{O}}{\hat{H}_{\rm RWA}} \hat{\rho} ] \\+ \Tr[ \hat{O} \left(\mathcal{D}_p \left(\hat{\rho}\right)+\mathcal{D}_i \left(\hat{\rho}\right)\right) ].
\end{multline}
The equations of motion for $\langle\hat{\bm{B}}\rangle$ and $\bar{\sigma}$ derived from this expression are~\cite{Gupta2025}
\begin{equation}
    \dv{ \langle\hat{\bm{B}}\rangle}{t} = \bar{A} \langle\hat{\bm{B}}\rangle,
\end{equation}
\begin{equation}
    \dv{}{t}\bar\sigma = \bar{A}\bar\sigma+\bar\sigma\bar{A}^T+\bar{C},
\end{equation}
with the drift matrix
\begin{multline}
    \bar{A}= \text{diag}[-\mathrm{i}\Omega_{p}',\mathrm{i}\Omega_{p}',-\mathrm{i}\Omega_{p}',\mathrm{i}\Omega_{p}']
    \\+\frac{1}{2}\left[\begin{array}{c|c}
       -\gamma_{i}\mathbb{1}_{2\times 2}  & -\mathrm{i}g' \mathbb{Z} \mathbb{J}_{2\times 2} \\ \hline
        -\mathrm{i}g' \mathbb{Z} \mathbb{J}_{2\times 2}  & -\gamma_{\rm gas}\mathbb{1}_{2\times 2}
    \end{array}\right],
\end{multline}
($\mathbb{Z}$ is the Pauli-Z matrix and $\mathbb{J}_{n\times n}$ is an $n\times n$ matrix whose entries are all equal to $1$) and the diffusion matrix
\begin{align}
    \bar{C}= \mqty[
    0 && \tilde{\Gamma}_{i} && 0 && 0 \\
    \tilde{\Gamma}_{i} && 0 && 0 && 0 \\
    0 && 0 && - \Gamma_{\rm td} && \tilde{\Gamma}_{p} \\
    0 && 0 && \tilde{\Gamma}_{p} && -\Gamma_{\rm td}
    ],
\end{align}
with $\tilde{\Gamma}_i=\Gamma_i+\gamma_i/2$ and $\tilde{\Gamma}_p=\Gamma_{\rm gas}+\Gamma_{\rm td}+\gamma_{\rm gas}/2$.
The vector of expectation values vanishes in the initial thermal state, so it stays zero throughout the evolution (note also that system stability requires all eigenvalues of $\bar{A}$ to have negative real parts, so even a non-zero initial $\langle\hat{\bm{B}}\rangle$ would decay). The full analytical solution can in principle be obtained for the covariance matrix (see Appendix of Ref.~\cite{Gupta2025}), but the complicated form of the expression makes it hard to extract physical insights from it. To get some physical insight we assume the steady state has negligible squeezing and two-mode correlations, $\langle\hat{a}_i^2\rangle\approx\langle\hat{a}_p^2\rangle\approx\langle\hat{a}_i\hat{a}_p\rangle\approx 0$, which leads us to the simplified expression
\begin{equation}
    \langle \hat{n}_p\rangle_{\rm ss}\approx \frac{(\Gamma_i+\Gamma_{\rm gas}+\Gamma_{\rm td})g'^2+(\Gamma_{\rm gas}+\Gamma_{\rm td})\gamma_i^2}{\gamma_ig'^2},
\end{equation}
which achieves a minimum of
\begin{equation}
    \langle \hat{n}_{p}\rangle_{\rm ss, opt}=\frac{2\sqrt{(\Gamma_i+\Gamma_{\rm gas}+\Gamma_{\rm td})(\Gamma_{\rm gas}+\Gamma_{\rm td})}}{|g'|}
\end{equation}
for $\gamma_{i, \rm opt}=|g'|\sqrt{(\Gamma_i+\Gamma_{\rm gas}+\Gamma_{\rm td})/(\Gamma_{\rm gas}+\Gamma_{\rm td})}$. If $\Gamma_{\rm gas}$ is the dominant noise rate as assumed in this paper, we have $\gamma_{i,\rm opt}\approx |g'|$, in line with the intuitive explanation and the full solution shown in main text Fig.~\ref{fig3}.

\section{Gas collisions and blackbody radiation}\label{A-GAS}

The gas dissipator in Eq.~\eqref{Dp} is the long-wavelength limit of the collisional master equation~\cite{Joos1985,Gallis1990,Hornberger2003,Vacchini2009}. Since the mass of the gas molecules is much smaller than the nanoparticle mass ($m_0/M_p\approx3\times10^{-9}$), collisions are described to leading order by the recoil-free collisional dissipator
\begin{equation}
    \mathcal{L}_{\rm coll}(\hat\rho)=\gamma_c\int d^3\mathbf{q}\,w(\mathbf{q})\Big(e^{\mathrm{i}\mathbf{q}\cdot\hat{\mathbf{r}}_p/\hbar}\hat\rho\,e^{-\mathrm{i}\mathbf{q}\cdot\hat{\mathbf{r}}_p/\hbar}-\hat\rho\Big),
    \label{Lcoll}
\end{equation}
where $\gamma_c$ is the collision rate, $w(\mathbf{q})$ is the probability distribution of the momentum $\mathbf{q}$ transferred in a collision, and $\hat{\mathbf{r}}_p$ is the nanoparticle position operator. In position representation, coherences between positions separated by $\Delta z$ along $z$ decay at the rate $\gamma_{\rm dec}(\Delta z)=\gamma_c\int d^3\mathbf{q}\,w(\mathbf{q})(1-e^{\mathrm{i}q_z\Delta z/\hbar})$. For $\Delta z\ll\ell_c=\hbar/\langle q_z^2\rangle^{1/2}$, this rate is $\gamma_{\rm dec}\approx\Lambda_{\rm gas}(\Delta z)^2$ with $\Lambda_{\rm gas}=\gamma_c\langle q_z^2\rangle/(2\hbar^2)$, while for $\Delta z\gg\ell_c$ it saturates at $\gamma_c$, since each collision can destroy the coherence only once. Expanding Eq.~\eqref{Lcoll} to second order in $\mathbf{q}$ gives momentum diffusion with $\hbar^2\Lambda_{\rm gas}=\gamma_{\rm gas}M_pk_BT$; together with the damping, which arises from the leading recoil correction to Eq.~\eqref{Lcoll}. This is exactly the quantum Brownian motion master equation~\cite{Vacchini2009}, which after the RWA becomes the gas term of Eq.~\eqref{Dp}, with
\begin{equation}
    \Gamma_{\rm gas}=\gamma_c\langle\xi^2\rangle,\quad \xi=q_zx_{\mathrm{zpm},p}/\hbar,
\end{equation}
i.e., the heating rate is the mean number of phonons transferred per collision times the collision rate. For hydrogen molecules at $T=300$~K and $P=5\times10^{-12}$~mbar, the collision rate is $\gamma_c=n_0\bar{v}\pi R_p^2\approx1.7\,\rm s^{-1}$, where $n_0$ is the gas density and $\bar v=\sqrt{8k_BT/\pi m_0}$ the mean speed. The mean squared momentum transferred per collision, $\langle q_z^2\rangle=\frac{8}{3}(1+\frac{\pi}{8})m_0k_BT$ for diffuse re-emission of the molecules, reproduces the damping rate of Appendix~\ref{A-EP} and corresponds to $\langle\xi^2\rangle\approx38$ phonons per collision. The resolution length $\ell_c\approx15$~pm is thus much smaller than $x_{\mathrm{zpm},p}\approx90$~pm, i.e., a single collision kicks the nanoparticle by more than its zero-point momentum, and the long-wavelength expansion is not valid for coherences between states separated by more than $\ell_c$.

In the frame rotating with $\Omega_p'$, the direction of the kick in phase space depends on the time of the collision, and averaging over it gives the dissipator
\begin{multline}
    \mathcal{D}_{\rm gas}(\hat\rho)=\gamma_c\int d^3\mathbf{q}\,w(\mathbf{q})\int_0^{2\pi}\frac{d\vartheta}{2\pi}\Big(\hat{\mathcal{T}}_{\xi\vartheta}\hat\rho\hat{\mathcal{T}}_{\xi\vartheta}^\dagger-\hat\rho\Big)\\+\frac{\gamma_{\rm gas}}{2}\big(\mathcal{L}_{\hat a_p,\hat a_p^\dagger}-\mathcal{L}_{\hat a_p^\dagger,\hat a_p}\big)(\hat\rho),
    \label{Dgas}
\end{multline}
where $\hat{\mathcal{T}}_{\xi\vartheta}=\exp[\mathrm{i}\xi(e^{\mathrm{i}\vartheta}\hat a_p^\dagger+e^{-\mathrm{i}\vartheta}\hat a_p)]$ displaces the nanoparticle by the kick $\xi$ in the phase-space direction $\vartheta$. Since $\xi\approx6$, $\hat{\mathcal{T}}_{\xi\vartheta}$ cannot be expanded in $\xi$. Nevertheless, for any distribution $w(\mathbf{q})$, the jump term contributes to the equations of motion of the first and second moments in the same way as $\Gamma_{\rm gas}(\mathcal{L}_{\hat a_p,\hat a_p^\dagger}+\mathcal{L}_{\hat a_p^\dagger,\hat a_p})$: averaged over $\vartheta$, a kick leaves $\langle\hat a_p\rangle$ and $\langle\hat a_p^2\rangle$ unchanged and increases $\langle\hat a_p^\dagger\hat a_p\rangle$ by $\xi^2$.

Since the equations of motion for the first and second moments contain only $\gamma_c\langle\xi^2\rangle=\Gamma_{\rm gas}$, the covariance matrices obtained with Eq.~\eqref{Dp}, and therefore $\langle\hat n_p\rangle_{\rm ss}$ and the squeezing in Secs.~\ref{sec3} and~\ref{sec4}, are exact, up to corrections of relative order $\gamma_{\rm gas}/\Gamma_{\rm gas}\sim10^{-10}$ from the slightly different damping terms in Eqs.~\eqref{Dp} and~\eqref{Dgas}. Since $\tilde\nu_-\geq1/2$ is necessary for separability of any state~\cite{Simon2000}, $\tilde\nu_-<1/2$ also certifies entanglement of the non-Gaussian state. The state itself, however, is not Gaussian. A collision displaces the nanoparticle, which shifts the mean values but leaves the covariance matrix unchanged. Between collisions, the evolution is generated by the remaining terms of the master equation, which are all quadratic, so that the displacement obeys the same linear equations as the mean values $\langle\hat{\bm B}\rangle$ in Appendix~\ref{A-GD}, which do not depend on $\Gamma_{\rm gas}$, and decays at the rates given by their eigenvalues. Each collision history thus produces a displaced copy of the Gaussian state $\hat\rho_{\rm nc}$ obtained without collisions ($\Gamma_{\rm gas}=0$), and the steady state is a mixture of such copies,
\begin{equation}
    \hat\rho_{\rm ss}=\int d^4\bm\beta\,\mathcal{P}(\bm\beta)\,\hat{\mathcal{T}}_{\bm\beta}\hat\rho_{\rm nc}\hat{\mathcal{T}}_{\bm\beta}^\dagger,
    \label{mixture}
\end{equation}
where $\hat{\mathcal{T}}_{\bm\beta}$ displaces the quadratures of both modes by the vector $\bm\beta$, and $\mathcal{P}(\bm\beta)$ is the distribution of the displacements left by previous collisions (for the two-tone drive, $\hat\rho_{\rm nc}$ and $\mathcal{P}$ are periodic in time). We sample $\mathcal{P}(\bm\beta)$ numerically from the Poisson statistics of the collisions, the distribution $w(\mathbf{q})$ for diffuse re-emission and the decay of $\langle\hat{\bm B}\rangle$. In the cooling steady state of Fig.~\ref{fig3} for $\Gamma_{\rm td}/2\pi=1$~Hz, a displacement decays at the rate $\approx2\pi\times90$~Hz, so that 98\% of the mixture has $|\bm\beta|<0.01$, while $\hat\rho_{\rm nc}$ has $\langle\hat n_p\rangle=0.03$; the steady-state occupation $\langle\hat n_p\rangle_{\rm ss}=0.09$ is dominated by the rare kicked states.

Quantities that are not determined by the covariance matrix alone, such as the purity and $E_N$, are in general not bounded by those of the Gaussian state with the same covariance matrix; for example, the state $(1-\epsilon)\ket{0}\bra{0}+\epsilon\ket{1}\bra{1}$ with $0<\epsilon<1/2$ has a lower purity than the thermal state with the same $\langle\hat n\rangle$. We therefore evaluate them for the mixture~\eqref{mixture}. The purity follows from the overlaps of displaced Gaussian states,
\begin{equation}
    \mathrm{Tr}\,\hat\rho_p^2=\int d^4\bm\beta\, d^4\bm\beta'\,\mathcal{P}(\bm\beta)\mathcal{P}(\bm\beta')\frac{e^{-\Delta\bm\beta_p^T(\bar\Sigma_p^{\rm nc})^{-1}\Delta\bm\beta_p/4}}{\sqrt{\det(2\bar\Sigma_p^{\rm nc})}},
\end{equation}
where $\Delta\bm\beta_p$ is the nanoparticle part of $\bm\beta-\bm\beta'$ and $\bar\Sigma_p^{\rm nc}$ the nanoparticle covariance matrix of $\hat\rho_{\rm nc}$. For $\Gamma_{\rm td}/2\pi=1$~Hz, we obtain a purity of 0.94, compared to 0.85 for the Gaussian state in Fig.~\ref{fig3}(c). For the negativity $\mathcal{N}_{\rm neg}=(e^{E_N}-1)/2$, the operator $\mathbb{1}-2\hat{\mathcal{E}}_-$, where $\hat{\mathcal{E}}_-$ is the projector on the negative eigenspace of the partial transpose of $\hat\rho_{\rm nc}$, gives the lower bound
\begin{equation}
    \mathcal{N}_{\rm neg}\geq\int d^4\bm\beta\,\mathcal{P}(\bm\beta)\,\frac{1}{2}\left(\frac{e^{-|\bm\beta_-|^2/2\tilde\nu_-}}{2\tilde\nu_-}-1\right),
    \label{Nbound}
\end{equation}
where $\tilde\nu_-$ is the smallest symplectic eigenvalue of the partially transposed covariance matrix of $\hat\rho_{\rm nc}$, and $\bm\beta_-$ is the part of the displacement along the normal mode of the partially transposed state that carries its negativity (with the nanoparticle momentum reversed by the partial transposition), so that only kicks that displace this mode reduce the bound. At the optimum of Fig.~\ref{fig4}(d) for $g_1'/2\pi=150$~Hz, this gives $E_N\geq0.49$ for $\Gamma_{\rm td}/2\pi=1$~Hz and $E_N\geq0.35$ for $\Gamma_{\rm td}/2\pi=10$~Hz, compared to $0.35$ and $0.24$ for the Gaussian state, and the bound exceeds the Gaussian value at all points we checked ($g_1'/2\pi=80$--$250$~Hz, $\Gamma_{\rm td}/2\pi\le10$~Hz). The purity in Fig.~\ref{fig3}(c) and the values of $E_N$ in Fig.~\ref{fig4}(d) are therefore conservative.

For non-Gaussian states the difference is significant. With the Gaussian dissipator, the Fock state $\ket{1_p}$ decays at the rate $3\Gamma_{\rm gas}\approx 200\,\rm s^{-1}$, and the coherence of a cat state with $\alpha=2$ at the rate $4\alpha^2\Gamma_{\rm gas}\approx10^3\,\rm s^{-1}$, while with Eq.~\eqref{Dgas} these rates cannot exceed $\gamma_c\approx1.7\,\rm s^{-1}$. Separating the histories with and without collisions, i.e., expanding the evolution in the jump term of Eq.~\eqref{Dgas}, the state at time $t$ is
\begin{equation}
    \hat\rho(t)=e^{-\gamma_ct}\hat\rho_{\rm nc}(t)+\big(1-e^{-\gamma_ct}\big)\hat\rho_{\rm col}(t),
\end{equation}
where $\hat\rho_{\rm nc}$ evolves without the gas term and $\hat\rho_{\rm col}$ is the normalized state conditioned on at least one collision. Since $\hat\rho_{\rm col}$ is a density matrix, $F_{\rm col}=\bra{\psi}\hat\rho_{\rm col}\ket{\psi}\geq0$, so that $F=e^{-\gamma_ct}F_{\rm nc}+(1-e^{-\gamma_ct})F_{\rm col}\geq e^{-\gamma_ct}F_{\rm nc}$. A kicked state has negligible overlap with the target state, $F_{\rm col}\approx0$, so the bound is nearly tight and we use it in Sec.~\ref{sec5} and Appendix~\ref{A-NH}. For the Wigner negativity volume, we plot $e^{-\gamma_ct}\delta(W)_{\rm nc}$, which is not a strict bound; the rigorous bound $\delta(W)\geq e^{-\gamma_ct}\delta(W)_{\rm nc}-(1-e^{-\gamma_ct})[1+\delta(W)_{\rm col}]$, with $\delta(W)_{\rm col}\leq0.5$, differs from it by less than $0.005$ at the end of the state transfer ($\gamma_ct_{\rm swap}\approx3\times10^{-3}$). Before the transfer, the nanoparticle is in the mixture~\eqref{mixture}. Since the beam-splitter interaction moves an initial displacement of the nanoparticle mostly to the ion (the amplitude remaining in the nanoparticle after $t_{\rm swap}$ is $\approx0.18$), averaging over $\mathcal{P}(\bm\beta)$ gives $F\approx0.955$ and $0.932$ for the Fock and cat states, which differ by less than 1\% from the values for the Gaussian initial state with $\langle\hat n_p\rangle_{\rm ss}$ used in Fig.~\ref{fig5} (0.953 and 0.931). The rigorous bound $F\geq0.98\,e^{-\gamma_ct}F_{\rm nc}$, which counts only the undisplaced part of the mixture, gives 0.94 and 0.92. With the Gaussian dissipator, the fidelities in Fig.~\ref{fig5} would be 0.83 and 0.66 for the Fock and cat states, instead of 0.95 and 0.93.

Finally, thermal photons are scattered, emitted and absorbed by the nanoparticle. Since their wavelength ($\sim10\,\rm\mu m$ at room temperature) is much larger than all relevant separations, they act as position diffusion with localization rate $\Lambda_{\rm bb}=\Lambda_{\rm bb,sc}+\Lambda_{\rm bb,e}+\Lambda_{\rm bb,a}$, where~\cite{RomeroIsart2011}
\begin{align}
    \Lambda_{\rm bb,sc}&=\frac{8!\,8\zeta(9)cR_p^6}{9\pi}\left(\frac{k_BT}{\hbar c}\right)^9\mathrm{Re}^2\left[\frac{\varepsilon_{\rm bb}-1}{\varepsilon_{\rm bb}+2}\right],\\
    \Lambda_{\rm bb,e(a)}&=\frac{16\pi^5cR_p^3}{189}\left(\frac{k_BT}{\hbar c}\right)^6\mathrm{Im}\left[\frac{\varepsilon_{\rm bb}-1}{\varepsilon_{\rm bb}+2}\right],
\end{align}
with $\zeta$ the Riemann zeta function, $\varepsilon_{\rm bb}$ the average dielectric constant in the thermal infrared, and equal internal and environment temperatures $T=300$~K, since the nanoparticle is not illuminated. For our parameters and $|(\varepsilon_{\rm bb}-1)/(\varepsilon_{\rm bb}+2)|\lesssim1$, we obtain $\Lambda_{\rm bb}\lesssim1\times10^{19}\,\rm m^{-2}s^{-1}$, dominated by emission and absorption. This corresponds to a heating rate $\Gamma_{\rm bb}=2\Lambda_{\rm bb}x_{\mathrm{zpm},p}^2\lesssim2\pi\times0.03\,\rm Hz$, more than an order of magnitude below the trap noise $\Gamma_{\rm td}$, and to a decoherence rate $\lesssim3\,\rm s^{-1}$ for a cat state with $\alpha=2$. Blackbody radiation is therefore negligible at room temperature.

\section{Second-order drive effects}\label{A-PR}
 
Here we derive the effective coupling~\eqref{gprime} and the additional terms that appear in Eq.~\eqref{H_RWA} when the fast-rotating terms of the Hamiltonian~\eqref{H_ab} are treated to second order. We denote $a=A_{\rm cl}/D_0$ and $\kappa_i=\Omega_i'^2-\Omega_i^2$, corresponding to the static value of $\mathcal{K}^2_i(t)$ in Eq.~\eqref{Freq_t}, so that we have $\delta\Omega_i'=-3a\kappa_i/(2\Omega_i')$. For our parameters, $\kappa_i/\Omega_i'^2\approx0.66$. In the frame rotating with $\omega_d$ for the ion operators, the terms neglected in the RWA oscillate at multiples of $\omega_d$ and can be written as $\sum_{\nu>0}(\hat{h}_\nu e^{-\mathrm{i}\nu t}+\mathrm{H.c.})$ with $\nu\in\{\omega_d,2\omega_d,3\omega_d\}$. To second order, their effect is described by the effective Hamiltonian~\cite{James2007}
\begin{equation}
    \hat{H}_{\rm eff}/\hbar=\sum_{\nu>0}\frac{[\hat{h}_\nu^\dagger,\hat{h}_\nu]}{\nu},
    \label{H_eff}
\end{equation}
which has to be added to Eq.~\eqref{H_RWA}. The dominant contribution comes from
\begin{equation}
    \hat{h}_{\omega_d}=g_0\hat{a}_i(\hat{a}_p+\hat{a}_p^\dagger)+\frac{\delta\Omega_i'}{4}\big(\hat{a}_i^2+2\hat{a}_i^\dagger\hat{a}_i\big),
\end{equation}
specifically from the cross terms between the static coupling and the frequency modulation. They produce a Hamiltonian $-(g_0\delta\Omega_i'/\omega_d)(\hat{a}_i+\hat{a}_i^\dagger)(\hat{a}_p+\hat{a}_p^\dagger)$, i.e., a correction of the same form as the coupling term in Eq.~\eqref{H_RWA}. Since $2g_0\delta\Omega_i'/\omega_d=\tilde{g}\kappa_i/(\Omega_i'\omega_d)$, the total ion-nanoparticle motional coupling rate is corrected to $g'=\tilde{g}[1-\kappa_i/(\Omega_i'\omega_d)]\approx\tilde{g}\,\Omega_i^2/\Omega_i'^2$, which turns into Eq.~\eqref{gprime} when the slow free evolution is also taken into account in the denominators. The correction is linear in $A_{\rm cl}$, like $\tilde{g}$, so it does not become negligible for weak driving.
 
The terms quadratic in $g_0$ and in $\delta\Omega_i'$, together with the shift of the mean distance, $\bar{D}_{\rm cl}\approx D_0(1+a^2/2)$, and the second harmonic of the classical motion, $D_{\rm cl}(t)\supset-D_0a^2[\kappa_i/(4\Omega_i'^2)]\cos(2\omega_dt)$, lead to the frequency shifts
\begin{equation}
    \Delta\nu_i=\frac{3a^2\kappa_i}{4\Omega_i'}\frac{\Omega_i^2}{\Omega_i'^2},\quad \Delta\nu_p=-\frac{2g_0^2}{\omega_d},
\end{equation}
for ion and nanoparticle respectively, so that the resonance condition becomes $\omega_d=\Omega_i'-\Omega_p'+\Delta_d$ with $\Delta_d=\Delta\nu_i-\Delta\nu_p$ ($\Delta_d/2\pi\approx0.31\,\rm kHz$ for $A_{\rm cl}=0.66\,\rm\mu m$ and $1.5\,\rm kHz$ for $A_{\rm cl}=1.5\,\rm\mu m$; the exact solution gives somewhat smaller values, see Appendix~\ref{A-FL}). They also produce the parametric term $\chi_i(\hat{a}_i^2+(\hat{a}_i^{\dagger})^2)$ with
\begin{equation}
    \chi_i=\frac{3a^2\kappa_i}{8\Omega_i'}\Big(1+\frac{\kappa_i}{4\Omega_i'^2}-\frac{3\kappa_i}{2\Omega_i'^2}\Big)\approx0.18\,\frac{3a^2\kappa_i}{8\Omega_i'}.
    \label{chi_i}
\end{equation}
The first two terms in the parentheses come from the second harmonic of the ion frequency, $\Omega_i'(t)\supset\delta\Omega_{i,2}'\cos(2\omega_d t)$, generated directly by the term $\propto a^2$ in the expansion of $D_{\rm cl}^{-3}$ and by the second harmonic of $D_{\rm cl}$, while the last term is the second-order contribution of the first harmonic $\delta\Omega_i'$. The corresponding term for the nanoparticle, $-(g_0^2/\omega_d)(\hat{a}_p^2+(\hat{a}_p^\dagger)^2)$, is negligible ($g_0^2/\omega_d\approx2\pi\times10\,\rm Hz$).
 
Including $\chi_i$, the Hamiltonian~\eqref{H_RWA} is modified as
\begin{multline}
    \hat{H}/\hbar=\Omega_p'\hat{a}_i^{\dag}\hat{a}_i+\Omega_p'\hat{a}_p^{\dag}\hat{a}_p+\frac{g'}{2}(\hat{a}_i\hat{a}_p+\hat{a}_i^{\dag}\hat{a}_p+\mathrm{H.c.})\\+\chi_i(\hat{a}_i^2+(\hat{a}_i^{\dagger})^2).
\end{multline}
This Hamiltonian is still exactly solvable, with the squared normal-mode frequencies
\begin{equation}
    \Omega_{\pm}^2=\Omega_p'^2-2\chi_i^2\pm\sqrt{4\chi_i^4+g'^2\Omega_p'(\Omega_p'-2\chi_i)}.
\end{equation}
For $U_d=40\,\rm\mu V$ without the compensation tone ($A_{\rm cl}\approx0.7\,\rm\mu m$), the exact numerical solution (Appendix~\ref{A-FL}) gives $\chi_i/2\pi\approx50$--$80\,\rm Hz$ -- in agreement with Eq.~\eqref{chi_i} -- which is no longer negligible compared to $g'/2\pi\approx140\,\rm Hz$. Since $\chi_i\propto A_{\rm cl}^2$ grows faster than $g'\propto A_{\rm cl}$, it dominates for stronger driving, where it squeezes the ion motion and degrades the cooling. The second harmonic of $D_{\rm cl}$ is generated by the nonlinearity of the Coulomb force acting on the driven ion. The compensation tone at $2\omega_d$ introduced in Sec.~\ref{sec2} drives the ion directly at $2\omega_d$, in antiphase with this second harmonic, and thereby cancels the second term in the parentheses of Eq.~\eqref{chi_i}; if it is canceled completely, only $1-3\kappa_i/(2\Omega_i'^2)\approx0.01$ remains, suppressing $\chi_i$ by a factor $\sim 18$. This near cancellation is not a coincidence of our parameters: for $M_p\Omega_p^2\gg M_i\Omega_i^2$ (here by a factor of 30), the nanoparticle barely moves and the force balance of the ion gives $Q_iQ_p/(4\pi\varepsilon_0D_0^3)\approx M_i\Omega_i^2$, so that $\kappa_i\approx2\Omega_i^2$, $\kappa_i/\Omega_i'^2\approx2/3$ and $1-3\kappa_i/(2\Omega_i'^2)\approx0$, independently of the other parameters. In practice, we chose the amplitude and phase of the tone to optimize the cooling, for which the second harmonic is suppressed by a factor $\approx3$, leaving $|\chi_i|/2\pi\approx0.1\,\rm kHz$. This barely affects $\langle\hat{n}_p\rangle_{\rm ss}$, but increases the ion occupation in the exact solution compared to Eq.~\eqref{H_RWA} with $\chi_i=0$ [Fig.~\ref{fig3}(c)]. Conversely, the parametric term can be enhanced by stronger driving, or by driving closer to resonance, $|\omega_d-\Omega_i'|<\Omega_p'$. Driving the parametric resonance directly will typically lead to an instability, but it can also be utilized in some protocols, such as the one discussed in Appendix~\ref{A-ND}.
 
For two-tone driving (Sec.~\ref{sec4}), the products of the first harmonics at $\omega_{d1}$ and $\omega_{d2}$ produce, in the same way, the term $\chi_{12}(\hat{a}_i^2+(\hat{a}_i^\dagger)^2)$ with
\begin{equation}
    \chi_{12}=\frac{3a_1a_2\kappa_i}{4\Omega_i'}\Big(1+\frac{\kappa_i}{4\Omega_i'^2}-\frac{3\kappa_i}{2\Omega_i'^2}\Big),
\end{equation}
where $a_{1,2}=A_{{\rm cl},1,2}/D_0$, or $|\chi_{12}|\approx|g_1'g_2'|/(2\pi\times125\,\rm Hz)$ in terms of the effective couplings. Since $\omega_{d1}+\omega_{d2}=2\Omega_i'$, this term is resonant in the frame of Eq.~\eqref{H_st}. It can be canceled by a third tone at $\omega_{d1}+\omega_{d2}$ that nulls the corresponding Fourier component of $D_{\rm cl}(t)$ (amplitude $\approx1\,\rm mV$ for $g_1'/2\pi\approx170\,\rm Hz$).
 
\section{Exact numerical solution}\label{A-FL}
 
\begin{figure}[t]
	\centering
    \includegraphics[width=86mm]{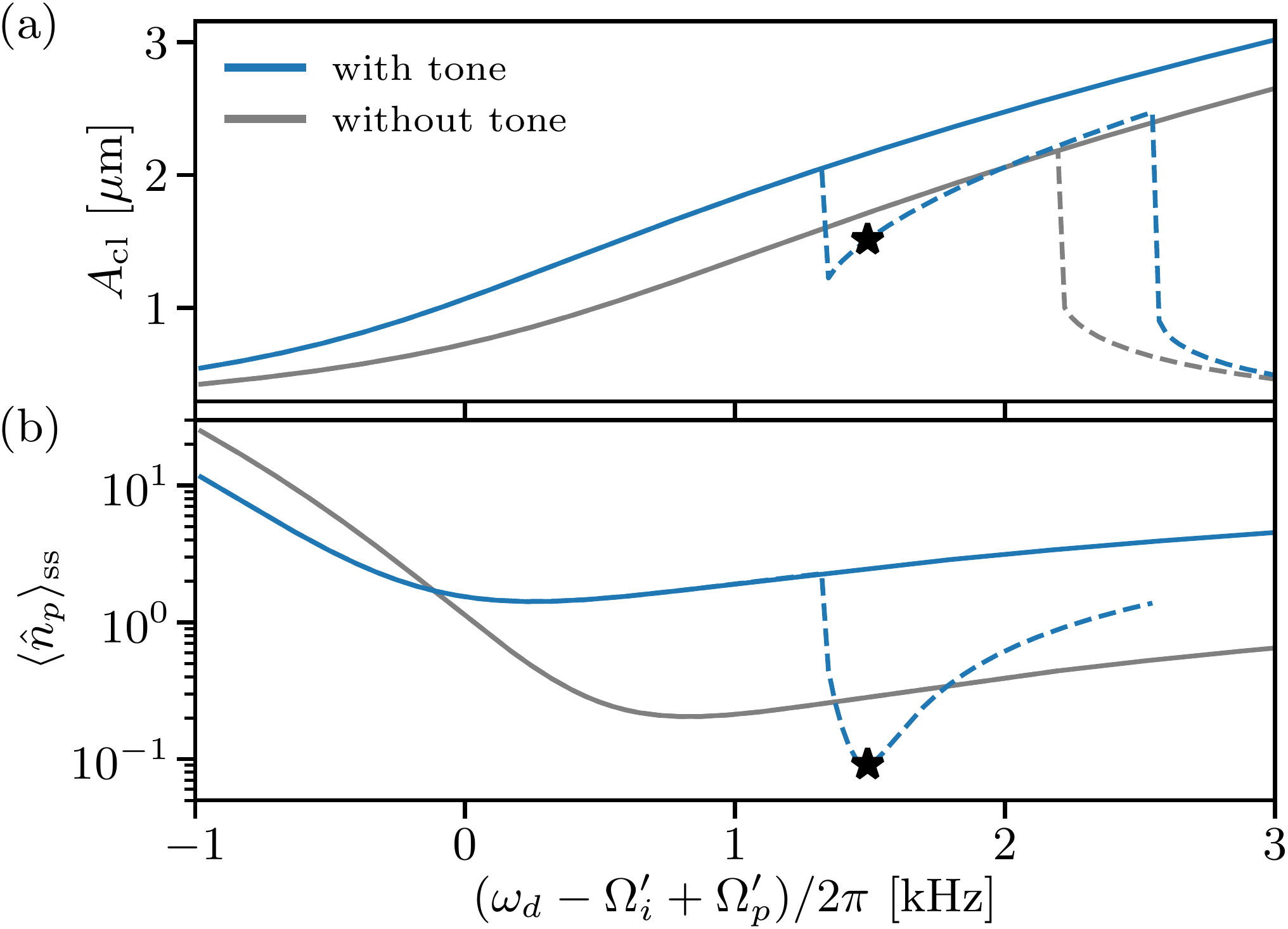}
    \caption{\textbf{Exact solution.} (a) Steady-state classical amplitude $A_{\rm cl}$ and (b) nanoparticle occupation $\langle\hat{n}_p\rangle_{\rm ss}$ as a function of the drive frequency for $U_d=60\,\rm\mu V$ and $\gamma_i/2\pi=350\,\rm Hz$, with (blue) and without (grey) the compensation tone at $2\omega_d$. Solid (dashed) lines: upward (downward) frequency sweep; the star marks the parameters of Fig.~\ref{fig2}. All results for $\Gamma_{\rm td}/2\pi=1\,\rm Hz$.}
    \label{fig6}
\end{figure}

To verify the results obtained with Eq.~\eqref{H_RWA}, we solved the quantum dynamics without the RWA and without expanding the coefficients of the Hamiltonian in $A_{\rm cl}/D_0$, i.e. we solve the master equation Eq.~\eqref{master-eq} with the Hamiltonian \eqref{Hdisplaced}. As in Sec.~\ref{sec2b}, the quantum fluctuations are treated to second order around the classical orbit, but the time-dependent Hessian is evaluated on the full numerical periodic orbit, including the compensation tone, and the resulting Gaussian dynamics is solved in the laboratory frame. We first find the exact periodic solution of the classical equations of motion~\eqref{di_cl} and~\eqref{dp_cl} (including the compensation tone) by a shooting method. The quantum fluctuations around it are then described, to second order in the fluctuations, by the full time-dependent Hessian of the potential, and the resulting Gaussian master equation in the laboratory frame is periodic in time with period $2\pi/\omega_d$. Its periodic steady state is obtained from the monodromy matrix (Floquet theory) by solving a discrete Lyapunov equation. For two-tone driving, we choose commensurate drive frequencies so that the dynamics is periodic with period $2\pi/(\Omega_p'-\delta)$.
 
The minimal splitting of the Floquet quasi-energies, which equals $|g'|$, agrees with Eq.~\eqref{gprime} within 5\% for $A_{\rm cl}\le1.2\,\rm\mu m$, and the frequency value at which it arises agrees with $\Delta_d$ within $\approx20\%$. In the main text, we take $g'$ as the $\omega_d$ component of the exact ion--nanoparticle coupling in the basis of the static normal modes, which is slightly larger than the splitting (e.g., $2\pi\times317\,\rm Hz$ compared to $2\pi\times311\,\rm Hz$ at the operating point of Fig.~\ref{fig2}); this difference changes $\langle\hat{n}_p\rangle_{\rm ss}$ obtained with Eq.~\eqref{H_RWA} by only $\approx1\%$. The nonlinear classical response for the single-tone drive is shown in Fig.~\ref{fig6}(a): since the resonance frequency of the driven ion increases with the amplitude, the response folds over and becomes bistable for large amplitudes. The compensation tone at $2\omega_d$ modifies the nonlinearity and the response becomes multistable already for $\omega_d-\Omega_i'+\Omega_p'\gtrsim2\pi\times1.3\,\rm kHz$. The operating point of Fig.~\ref{fig2} lies on the branch that is reached by switching on the drive at fixed frequency [Fig.~\ref{fig2}(a)] or by a downward frequency sweep, while an upward sweep ends on a branch with larger amplitude and worse cooling. Since switching on the drive at a detuning only $\approx10\,\rm Hz$ smaller already ends on the large-amplitude branch, it is safer to switch on the drive at a slightly larger $\omega_d$ and sweep it down to the operating point (Sec.~\ref{sec_exp}). The branch is then followed down to $\omega_d-\Omega_i'+\Omega_p'\approx2\pi\times1.35\,\rm kHz$. Figure~\ref{fig6}(b) shows that the tone strongly improves the cooling, and that the exact solution reproduces $\langle\hat{n}_p\rangle_{\rm ss}$ obtained with Eq.~\eqref{H_RWA} (see also Fig.~\ref{fig3}). Without the compensation, increasing $U_d$ beyond $\approx50\,\rm\mu V$ makes the cooling worse even if $\gamma_i$ is increased into the kHz range, because of the parametric term $\chi_i$. With the compensation tone, the lowest value $\langle\hat{n}_p\rangle_{\rm ss}\approx0.09$ is obtained for $U_d\approx60\,\rm\mu V$ (the parameters of Fig.~\ref{fig2}), and $\langle\hat{n}_p\rangle_{\rm ss}\lesssim0.1$ over a wide range of $U_d$, limited mainly by $\Gamma_{\rm gas}$. All obtained values are time-averaged, ion phonon occupations fluctuate within $\pm 15$\% of the mean value, while nanoparticle phonon occupations are constant to 4 digits. Since the tone modulates the ion frequency at $2\omega_d$, it also acts as a parametric drive of the classical motion. For the amplitudes used here, the oscillation at $\omega_d$ is sustained even for small $U_d$, which mainly fixes its phase, so that $A_{\rm cl}\approx1.2\,\rm\mu m$ is obtained already for $U_d=10\,\rm\mu V$ after optimizing $\omega_d$, $\gamma_i$ and $U_d^{(2)}$, $\varphi_2$ of the compensation tone.
 
For two-tone driving, we followed the stationary solution while increasing $U_{d1}$, adjusting $U_{d2}$ to keep $g_2'/g_1'\approx0.72$. Because of the nonlinear response, this required $U_{d2}>U_{d1}$. With the compensating third tone and the center frequency $(\omega_{d1}+\omega_{d2})/2$ optimized, the squeezing for $\delta_i=\chi=0$ agrees with the results of Sec.~\ref{sec4} within $0.3$~dB for $g_1'/2\pi\lesssim170\,\rm Hz$. The additional parameters of Sec.~\ref{sec4} are implemented by shifting the center frequency by $-\delta_i$ with respect to the drive-induced frequency shift, and, for squeezing, by detuning the amplitude and phase of the third tone from the point where it cancels $\chi_{12}$. Optimizing these parameters in the exact model gives $3.0$~dB of squeezing and $E_N\approx0.29$ for $g_1'/2\pi=150\,\rm Hz$ and $\Gamma_{\rm td}/2\pi=1\,\rm Hz$, compared to $3.2$~dB and $0.35$ obtained from the effective model, Eq.~\eqref{H_RWA2}. The effective model does not include the modulation of the ion frequency at $\omega_{d2}-\omega_{d1}=2(\Omega_p'-\delta)$, which the two tones generate in second order, with an amplitude of order kHz at these optima. The exact model includes it, and the agreement above is obtained after optimizing the residual parametric term in the exact model; for the same parameters, the two models can differ by several tenths of a dB. The exact solution is integrated with 96 Runge--Kutta steps per ion period, which is converged. The entanglement optimum lies very close to the boundary of stability of the exact periodic solution, so that the exact value of $E_N$ is sensitive to the drive parameters. For stronger driving, the exact squeezing and entanglement do not increase significantly, while Eq.~\eqref{H_st} increasingly overestimates them, and no stable periodic solution is found for $g_1'/2\pi\gtrsim210$--$250\,\rm Hz$, depending on the detunings. Without the third tone, the system is unstable already for $g_1'/2\pi\approx100\,\rm Hz$ (for $\gamma_i/2\pi=100\,\rm Hz$) due to the cross-term $\propto \chi_{12}$. Additional tones at $2\omega_{d1}$ and $2\omega_{d2}$, which cancel the corresponding harmonics of $D_{\rm cl}(t)$, reduce the ion occupation but slightly decrease the squeezing (e.g., from $2.1$ to $1.7$~dB at $g_1'/2\pi=150\,\rm Hz$), and increase $E_N$ only by $\approx15\%$ (e.g., from $0.28$ to $0.32$ at $g_1'/2\pi=190\,\rm Hz$) without extending the range of stable driving, so we do not include them.
 
\section{Higher-order interaction terms}\label{A-HO}
 
In the main text, the Coulomb interaction was expanded to second order in fluctuations around the equilibrium positions. Here we show that higher-order terms in the expansion are negligible for all our proposed protocols. The full Coulomb interaction is given by
\begin{equation}
    \hat{V}=\frac{Q_iQ_p}{4\pi\varepsilon_0}\frac{1}{|\hat{\mathbf{r}}_i-\hat{\mathbf{r}}_p|},\quad \hat{\mathbf{r}}_j=[\hat{x}_j, \hat{y}_j, \hat{z}_j],\quad (j=i,p).
\end{equation}
We expand $\hat{V}$ around the classical positions $\mathbf{r}_{j,\rm cl}$, with $\mathbf{r}_{i,\rm cl}-\mathbf{r}_{p,\rm cl}=[0, 0, D_{\rm cl}]$~\cite{Gupta2025}. Denoting $\hat{\mathbf{r}}_j\to \hat{\mathbf{r}}_j-\mathbf{r}_{j,\rm cl}$, we can write the two expansion parameters as
\begin{equation}
    \hat{\alpha}=\frac{\hat{z}_i-\hat{z}_p}{D_{\rm cl}}, \quad \hat{\beta}=\frac{1}{D_{\rm cl}}\sqrt{(\hat{x}_i-\hat{x}_p)^2+(\hat{y}_i-\hat{y}_p)^2}.
\end{equation}
Up to the third order we obtain
\begin{equation}
    \hat{V}=\frac{Q_iQ_p}{4\pi\varepsilon_0}\frac{1}{D_{\rm cl}}\Big(1-\hat{\alpha}+\hat{\alpha}^2-\hat{\alpha}^3+\frac{3}{2}\hat{\alpha}\hat{\beta}^2-\frac{1}{2}\hat{\beta}^2\Big).
    \label{V_order3}
\end{equation}
The first three terms were included in the main-text analysis, and the last term couples only radial modes and is irrelevant for the dynamics along $z$ axis; we are thus interested in the terms $\hat{\alpha}^3$ and $\hat{\alpha}\hat{\beta}^2$. Writing the position operators as $\hat{x}_j=x_{\mathrm{zpm}, xj}(\hat{a}_{xj}+\hat{a}_{xj}^\dagger)$ (analogously in the $y$ direction, while along $z$ we keep the notation $\hat{z}_j=x_{\mathrm{zpm},j}(\hat{a}_j+\hat{a}_j^\dagger)$ of the main text), it becomes clear that the magnitude of each further order in the expansion is suppressed by $x_{\mathrm{zpm},j}/D_{\rm cl}$. This is of the order of magnitude $\sim 10^{-3}$ for the ion and $\sim 10^{-6}$ for the nanoparticle, so all terms higher than third order are completely negligible, while third-order terms that contain only ion operators can be non-negligible compared to the second-order mixed (interaction) terms.
 
The first such term is proportional to $\hat{z}_i^3$, which arises from $\alpha^3$ in Eq.~\eqref{V_order3}. Written in terms of ladder operators, we have
\begin{equation}
    \hat{H}_{3,z}=-\frac{Q_iQ_p}{4\pi\varepsilon_0}\frac{x_{\mathrm{zpm},i}^3}{D_{\rm cl}^4}(\hat{a}_{i}+\hat{a}_{i}^\dagger)^3.
\end{equation}
As before, we move to a frame rotating with $\omega_d=\Omega_i'-\Omega_p'+\Delta_d$ and keep in mind that classical distance is time-dependent, $D_{\rm cl}(t)=D_0+A_{\rm cl}\cos(\omega_d t)$. After a RWA, we are left with
\begin{equation}
    \hat{H}_{3,z,\rm RWA}/\hbar=\frac{G'}{2}(3\hat{a}_{i}^\dagger\hat{a}_{i}\hat{a}_{i}+3\hat{a}_{i}+\mathrm{H.c.}),
    \label{H3RWA}
\end{equation}
with
\begin{equation}
    G'=4\frac{Q_iQ_p}{4\pi\varepsilon_0}\frac{x_{\mathrm{zpm},i}^3A_{\rm cl}}{\hbar D_0^5}=\frac{2x_{\mathrm{zpm},i}^2}{3x_{\mathrm{zpm},p}D_0}\tilde{g}.
\end{equation}
For the parameters used here, $G'/2\pi\approx 41\,\mathrm{Hz}\approx 0.044 \tilde{g}/2\pi$. Like the coupling, this term is reduced by second-order contributions of the fast-rotating terms (Appendix~\ref{A-PR}). An exact classical calculation of the phonon-number-dependent displacement of the driven ion, which is the main effect of the cubic term, gives an effective value $G'_{\rm eff}\approx0.14\,G'\approx2\pi\times6\,\rm Hz$. Terms linear in ladder operators in Eq.~\eqref{H3RWA} slightly displace the equilibrium position by $\alpha_{\rm disp}=-3G'_{\rm eff}/(2\Omega_p')\approx -0.008$ and can be canceled by an appropriate frame displacement without affecting the physics. The cubic terms can be treated perturbatively since $G'_{\rm eff}\ll \Omega_p'$. A rigorous treatment can be done using a Schrieffer--Wolff transformation, but for our purposes it is enough to note that the first-order correction to the energies vanishes, and that the second-order correction is proportional to $G_{\rm eff}'^2n^2/\Omega_p'\sim4\times 10^{-5}n^2\,\Omega_p'$ for the $n$-th number state of the ion. Therefore, the effect on the dynamics of low-lying states is small. This was further confirmed by the state-transfer simulations of Fig.~\ref{fig5}, which include the cubic term~\eqref{H3RWA}: compared to simulations without it, the transfer fidelity changes by less than 1\% for the states considered in the main text.
 
 The effects of the cross-term $\hat{\alpha}\hat{\beta}^2$ are even smaller, because the ion's mechanical frequencies are much higher in the transverse plane (see Appendix~\ref{A-EP}), leading to a coupling smaller by an order of magnitude. Also, terms that survive the RWA are e.g. $\hat{a}_{xi}^\dagger\hat{a}_{xi}\hat{a}_{i}$ which almost vanish if the ion's motion along $x$ (equivalently $y$) direction is cooled very close to the ground state.
 
Despite the fact that nonlinear terms do not affect the dynamics strongly for the system parameters we used, we note that the nonlinear terms can be enhanced if one wants to utilize them. This can be done just by driving the classical motion at $\omega_d\sim \Omega_i'$ strongly, but also by introducing another driving term at $\sim 3\Omega_i'$ that would make $\hat{a}_i^3+(\hat{a}_i^\dagger)^3$ terms resonant, opening the possibility to prepare non-Gaussian states of motion without coupling to auxiliary degrees of freedom.
 
\section{Nonlinear dissipation}\label{A-ND}

An alternative route toward non-Gaussian state preparation of the nanoparticle is nonlinear dissipation.
It is possible to engineer nonlinear motional dissipation of the ion~\cite{Poyatos1996, Mirrahimi2014, Bartolo2016} by driving the second red sideband of the 729~nm optical qubit transition at the frequency $\Omega_q-2\Omega_i'$, where $\Omega_q$ is the qubit transition frequency, combined with rapid repumping of the excited state (via the 854~nm transition). This drives the transition $\ket{\downarrow, n}\to\ket{\uparrow, n-2}\to\ket{\downarrow, n-2}$, and the process enters the master equation with the dissipator $\gamma_2\mathcal{L}_{\hat{a}_i^2,(\hat{a}_i^\dagger)^2} \left(\hat{\rho}\right)$. The achievable rate $\gamma_2$ is suppressed compared to the sideband cooling rate $\gamma_i$, since the second-sideband Rabi frequency is smaller than the first-sideband one by a factor $\sim\eta/2$, with the Lamb--Dicke parameter $\eta=k_zx_{\mathrm{zpm},i}\approx 0.09$; depending on the drive strength, $\gamma_2/\gamma_i$ thus lies between $(\eta/2)^2$ and $\eta/2$. Achievable sideband cooling rates are in the kHz range, so $\gamma_2/2\pi\sim 10\,\rm Hz$ is realistic, which is much bigger than the ion heating rate $\Gamma_i$.
 
Combining the engineered nonlinear dissipation with a parametric drive at frequency $\approx2\Omega_i'$ allows the stabilization of cat states with amplitude $|\alpha|=\sqrt{2\chi_2/\gamma_2}$ -- where $\chi_2$ is the strength of the resulting parametric term -- which are the steady states of the master equation
\begin{equation}
    0=\dot{\hat{\rho}}_{\rm ss}=-\mathrm{i}[\chi_2(\hat{a}_i^2+(\hat{a}_i^\dagger)^2), \hat{\rho}_{\rm ss}]+\gamma_2\mathcal{L}_{\hat{a}_i^2,(\hat{a}_i^\dagger)^2} \left(\hat{\rho}_{\rm ss}\right).
\end{equation}
The coherence time is limited by the linear dissipation due to ion heating with rate $\Gamma_i$. If the ion and the nanoparticle are coupled, the beam-splitter term can transfer excitations between the modes, and transient Wigner negativity of the nanoparticle state can be created. However, the steady state is no longer pure, and does not have a cat structure, making the state preparation protocol presented in the main text a cleaner way to generate non-Gaussian states of the nanoparticle.
 
\section{Estimate of the Nimmrichter--Hornberger macroscopicity}\label{A-NH}
 
We estimate the macroscopicity of the nanoparticle state using the measure introduced by Nimmrichter and Hornberger \cite{Nimmrichter2013},
\begin{equation}
\mu = \log_{10}\left(\frac{\tau_e}{1~\mathrm{s}}\right),
\end{equation}
where $\tau_e$ denotes the largest classicalization timescale excluded by the experiment.
 
For a homogeneous sphere of mass $M_p$ and radius $R_p$, the minimal-modification model predicts the decoherence rate of a spatial superposition separated by $\Delta x$ to be
\begin{equation}
\Gamma(r_c)=
\frac{1}{\tau_e m_e^2}
\int d^3\mathbf q\, g_e(q)\,
|\tilde{\rho}(q)|^2
\left[
1-e^{\mathrm{i}\mathbf q\cdot\Delta\mathbf x/\hbar}
\right],
\end{equation}
where $m_e$ is the electron mass, $r_c=\hbar/\sigma_q$ is the localization length of the modification, $g_e(q)$ is the Gaussian momentum-transfer distribution defined in Ref.~\cite{Nimmrichter2013}, and
\begin{equation}
\tilde{\rho}(q)=M_p\,\phi(qR_p/\hbar), \quad
\phi(x)=\frac{3}{x^3}(\sin x-x\cos x)
\end{equation}
is the Fourier transform of the mass density. Observing interference with visibility $f$ after an evolution time $t$ implies $\Gamma t\lesssim |\ln f|$, i.e., the whole reduction of the visibility is conservatively attributed to the modification~\cite{Schrinski2019}. For a cat state $\ket{\psi}\propto\ket{\alpha}+\ket{-\alpha}$, we define the visibility as the coherence between the two branches,
\begin{equation}
    f=\frac{2\lvert\bra{\alpha}\hat{\rho}\ket{-\alpha}\rvert}{\bra{\alpha}\hat{\rho}\ket{\alpha}+\bra{-\alpha}\hat{\rho}\ket{-\alpha}},
    \label{f_cat}
\end{equation}
which equals 1 for the pure cat state and 0 for a mixture of the two branches. The excluded timescale is obtained by minimizing $\Gamma$ (or equivalently maximizing $\tau_e$) over $r_c$.
 
Typically, $\Delta x\ll R_p$, so we expand $\tau_e$ in $\Delta x/R_p$:
\begin{equation}
\tau_e(r_c)\simeq
\frac{t}{|\ln f|}
\left(\frac{M_p}{m_e}\right)^2
\frac{\Delta x^2}{2R_p^2}
H\!\left(\frac{R_p}{r_c}\right),
\end{equation}
where
\begin{equation}
H(y)=
\frac{y^2}{3\sqrt{\pi/2}}
\int_0^\infty
k^4e^{-k^2/2}
\phi^2(ky)\,dk.
\end{equation}
Numerical maximization gives $H_{\max}\approx0.42$ at $R_p/r_c\simeq1.16$, resulting in
\begin{equation}
\tau_e\simeq
0.42\,
\frac{t}{|\ln f|}
\left(\frac{M_p}{m_e}\right)^2
\frac{\Delta x^2}{2R_p^2}.
\label{tau_e}
\end{equation}
For the given nanoparticle parameters and a cat state with $\alpha=2$, the branch separation oscillates in the laboratory frame as $\Delta x(t)=4\alpha x_{\mathrm{zpm},p}\cos(\Omega_p't)$, with $4\alpha x_{\mathrm{zpm},p}=0.72~\mathrm{nm}$, so $\Delta x^2$ in Eq.~\eqref{tau_e} is replaced by its time average $8\alpha^2x_{\mathrm{zpm},p}^2$. The cat state prepared in Fig.~\ref{fig5} ($\Gamma_{\rm td}/2\pi=1$~Hz) has $f\approx0.91$ directly after the transfer, and its free evolution reduces the visibility to $f=0.05$ after $t\approx35\,$ms. The decoherence during this time is dominated by the trap noise, which reduces the coherence at the rate $4\alpha^2\Gamma_{\rm td}\approx100\,\rm s^{-1}$, while gas collisions contribute only $\gamma_c\approx1.7\,\rm s^{-1}$ (Appendix~\ref{A-GAS}). From Eq.~\eqref{tau_e} we obtain $\tau_e\approx3\times10^{17}~\mathrm{s}$, corresponding to a Nimmrichter--Hornberger macroscopicity of $\mu\approx17.5$.

Equation~\eqref{tau_e} follows from the expansion of $\Gamma$ to second order in $\Delta x$, which requires $\Delta x\ll R_p$ and, since the optimal localization length $r_c\approx R_p/1.16$ is comparable to $R_p$, also $\Delta x\ll r_c$. In this limit, the decoherence rate is $\Gamma\approx\Lambda_{\rm mod}\Delta x^2$ with $\Lambda_{\rm mod}=(M_p/m_e)^2H_{\max}/(2R_p^2\tau_e)$, i.e., the modification acts on the nanoparticle as position diffusion, $-\Lambda_{\rm mod}[\hat z_p,[\hat z_p,\hat\rho]]$, which has the same form as the trap-displacement term in Eq.~\eqref{Dp}, with the heating rate
\begin{equation}
    \Gamma_x=2\Lambda_{\rm mod}x_{\mathrm{zpm},p}^2=\left(\frac{M_p}{m_e}\right)^2\frac{H_{\max}x_{\mathrm{zpm},p}^2}{R_p^2\tau_e}.
    \label{Gamma_x}
\end{equation}
Equation~\eqref{Gamma_x} applies to arbitrary states and also allows us to include the known decoherence sources. For our parameters, $\mu=17.4-\log_{10}(\Gamma_x/2\pi\,\rm [Hz])$.
In the conservative estimate, $\Gamma_x$ is the rate that, acting alone on the ideal state for the time $t$, reproduces the observed visibility $f$ (cat states) or the Wigner negativity volume $\delta(W)$ (Fock states). For cat states, this reduces to Eq.~\eqref{tau_e} for small $\Gamma_xt$, and gives $\mu\approx17.4$ for the cat state above, the same value as for the Fock state $\ket{1_p}$. Since the trap noise acts in the same way as the modification, the conservative estimate approaches
\begin{equation}
    \mu\approx \log_{10}\Big(\frac{M_p^2}{m_e^2}H_{\max}\frac{x_{\mathrm{zpm},p}^2}{R_p^2\Gamma_{\rm td}}\Big)
    \label{mu_est}
\end{equation}
for long evolution times, and it changes by less than 0.05 for cat states with $\alpha=1.5$--4 and Fock states with $N=1$--8. If the known decoherence is characterized independently, so that only additional decoherence that would halve $f$ or $\delta(W)$ has to be excluded, we obtain $\mu\approx17.9$ for the cat state and $\mu\approx18.2$ for Fock states. In all estimates we required $f\geq0.05$ and $\delta(W)\geq0.02$ at the time $t$. We note that this is an order-of-magnitude estimate, while the precise value would depend on experimental details and choice of coherence witness.

\begin{figure}[t]
	\centering
    \includegraphics[width=86mm]{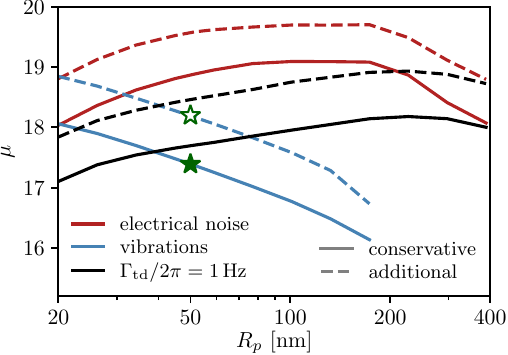}
    \caption{\textbf{Optimized macroscopicity.} Macroscopicity $\mu$ as a function of the nanoparticle radius $R_p$ for trap noise dominated by electrical noise (red), mechanical vibrations (blue), and for fixed $\Gamma_{\rm td}/2\pi=1$~Hz (black), optimized over the nanoparticle charge, $U_{0z}\le600$~V and $\Omega_p'\ge2\pi\times500$~Hz, and over cat and Fock states. Solid (dashed) lines: conservative estimate (only additional decoherence excluded). The filled (open) star marks the conservative (additional-decoherence) estimate for the parameters of Table~\ref{Table1}.}
    \label{fig7}
\end{figure}

The macroscopicity can be increased by choosing the nanoparticle parameters. Since $\mu$ is set mainly by the trap noise [Eq.~\eqref{mu_est}], and $M_px_{\mathrm{zpm},p}^2\propto1/\Omega_p'$, it scales as $\log_{10}[R_p/(\Omega_p'\Gamma_{\rm td})]$, and the optimal parameters depend on how $\Gamma_{\rm td}$ scales with the nanoparticle size and charge (Sec.~\ref{sec_exp}). In Fig.~\ref{fig7}, we consider trap noise with fixed $\Gamma_{\rm td}/2\pi=1$~Hz, trap noise dominated by mechanical vibrations ($\Gamma_{\rm td}\propto M_p/\Omega_p'$) or by electrical noise ($\Gamma_{\rm td}\propto Q_p^2/M_p\Omega_p'$), both equal to $2\pi\times1$~Hz for the parameters of Table~\ref{Table1}. For each radius, we optimize the charge $Q_p$ (below the surface-charge limit $\approx750\,e\,(R_p/135\,{\rm nm})^2$ of Ref.~\cite{Bykov2025}), the voltage $U_{0z}\leq600$~V and the frequency $\Omega_p'\geq2\pi\times500$~Hz, with the drive amplitude limited so that $|g'|\le\Omega_p'/3$. The residual ion parametric term, estimated from Eq.~\eqref{chi_i} reduced by the compensation tone, is assumed not to exceed its estimated value for the parameters of Table~\ref{Table1}, $|\chi_i|\leq0.14\,\Omega_p'$. The frequencies are calculated from the trap parameters. The cooling is calculated as in Sec.~\ref{sec3}, the state transfer as in Sec.~\ref{sec5} (neglecting the cubic term), and the free evolution includes gas collisions and blackbody radiation (Appendix~\ref{A-GAS}). For cat states, we multiply $f$ by $e^{-\gamma_ct}$, which is conservative, since the normalization in Eq.~\eqref{f_cat} largely removes the contribution of the collided part. Larger $U_{0z}$ requires a stronger radial confinement of the ion; for $U_{0z}=600$~V and $\Omega_p'\approx2\pi\times500$~Hz, a stable configuration of the two-frequency trap requires a fast RF amplitude of $\approx3$~kV at $\approx27$~MHz with the electrode geometry of Table~\ref{Table1}. For vibrational noise, smaller particles are favorable, with $\mu\propto-\log_{10}R_p^2$, reaching $\mu\approx17.4$ at $R_p=50$~nm and $18.1$ at $20$~nm. For electrical noise, a small charge of a few $e$ at large $U_{0z}$ makes $\Gamma_{\rm td}$ negligible, so that the decoherence is limited by gas collisions and blackbody radiation, and we obtain $\mu\approx19.1$ for $R_p\approx80$--$170$~nm. For fixed $\Gamma_{\rm td}$, $\mu$ increases with $R_p$ until the collision rate $\gamma_c\propto R_p^2$ becomes relevant, reaching $\mu\approx18.2$ for $R_p\approx230$~nm. The choice of the state is of minor importance as commented above. For $\Gamma_{\rm td}$ fixed independently of particle parameters, larger states are slightly favorable: for fixed $\Gamma_{\rm td}$ the optimum is $\alpha\approx2.5$--3 for cat states, limited by the transfer fidelity at larger $\alpha$, and $N\geq4$ for Fock states, while for electrical noise it is $\alpha\approx5$ ($\mu\approx19.1$, compared to $18.8$ for $\alpha=2$) and $N\approx8$ ($\mu\approx18.9$). If vibrational noise dominates, $\mu$ is almost independent of the state. If only the additional decoherence is attributed to the modification, Fock states with $N=3$--4 give values larger than those of cat states by $\approx0.2$.

\bibliography{ref}

@article{Gupta2025,
  title         = {Quantum theory of electrically levitated nanoparticle-ion systems: Motional dynamics and sympathetic cooling},
  author        = {Gupta, Saurabh and Faulend, Bernard and Bykov, Dmitry S. and Northup, Tracy E. and Gonzalez-Ballestero, Carlos},
  journal       = {Phys. Rev. Res.},
  year          = {2026},
  note          = {to be published},
  doi           = {10.1103/t9fh-4bsq},
  url           = {https://doi.org/10.1103/t9fh-4bsq},
  eprint        = {2511.21495},
  archivePrefix = {arXiv},
  primaryClass  = {quant-ph}
}

@article{Bykov2025,
  title = {Nanoparticle Stored with an Atomic Ion in a Linear {P}aul Trap},
  author = {Bykov, Dmitry S. and Dania, Lorenzo and Goschin, Florian and Northup, Tracy E.},
  journal = {Phys. Rev. Lett.},
  volume = {135},
  issue = {21},
  pages = {213602},
  numpages = {6},
  year = {2025},
  month = {Nov},
  publisher = {American Physical Society},
  doi = {10.1103/5lzj-f61x},
  url = {https://link.aps.org/doi/10.1103/5lzj-f61x}
}

@article{Roos1999,
  title = {Quantum State Engineering on an Optical Transition and Decoherence in a {P}aul Trap},
  author = {Roos, Ch. and Zeiger, Th. and Rohde, H. and N\"agerl, H. C. and Eschner, J. and Leibfried, D. and Schmidt-Kaler, F. and Blatt, R.},
  journal = {Phys. Rev. Lett.},
  volume = {83},
  issue = {23},
  pages = {4713--4716},
  numpages = {0},
  year = {1999},
  month = {Dec},
  publisher = {American Physical Society},
  doi = {10.1103/PhysRevLett.83.4713},
  url = {https://link.aps.org/doi/10.1103/PhysRevLett.83.4713}
}

@article{Nimmrichter2013,
  title = {Macroscopicity of Mechanical Quantum Superposition States},
  author = {Nimmrichter, Stefan and Hornberger, Klaus},
  journal = {Phys. Rev. Lett.},
  volume = {110},
  issue = {16},
  pages = {160403},
  numpages = {5},
  year = {2013},
  month = {Apr},
  publisher = {American Physical Society},
  doi = {10.1103/PhysRevLett.110.160403},
  url = {https://link.aps.org/doi/10.1103/PhysRevLett.110.160403}
}

@article{Dinter2026,
  title = {Three-dimensional and selective displacement sensing of a levitated nanoparticle via spatial mode decomposition},
  author = {Dinter, Thomas J. and Roberts, Reece and Volz, Thomas and Schmidt, Mikolaj K. and Laplane, Cyril},
  journal = {Phys. Rev. Res.},
  volume = {8},
  issue = {2},
  pages = {023220},
  numpages = {13},
  year = {2026},
  month = {May},
  publisher = {American Physical Society},
  doi = {10.1103/9mnc-ddvw},
  url = {https://link.aps.org/doi/10.1103/9mnc-ddvw}
}

@article{Fiurasek2013,
  title = {Witnessing negativity of {W}igner function by estimating fidelities of catlike states from homodyne measurements},
  author = {Fiur\'a\ifmmode \check{s}\else \v{s}\fi{}ek, Jarom\'{\i}r and Je\ifmmode \check{z}\else \v{z}\fi{}ek, Miroslav},
  journal = {Phys. Rev. A},
  volume = {87},
  issue = {6},
  pages = {062115},
  numpages = {9},
  year = {2013},
  month = {Jun},
  publisher = {American Physical Society},
  doi = {10.1103/PhysRevA.87.062115},
  url = {https://link.aps.org/doi/10.1103/PhysRevA.87.062115}
}

@article{Chabaud2021,
  doi = {10.22331/q-2021-06-08-471},
  url = {https://doi.org/10.22331/q-2021-06-08-471},
  title = {Witnessing {W}igner negativity},
  author = {Chabaud, Ulysse and Emeriau, Pierre-Emmanuel and Grosshans, Fr{\'{e}}d{\'{e}}ric},
  journal = {{Quantum}},
  issn = {2521-327X},
  publisher = {{Verein zur F{\"{o}}rderung des Open Access Publizierens in den Quantenwissenschaften}},
  volume = {5},
  pages = {471},
  month = jun,
  year = {2021}
}

@article{Li2011,
	author = {Li, Tongcang and Kheifets, Simon and Raizen, Mark G.},
	date = {2011/07/01},
	doi = {10.1038/nphys1952},
	id = {Li2011},
	issn = {1745-2481},
	journal = {Nat. Phys.},
	number = {7},
	pages = {527--530},
	title = {Millikelvin cooling of an optically trapped microsphere in vacuum},
	url = {https://doi.org/10.1038/nphys1952},
	volume = {7},
	year = {2011}
}

@article{Beresnev1990, 
    title={Motion of a spherical particle in a rarefied gas. {P}art 2. {D}rag and thermal polarization}, 
    volume={219},
    DOI={10.1017/S0022112090003007}, 
    journal={J. Fluid Mech.}, 
    author={Beresnev, S. A. and Chernyak, V. G. and Fomyagin, G. A.}, 
    year={1990},
    pages={405--421}
}

@article{Schrodinger1935,
  author = {Erwin Schr{\"o}dinger},
  title = {Die gegenw{\"a}rtige {S}ituation in der {Q}uantenmechanik},
  journal = {Naturwissenschaften},
  volume = {23},
  number = {48},
  pages = {807--812},
  year = {1935},
  doi = {10.1007/BF01491891}
}

@book{DeWitt2011,
  editor    = {C{\'e}cile M. DeWitt and Dean Rickles},
  title     = {The Role of Gravitation in Physics: Report from the 1957 Chapel Hill Conference},
  series    = {Max Planck Research Library for the History and Development of Knowledge: Sources},
  volume    = {5},
  publisher = {Edition Open Access},
  address   = {Berlin},
  year      = {2011},
  isbn       = {9783869319636},
  doi        = {10.34663/9783945561294-00}
}

@book{Feynman1995,
  author    = {Richard P. Feynman and Fernando B. Morinigo and William G. Wagner},
  title     = {Feynman Lectures on Gravitation},
  editor    = {Brian Hatfield},
  publisher = {Addison-Wesley},
  address   = {Reading, MA},
  year      = {1995},
  isbn      = {9780201627343}
}

@article{Penrose1996,
  author = {Roger Penrose},
  title = {On Gravity's Role in Quantum State Reduction},
  journal = {Gen. Relativ. Gravit.},
  volume = {28},
  number = {5},
  pages = {581--600},
  year = {1996},
  doi = {10.1007/BF02105068}
}

@article{Bassi2013,
  author = {Angelo Bassi and Kinjalk Lochan and Seema Satin and Tejinder P. Singh and Hendrik Ulbricht},
  title = {Models of Wave-function Collapse, Underlying Theories, and Experimental Tests},
  journal = {Rev. Mod. Phys.},
  volume = {85},
  pages = {471--527},
  year = {2013},
  doi = {10.1103/RevModPhys.85.471}
}

@article{Arndt1999,
  author  = {Markus Arndt and Olaf Nairz and Julian Vos-Andreae and Claudia Keller and Gerbrand van der Zouw and Anton Zeilinger},
  title   = {Wave-Particle Duality of {C$_{60}$} Molecules},
  journal = {Nature},
  volume  = {401},
  number  = {6754},
  pages   = {680--682},
  year    = {1999},
  doi     = {10.1038/44348}
}

@article{OConnell2010,
  author  = {A. D. O'Connell and M. Hofheinz and M. Ansmann and Radoslaw C. Bialczak and M. Lenander and Erik Lucero and M. Neeley and D. Sank and H. Wang and M. Weides and J. Wenner and John M. Martinis and A. N. Cleland},
  title   = {Quantum Ground State and Single-Phonon Control of a Mechanical Resonator},
  journal = {Nature},
  volume  = {464},
  pages   = {697--703},
  year    = {2010},
  doi     = {10.1038/nature08967}
}

@article{Fein2019,
  author  = {Yaakov Y. Fein and Philipp Geyer and Patrick Zwick and Filip Kia{\l}ka and Sebastian Pedalino and Marcel Mayor and Stefan Gerlich and Markus Arndt},
  title   = {Quantum Superposition of Molecules Beyond 25 k{D}a},
  journal = {Nat. Phys.},
  volume  = {15},
  pages   = {1242--1245},
  year    = {2019},
  doi     = {10.1038/s41567-019-0663-9}
}

@article{Chu2018,
  author  = {Yiwen Chu and Prasanna Kharel and Tyler Yoon and Leo Frunzio and Peter T. Rakich and Robert J. Schoelkopf},
  title   = {Creation and Control of Multi-Phonon {F}ock States in a Bulk Acoustic-Wave Resonator},
  journal = {Nature},
  volume  = {563},
  pages   = {666--670},
  year    = {2018},
  doi     = {10.1038/s41586-018-0717-7}
}

@article{Bild2023,
  author  = {Marius Bild and Matteo Fadel and Yu Yang and Uwe von L{\"u}pke and Phillip Martin and Alessandro Bruno and Yiwen Chu},
  title   = {Schr{\"o}dinger Cat States of a 16-Microgram Mechanical Oscillator},
  journal = {Science},
  volume  = {380},
  number  = {6642},
  pages   = {274--278},
  year    = {2023},
  doi     = {10.1126/science.adf7553}
}

@article{Pedalino2026,
  author  = {Sebastian Pedalino and Bruno E. Ram{\'i}rez-Galindo and Richard Ferstl and Klaus Hornberger and Markus Arndt and Stefan Gerlich},
  title   = {Probing Quantum Mechanics with Nanoparticle Matter-Wave Interferometry},
  journal = {Nature},
  volume  = {649},
  number  = {8098},
  pages   = {866--870},
  year    = {2026},
  doi     = {10.1038/s41586-025-09917-9}
}

@article{RomeroIsart2010,
  author  = {Oriol Romero-Isart and Mathieu L. Juan and Romain Quidant and J. Ignacio Cirac},
  title   = {Toward Quantum Superposition of Living Organisms},
  journal = {New J. Phys.},
  volume  = {12},
  pages   = {033015},
  year    = {2010},
  doi     = {10.1088/1367-2630/12/3/033015}
}

@article{RomeroIsart2011,
  author  = {Oriol Romero-Isart},
  title   = {Quantum Superposition of Massive Objects and Collapse Models},
  journal = {Phys. Rev. A},
  volume  = {84},
  number  = {5},
  pages   = {052121},
  year    = {2011},
  doi     = {10.1103/PhysRevA.84.052121}
}

@article{RomeroIsart2011b,
  title = {Large Quantum Superpositions and Interference of Massive Nanometer-Sized Objects},
  author = {Romero-Isart, O. and Pflanzer, A. C. and Blaser, F. and Kaltenbaek, R. and Kiesel, N. and Aspelmeyer, M. and Cirac, J. I.},
  journal = {Phys. Rev. Lett.},
  volume = {107},
  issue = {2},
  pages = {020405},
  numpages = {4},
  year = {2011},
  month = {Jul},
  publisher = {American Physical Society},
  doi = {10.1103/PhysRevLett.107.020405},
  url = {https://link.aps.org/doi/10.1103/PhysRevLett.107.020405}
}

@article{Bose2017,
  author  = {Sougato Bose and Anupam Mazumdar and Gavin W. Morley and Hendrik Ulbricht and Markus Toro{\v{s}} and Mauro Paternostro and Andrew A. Geraci and Peter F. Barker and M. S. Kim and Gerard Milburn},
  title   = {Spin Entanglement Witness for Quantum Gravity},
  journal = {Phys. Rev. Lett.},
  volume  = {119},
  pages   = {240401},
  year    = {2017},
  doi     = {10.1103/PhysRevLett.119.240401}
}

@article{MarlettoVedral2017,
  author  = {Chiara Marletto and Vlatko Vedral},
  title   = {Gravitationally Induced Entanglement between Two Massive Particles Is Sufficient Evidence of Quantum Effects in Gravity},
  journal = {Phys. Rev. Lett.},
  volume  = {119},
  pages   = {240402},
  year    = {2017},
  doi     = {10.1103/PhysRevLett.119.240402}
}

@article{GonzalezBallestero2021,
  author  = {Carlos Gonzalez-Ballestero and Markus Aspelmeyer and Lukas Novotny and Romain Quidant and Oriol Romero-Isart},
  title   = {Levitodynamics: Levitation and Control of Microscopic Objects in Vacuum},
  journal = {Science},
  volume  = {374},
  number  = {6564},
  pages   = {eabg3027},
  year    = {2021},
  doi     = {10.1126/science.abg3027}
}

@article{Dania2024,
  author  = {Lorenzo Dania and Dmitry S. Bykov and Florian Goschin and Markus Teller and Abderrahmane Kassid and Tracy E. Northup},
  title   = {Ultrahigh Quality Factor of a Levitated Nanomechanical Oscillator},
  journal = {Phys. Rev. Lett.},
  volume  = {132},
  number  = {13},
  pages   = {133602},
  year    = {2024},
  doi     = {10.1103/PhysRevLett.132.133602}
}

@article{Delic2020,
author = {Uro{\v{s}} Deli{\'c} and Manuel Reisenbauer and Kahan Dare and David Grass and Vladan Vuleti{\'c} and Nikolai Kiesel and Markus Aspelmeyer},
title = {Cooling of a levitated nanoparticle to the motional quantum ground state},
journal = {Science},
volume = {367},
number = {6480},
pages = {892-895},
year = {2020},
doi = {10.1126/science.aba3993},
URL = {https://www.science.org/doi/abs/10.1126/science.aba3993}}

@article{Tebbenjohanns2021,
	author = {Tebbenjohanns, Felix and Mattana, M. Luisa and Rossi, Massimiliano and Frimmer, Martin and Novotny, Lukas},
	date = {2021/07/01},
	id = {Tebbenjohanns2021},
	issn = {1476-4687},
	journal = {Nature},
	number = {7867},
	pages = {378--382},
	title = {Quantum control of a nanoparticle optically levitated in cryogenic free space},
	url = {https://doi.org/10.1038/s41586-021-03617-w},
	volume = {595},
	doi = {10.1038/s41586-021-03617-w},
	year = {2021}}

@article{Rossi2025,
  title = {Quantum Delocalization of a Levitated Nanoparticle},
  author = {Rossi, M. and Militaru, A. and Carlon Zambon, N. and Riera-Campeny, A. and Romero-Isart, O. and Frimmer, M. and Novotny, L.},
  journal = {Phys. Rev. Lett.},
  volume = {135},
  issue = {8},
  pages = {083601},
  numpages = {7},
  year = {2025},
  month = {Aug},
  publisher = {American Physical Society},
  doi = {10.1103/2yzc-fsm3},
  url = {https://link.aps.org/doi/10.1103/2yzc-fsm3}
}

@article{
Kamba2025,
author = {Mitsuyoshi Kamba  and Naoki Hara  and Kiyotaka Aikawa },
title = {Quantum squeezing of a levitated nanomechanical oscillator},
journal = {Science},
volume = {389},
number = {6766},
pages = {1225-1228},
year = {2025},
doi = {10.1126/science.ady4652},
URL = {https://www.science.org/doi/abs/10.1126/science.ady4652}}

@article{Jain2016,
  title = {Direct Measurement of Photon Recoil from a Levitated Nanoparticle},
  author = {Jain, Vijay and Gieseler, Jan and Moritz, Clemens and Dellago, Christoph and Quidant, Romain and Novotny, Lukas},
  journal = {Phys. Rev. Lett.},
  volume = {116},
  issue = {24},
  pages = {243601},
  numpages = {5},
  year = {2016},
  month = {Jun},
  publisher = {American Physical Society},
  doi = {10.1103/PhysRevLett.116.243601},
  url = {https://link.aps.org/doi/10.1103/PhysRevLett.116.243601}
}

@article{Bykov2019,
  title={Direct loading of nanoparticles under high vacuum into a {P}aul trap for levitodynamical experiments},
  author={Bykov, Dmitry S and Mestres, Pau and Dania, Lorenzo and Schm{\"o}ger, Lisa and Northup, Tracy E},
  journal={Appl. Phys. Lett.},
url={https://pubs.aip.org/aip/apl/article/115/3/034101/37962/Direct-loading-of-nanoparticles-under-high-vacuum},
  volume={115},
  pages={034101},
  doi={10.1063/1.5109645},
  number={3},
  year={2019},
  publisher={AIP Publishing}
}

@article{Goldwater2019,
	author = {Goldwater, Daniel and Stickler, Benjamin A and Martinetz, Lukas and Northup, Tracy E and Hornberger, Klaus and Millen, James},
	doi = {10.1088/2058-9565/aaf5f3},
	journal = {Quantum Sci. Technol.},
	month = {jan},
	number = {2},
	pages = {024003},
	publisher = {IOP Publishing},
	title = {Levitated electromechanics: all-electrical cooling of charged nano- and micro-particles},
	url = {https://dx.doi.org/10.1088/2058-9565/aaf5f3},
	volume = {4},
	year = {2019}}

@article{Schrinski2023,
  title = {Macroscopic Quantum Test with Bulk Acoustic Wave Resonators},
  author = {Schrinski, Bj\"orn and Yang, Yu and von L\"upke, Uwe and Bild, Marius and Chu, Yiwen and Hornberger, Klaus and Nimmrichter, Stefan and Fadel, Matteo},
  journal = {Phys. Rev. Lett.},
  volume = {130},
  issue = {13},
  pages = {133604},
  numpages = {6},
  year = {2023},
  month = {Mar},
  publisher = {American Physical Society},
  doi = {10.1103/PhysRevLett.130.133604},
  url = {https://link.aps.org/doi/10.1103/PhysRevLett.130.133604}
}

@article{Leibfried2003,
  title = {Quantum dynamics of single trapped ions},
  author = {Leibfried, D. and Blatt, R. and Monroe, C. and Wineland, D.},
  journal = {Rev. Mod. Phys.},
  volume = {75},
  issue = {1},
  pages = {281--324},
  numpages = {0},
  year = {2003},
  month = {Mar},
  publisher = {American Physical Society},
  doi = {10.1103/RevModPhys.75.281},
  url = {https://link.aps.org/doi/10.1103/RevModPhys.75.281}
}

@article{Poulsen2012,
  author = {Poulsen, G. and Miroshnychenko, Y. and Drewsen, M.},
  title = {Efficient ground-state cooling of an ion in a large room-temperature linear {P}aul trap with a sub-{H}ertz heating rate},
  journal = {Phys. Rev. A},
  volume = {86},
  pages = {051402},
  numpages = {4},
  year = {2012},
  doi = {10.1103/PhysRevA.86.051402}
}

@article{Meekhof1996,
  author = {Meekhof, D. M. and Monroe, C. and King, B. E. and Itano, W. M. and Wineland, D. J.},
  title = {Generation of Nonclassical Motional States of a Trapped Atom},
  journal = {Phys. Rev. Lett.},
  volume = {76},
  pages = {1796--1799},
  year = {1996},
  doi = {10.1103/PhysRevLett.76.1796}
}

@article{Monroe1996,
  author = {Monroe, C. and Meekhof, D. M. and King, B. E. and Wineland, D. J.},
  title = {A ``{S}chr{\"o}dinger {C}at'' Superposition State of an Atom},
  journal = {Science},
  volume = {272},
  number = {5265},
  pages = {1131--1136},
  year = {1996},
  doi = {10.1126/science.272.5265.1131}
}

@article{Poyatos1996,
  author = {Poyatos, J. F. and Cirac, J. I. and Zoller, P.},
  title = {Quantum Reservoir Engineering with Laser Cooled Trapped Ions},
  journal = {Phys. Rev. Lett.},
  volume = {77},
  pages = {4728--4731},
  year = {1996},
  doi = {10.1103/PhysRevLett.77.4728}
}

@article{Mirrahimi2014,
  author  = {Mazyar Mirrahimi and Zaki Leghtas and Victor V. Albert and Steven Touzard and Robert J. Schoelkopf and Liang Jiang and Michel H. Devoret},
  title   = {Dynamically Protected Cat-Qubits: A New Paradigm for Universal Quantum Computation},
  journal = {New J. Phys.},
  volume  = {16},
  pages   = {045014},
  year    = {2014},
  doi     = {10.1088/1367-2630/16/4/045014}
}

@article{Bartolo2016,
  author  = {Bartolo, Nicola and Minganti, Fabrizio and Casteels, Wim and Ciuti, Cristiano},
  title   = {Exact steady state of a {K}err resonator with one- and two-photon driving and dissipation: {C}ontrollable {W}igner-function multimodality and dissipative phase transitions},
  journal = {Phys. Rev. A},
  volume  = {94},
  number  = {3},
  pages   = {033841},
  year    = {2016},
  doi     = {10.1103/PhysRevA.94.033841}
}

@article{Pflanzer2013,
  title = {Optomechanics assisted by a qubit: From dissipative state preparation to many-partite systems},
  author = {Pflanzer, Anika C. and Romero-Isart, Oriol and Cirac, J. Ignacio},
  journal = {Phys. Rev. A},
  volume = {88},
  issue = {3},
  pages = {033804},
  numpages = {8},
  year = {2013},
  month = {Sep},
  publisher = {American Physical Society},
  doi = {10.1103/PhysRevA.88.033804},
  url = {https://link.aps.org/doi/10.1103/PhysRevA.88.033804}
}

@article{Toros2021,
  title = {Creating atom-nanoparticle quantum superpositions},
  author = {Toro\ifmmode \check{s}\else \v{s}\fi{}, M. and Bose, S. and Barker, P. F.},
  journal = {Phys. Rev. Res.},
  volume = {3},
  issue = {3},
  pages = {033218},
  numpages = {11},
  year = {2021},
  month = {Sep},
  publisher = {American Physical Society},
  doi = {10.1103/PhysRevResearch.3.033218},
  url = {https://link.aps.org/doi/10.1103/PhysRevResearch.3.033218}
}

@misc{Schut2025,
      title={Proposal for creating spatial superposition of a large mass in a {RF} trap}, 
      author={Martine Schut and Jackson Tiong and Valerio Scarani},
      year={2025},
      eprint={2509.17081},
      archivePrefix={arXiv},
      primaryClass={quant-ph},
      url={https://arxiv.org/abs/2509.17081}, 
}

@article{Wilson-Rae2008,
doi = {10.1088/1367-2630/10/9/095007},
url = {https://doi.org/10.1088/1367-2630/10/9/095007},
year = {2008},
month = {sep},
publisher = {},
volume = {10},
number = {9},
pages = {095007},
author = {Wilson-Rae, I and Nooshi, N and Dobrindt, J and Kippenberg, T J and Zwerger, W},
title = {Cavity-assisted backaction cooling of mechanical resonators},
journal = {New J. Phys.}
}

@article{Aldous2017,
author = {Matthew Aldous and Jonathan Woods and Andrei Dragomir and Ritayan Roy and Matt Himsworth},
journal = {Opt. Express},
number = {11},
pages = {12830--12838},
publisher = {Optica Publishing Group},
title = {Carrier frequency modulation of an acousto-optic modulator for laser stabilization},
volume = {25},
month = {May},
year = {2017},
url = {https://opg.optica.org/oe/abstract.cfm?URI=oe-25-11-12830},
doi = {10.1364/OE.25.012830},
}

@article{Peres1996,
  title = {Separability Criterion for Density Matrices},
  author = {Peres, Asher},
  journal = {Phys. Rev. Lett.},
  volume = {77},
  issue = {8},
  pages = {1413--1415},
  numpages = {0},
  year = {1996},
  month = {Aug},
  publisher = {American Physical Society},
  doi = {10.1103/PhysRevLett.77.1413},
  url = {https://link.aps.org/doi/10.1103/PhysRevLett.77.1413}
}

@article{Horodecki1996,
  title = {Separability of mixed states: necessary and sufficient conditions},
  author = {Horodecki, Micha{\l} and Horodecki, Pawe{\l} and Horodecki, Ryszard},
  journal = {Phys. Lett. A},
  volume = {223},
  number = {1},
  pages = {1--8},
  year = {1996},
  issn = {0375-9601},
  doi = {10.1016/S0375-9601(96)00706-2},
  url = {https://www.sciencedirect.com/science/article/pii/S0375960196007062}
}

@article{Millen2020,
  author        = {Millen, James and Monteiro, Tania S. and Pettit, Robert and Vamivakas, A. Nick},
  title         = {Optomechanics with levitated particles},
  journal       = {Rep. Prog. Phys.},
  volume        = {83},
  number        = {2},
  pages         = {026401},
  year          = {2020},
  publisher     = {IOP Publishing},
  doi           = {10.1088/1361-6633/ab6100},
}

@article{RomeroIsart2017,
  author  = {Romero-Isart, Oriol},
  title   = {Coherent inflation for large quantum superpositions of levitated microspheres},
  journal = {New J. Phys.},
  volume  = {19},
  number  = {12},
  pages   = {123029},
  year    = {2017},
  doi     = {10.1088/1367-2630/aa99bf}
}

@article{Pino2018,
  author  = {Pino, H. and Prat-Camps, J. and Sinha, K. and Venkatesh, B. Prasanna and Romero-Isart, O.},
  title   = {On-chip quantum interference of a superconducting microsphere},
  journal = {Quantum Sci. Technol.},
  volume  = {3},
  number  = {2},
  pages   = {025001},
  year    = {2018},
  doi     = {10.1088/2058-9565/aa9d15}
}

@article{Magrini2021,
  author  = {Magrini, Lorenzo and Rosenzweig, Philipp and Bach, Constanze and Deutschmann-Olek, Andreas and Hofer, Sebastian G. and Hong, Sungkun and Kiesel, Nikolai and Kugi, Andreas and Aspelmeyer, Markus},
  title   = {Real-time optimal quantum control of mechanical motion at room temperature},
  journal = {Nature},
  volume  = {595},
  number  = {7867},
  pages   = {373--377},
  year    = {2021},
  doi     = {10.1038/s41586-021-03602-3}
}

@article{Weiss2021,
  author  = {Weiss, T. and Roda-Llordes, M. and Torrontegui, E. and Aspelmeyer, M. and Romero-Isart, O.},
  title   = {Large Quantum Delocalization of a Levitated Nanoparticle Using Optimal Control: Applications for Force Sensing and Entangling via Weak Forces},
  journal = {Phys. Rev. Lett.},
  volume  = {127},
  number  = {2},
  pages   = {023601},
  year    = {2021},
  doi     = {10.1103/PhysRevLett.127.023601}
}

@article{Neumeier2024,
  author  = {Neumeier, Lukas and Ciampini, Mario A. and Romero-Isart, Oriol and Aspelmeyer, Markus and Kiesel, Nikolai},
  title   = {Fast quantum interference of a nanoparticle via optical potential control},
  journal = {Proc. Natl. Acad. Sci. U.S.A.},
  volume  = {121},
  number  = {4},
  pages   = {e2306953121},
  year    = {2024},
  doi     = {10.1073/pnas.2306953121}
}

@article{RodaLlordes2024,
  author  = {Roda-Llordes, M. and Riera-Campeny, A. and Candoli, D. and Grochowski, P. T. and Romero-Isart, O.},
  title   = {Macroscopic Quantum Superpositions via Dynamics in a Wide Double-Well Potential},
  journal = {Phys. Rev. Lett.},
  volume  = {132},
  number  = {2},
  pages   = {023601},
  year    = {2024},
  doi     = {10.1103/PhysRevLett.132.023601}
}

@article{Bose2025,
  author  = {Bose, Sougato and Fuentes, Ivette and Geraci, Andrew A. and Khan, Saba Mehsar and Qvarfort, Sofia and Rademacher, Markus and Rashid, Muddassar and Toro{\v{s}}, Marko and Ulbricht, Hendrik and Wanjura, Clara C.},
  title   = {Massive quantum systems as interfaces of quantum mechanics and gravity},
  journal = {Rev. Mod. Phys.},
  volume  = {97},
  number  = {1},
  pages   = {015003},
  year    = {2025},
  doi     = {10.1103/RevModPhys.97.015003}
}

@article{CunillSubiranas2026,
  author        = {Cunill-Subiranas, Jaume and Resare, Fabian and Zhao, Suocheng and Qvarfort, Sofia and Metelmann, A. and Navau, Carles and Wieczorek, Witlef},
  title         = {Macroscopic Quantum Interference of the Center-of-Mass Motion of Levitated Superconducting Microparticles enabled by Magnetic Higher-Order Traps},
  journal       = {Quantum Sci. Technol.},
  volume        = {11},
  number        = {4},
  pages         = {045072},
  doi           = {10.1088/2058-9565/aeae6f},
  year          = {2026}
}

@misc{SchmittKremer2026,
  author        = {Schmitt Kremer, Oscar and Dania, Lorenzo and Novotny, Lukas and Frimmer, Martin},
  title         = {Fast quantum squeezing of a nanomechanical oscillator with an inverted potential},
  year          = {2026},
  eprint        = {2609.12833},
  archivePrefix = {arXiv},
  primaryClass  = {quant-ph}
}

@article{Zaw2022,
  author  = {Zaw, Lin Htoo and Aw, Clive Cenxin and Lasmar, Zakarya and Scarani, Valerio},
  title   = {Detecting quantumness in uniform precessions},
  journal = {Phys. Rev. A},
  volume  = {106},
  pages   = {032222},
  year    = {2022},
  doi     = {10.1103/PhysRevA.106.032222},
}

@article{Chen2024,
  title = {Even-parity precession protocol for detecting nonclassicality and entanglement},
  author = {Chen, Jinyan and Tiong, Jackson and Zaw, Lin Htoo and Scarani, Valerio},
  journal = {Phys. Rev. A},
  volume = {110},
  issue = {6},
  pages = {062408},
  numpages = {11},
  year = {2024},
  month = {Dec},
  publisher = {American Physical Society},
  doi = {10.1103/PhysRevA.110.062408},
  url = {https://link.aps.org/doi/10.1103/PhysRevA.110.062408}
}

@article{James2007,
  author = {James, Daniel F. V. and Jerke, Jonathan},
  title = {Effective {H}amiltonian theory and its applications in quantum information},
  journal = {Can. J. Phys.},
  volume = {85},
  number = {6},
  pages = {625--632},
  year = {2007},
  doi = {10.1139/p07-060}
}

@book{Landau1976,
  author    = {Landau, L. D. and Lifshitz, E. M.},
  title     = {Mechanics},
  series    = {Course of Theoretical Physics},
  volume    = {1},
  edition   = {3rd},
  publisher = {Pergamon Press},
  address   = {Oxford},
  year      = {1976}
}

@article{Millen2015,
  title = {Cavity Cooling a Single Charged Levitated Nanosphere},
  author = {Millen, J. and Fonseca, P. Z. G. and Mavrogordatos, T. and Monteiro, T. S. and Barker, P. F.},
  journal = {Phys. Rev. Lett.},
  volume = {114},
  issue = {12},
  pages = {123602},
  numpages = {5},
  year = {2015},
  month = {Mar},
  publisher = {American Physical Society},
  doi = {10.1103/PhysRevLett.114.123602},
  url = {https://link.aps.org/doi/10.1103/PhysRevLett.114.123602}
}

@article{Fonseca2016,
  title = {Nonlinear Dynamics and Strong Cavity Cooling of Levitated Nanoparticles},
  author = {Fonseca, P. Z. G. and Aranas, E. B. and Millen, J. and Monteiro, T. S. and Barker, P. F.},
  journal = {Phys. Rev. Lett.},
  volume = {117},
  issue = {17},
  pages = {173602},
  numpages = {5},
  year = {2016},
  month = {Oct},
  publisher = {American Physical Society},
  doi = {10.1103/PhysRevLett.117.173602},
  url = {https://link.aps.org/doi/10.1103/PhysRevLett.117.173602}
}

@article{Conangla2018,
  author  = {Conangla, Gerard P. and Schell, Andreas W. and
             Rica, Ra{\'u}l A. and Quidant, Romain},
  title   = {Motion Control and Optical Interrogation of a
             Levitating Single Nitrogen Vacancy in Vacuum},
  journal = {Nano Lett.},
  volume  = {18},
  number  = {6},
  pages   = {3956--3961},
  year    = {2018},
  doi     = {10.1021/acs.nanolett.8b01414}
}

@article{Conagla2019,
  title = {Optimal Feedback Cooling of a Charged Levitated Nanoparticle with Adaptive Control},
  author = {Conangla, Gerard P. and Ricci, Francesco and Cuairan, Marc T. and Schell, Andreas W. and Meyer, Nadine and Quidant, Romain},
  journal = {Phys. Rev. Lett.},
  volume = {122},
  issue = {22},
  pages = {223602},
  numpages = {6},
  year = {2019},
  month = {Jun},
  publisher = {American Physical Society},
  doi = {10.1103/PhysRevLett.122.223602},
  url = {https://link.aps.org/doi/10.1103/PhysRevLett.122.223602}
}

@article{Gorman2014,
  author  = {Gorman, Dylan J. and Schindler, Philipp and Selvarajan, Sankaranarayanan and Daniilidis, Nikos and H{\"a}ffner, Hartmut},
  title   = {Two-mode coupling in a single-ion oscillator via parametric resonance},
  journal = {Phys. Rev. A},
  volume  = {89},
  pages   = {062332},
  year    = {2014},
  doi     = {10.1103/PhysRevA.89.062332}
}

@article{Hou2024,
  author  = {Hou, Pan-Yu and Wu, Jenny J. and Erickson, Stephen D. and Cole, Daniel C. and Zarantonello, Giorgio and Brandt, Adam D. and Geller, Shawn and Kwiatkowski, Alex and Glancy, Scott and Knill, Emanuel and Wilson, Andrew C. and Slichter, Daniel H. and Leibfried, Dietrich},
  title   = {Coherent coupling and non-destructive measurement of trapped-ion mechanical oscillators},
  journal = {Nat. Phys.},
  volume  = {20},
  number  = {10},
  pages   = {1636--1641},
  year    = {2024},
  doi     = {10.1038/s41567-024-02585-y}
}

@article{Nikolova2023,
  author  = {Nikolova, Bogomila S. and Ivanov, Peter A.},
  title   = {Laser-free method for creation of two-mode squeezed state and beam-splitter transformation with trapped ions},
  journal = {Phys. Scr.},
  volume  = {98},
  pages   = {065116},
  year    = {2023},
  doi     = {10.1088/1402-4896/acd72c}
}

@article{RodaLlordes2024b,
  title = {Numerical simulation of large-scale nonlinear open quantum mechanics},
  author = {Roda-Llordes, M. and Candoli, D. and Grochowski, P. T. and Riera-Campeny, A. and Agrenius, T. and Garc{\'{\i}}a-Ripoll, J. J. and Gonzalez-Ballestero, C. and Romero-Isart, O.},
  journal = {Phys. Rev. Res.},
  volume = {6},
  issue = {1},
  pages = {013262},
  numpages = {10},
  year = {2024},
  month = {Mar},
  publisher = {American Physical Society},
  doi = {10.1103/PhysRevResearch.6.013262},
  url = {https://link.aps.org/doi/10.1103/PhysRevResearch.6.013262}
}

@article{Kienzler2016,
  title = {Observation of Quantum Interference between Separated Mechanical Oscillator Wave Packets},
  author = {Kienzler, D. and Fl{\"u}hmann, C. and Negnevitsky, V. and Lo, H.-Y. and Marinelli, M. and Nadlinger, D. and Home, J. P.},
  journal = {Phys. Rev. Lett.},
  volume = {116},
  issue = {14},
  pages = {140402},
  numpages = {5},
  year = {2016},
  month = {Apr},
  publisher = {American Physical Society},
  doi = {10.1103/PhysRevLett.116.140402},
  url = {https://link.aps.org/doi/10.1103/PhysRevLett.116.140402}
}

@article{Johnson2017,
  author  = {Johnson, K. G. and Wong-Campos, J. D. and Neyenhuis, B. and Mizrahi, J. and Monroe, C.},
  title   = {Ultrafast creation of large {S}chr{\"o}dinger cat states of an atom},
  journal = {Nat. Commun.},
  volume  = {8},
  number  = {1},
  pages   = {697},
  year    = {2017},
  doi     = {10.1038/s41467-017-00682-6}
}

@article{Fluhmann2019,
  author  = {Fl{\"u}hmann, C. and Nguyen, T. L. and Marinelli, M. and Negnevitsky, V. and Mehta, K. and Home, J. P.},
  title   = {Encoding a qubit in a trapped-ion mechanical oscillator},
  journal = {Nature},
  volume  = {566},
  number  = {7745},
  pages   = {513--517},
  year    = {2019},
  doi     = {10.1038/s41586-019-0960-6}
}

@article{Schrinski2019,
  author  = {Schrinski, Bj\"orn and Nimmrichter, Stefan and Stickler, Benjamin A. and Hornberger, Klaus},
  title   = {Macroscopicity of quantum mechanical superposition tests via hypothesis falsification},
  journal = {Phys. Rev. A},
  volume  = {100},
  pages   = {032111},
  year    = {2019},
  doi     = {10.1103/PhysRevA.100.032111}
}

@article{Pedernales2020,
  author  = {Pedernales, Julen S. and Morley, Gavin W. and Plenio, Martin B.},
  title   = {Motional Dynamical Decoupling for Interferometry with Macroscopic Particles},
  journal = {Phys. Rev. Lett.},
  volume  = {125},
  pages   = {023602},
  year    = {2020},
  doi     = {10.1103/PhysRevLett.125.023602}
}

@article{Pedernales2022,
  author  = {Pedernales, Julen S. and Streltsov, Kirill and Plenio, Martin B.},
  title   = {Enhancing Gravitational Interaction between Quantum Systems by a Massive Mediator},
  journal = {Phys. Rev. Lett.},
  volume  = {128},
  pages   = {110401},
  year    = {2022},
  doi     = {10.1103/PhysRevLett.128.110401}
}

@article{Streltsov2021,
  author  = {Streltsov, Kirill and Pedernales, Julen S. and Plenio, Martin B.},
  title   = {Ground-State Cooling of Levitated Magnets in Low-Frequency Traps},
  journal = {Phys. Rev. Lett.},
  volume  = {126},
  pages   = {193602},
  year    = {2021},
  doi     = {10.1103/PhysRevLett.126.193602}
}

@article{Martinetz2020,
  author  = {Martinetz, Lukas and Hornberger, Klaus and Millen, James and Kim, M. S. and Stickler, Benjamin A.},
  title   = {Quantum electromechanics with levitated nanoparticles},
  journal = {npj Quantum Inf.},
  volume  = {6},
  pages   = {101},
  year    = {2020},
  doi     = {10.1038/s41534-020-00333-7}
}

@article{Kustura2022,
  title = {Mechanical Squeezing via Unstable Dynamics in a Microcavity},
  author = {Kustura, Katja and Gonzalez-Ballestero, Carlos and Sommer, Andr{\'e}s de los R{\'{\i}}os and Meyer, Nadine and Quidant, Romain and Romero-Isart, Oriol},
  journal = {Phys. Rev. Lett.},
  volume = {128},
  issue = {14},
  pages = {143601},
  numpages = {7},
  year = {2022},
  month = {Apr},
  publisher = {American Physical Society},
  doi = {10.1103/PhysRevLett.128.143601},
  url = {https://link.aps.org/doi/10.1103/PhysRevLett.128.143601}
}

@article{Casabone2015,
  title = {Enhanced Quantum Interface with Collective Ion-Cavity Coupling},
  author = {Casabone, B. and Friebe, K. and Brandst{\"a}tter, B. and Sch{\"u}ppert, K. and Blatt, R. and Northup, T. E.},
  journal = {Phys. Rev. Lett.},
  volume = {114},
  issue = {2},
  pages = {023602},
  numpages = {5},
  year = {2015},
  month = {Jan},
  publisher = {American Physical Society},
  doi = {10.1103/PhysRevLett.114.023602},
  url = {https://link.aps.org/doi/10.1103/PhysRevLett.114.023602}
}

@article{Joos1985,
  author  = {Joos, E. and Zeh, H. D.},
  title   = {The Emergence of Classical Properties Through Interaction with the Environment},
  journal = {Z. Phys. B},
  volume  = {59},
  pages   = {223--243},
  year    = {1985},
  doi     = {10.1007/BF01725541}
}

@article{Gallis1990,
  author  = {Gallis, Michael R. and Fleming, Gordon N.},
  title   = {Environmental and Spontaneous Localization},
  journal = {Phys. Rev. A},
  volume  = {42},
  pages   = {38--48},
  year    = {1990},
  doi     = {10.1103/PhysRevA.42.38}
}

@article{Hornberger2003,
  author  = {Hornberger, Klaus and Sipe, John E.},
  title   = {Collisional Decoherence Reexamined},
  journal = {Phys. Rev. A},
  volume  = {68},
  pages   = {012105},
  year    = {2003},
  doi     = {10.1103/PhysRevA.68.012105}
}

@article{Vacchini2009,
  author  = {Vacchini, Bassano and Hornberger, Klaus},
  title   = {Quantum Linear {B}oltzmann Equation},
  journal = {Phys. Rep.},
  volume  = {478},
  pages   = {71--120},
  year    = {2009},
  doi     = {10.1016/j.physrep.2009.06.001}
}

@article{Simon2000,
  author  = {Simon, R.},
  title   = {{P}eres-{H}orodecki Separability Criterion for Continuous Variable Systems},
  journal = {Phys. Rev. Lett.},
  volume  = {84},
  pages   = {2726--2729},
  year    = {2000},
  doi     = {10.1103/PhysRevLett.84.2726}
}

@article{Johnson2016,
  author  = {Johnson, K. G. and Wong-Campos, J. D. and Restelli, A. and Landsman, K. A. and Neyenhuis, B. and Mizrahi, J. and Monroe, C.},
  title   = {Active Stabilization of Ion Trap Radiofrequency Potentials},
  journal = {Rev. Sci. Instrum.},
  volume  = {87},
  pages   = {053110},
  year    = {2016},
  doi     = {10.1063/1.4948734}
}

@misc{Zhang2022,
  author        = {Zhang, Jie and Chow, B. T. and Haljan, P. C.},
  title         = {Stabilization of Ion-Trap Secular Frequencies for a Quantum Phase Transition Study},
  year          = {2022},
  eprint        = {2210.14478},
  archivePrefix = {arXiv},
  primaryClass  = {quant-ph}
}

@article{Osada2022,
  title = {Feasibility study on ground-state cooling and single-phonon readout of trapped electrons using hybrid quantum systems},
  author = {Osada, Alto and Taniguchi, Kento and Shigefuji, Masato and Noguchi, Atsushi},
  journal = {Phys. Rev. Res.},
  volume = {4},
  issue = {3},
  pages = {033245},
  numpages = {10},
  year = {2022},
  month = {Sep},
  publisher = {American Physical Society},
  doi = {10.1103/PhysRevResearch.4.033245},
  url = {https://link.aps.org/doi/10.1103/PhysRevResearch.4.033245}
}

@article{Leontica2021,
  title = {Schr\"odinger cat states of a macroscopic charged particle co-trapped with an ion},
  author = {Leontica, S. and Foot, C. J.},
  journal = {Phys. Rev. A},
  volume = {105},
  issue = {3},
  pages = {033109},
  numpages = {14},
  year = {2022},
  month = {Mar},
  publisher = {American Physical Society},
  doi = {10.1103/PhysRevA.105.033109},
  url = {https://link.aps.org/doi/10.1103/PhysRevA.105.033109}
}

@article{Trypogeorgos2016,
  title = {Cotrapping different species in ion traps using multiple radio frequencies},
  author = {Trypogeorgos, Dimitris and Foot, Christopher J.},
  journal = {Phys. Rev. A},
  volume = {94},
  issue = {2},
  pages = {023609},
  numpages = {9},
  year = {2016},
  month = {Aug},
  publisher = {American Physical Society},
  doi = {10.1103/PhysRevA.94.023609},
  url = {https://link.aps.org/doi/10.1103/PhysRevA.94.023609}
}

@article{Dehmelt1995,
  author  = {Dehmelt, H.},
  title   = {Economic synthesis and precision spectroscopy of
             anti-molecular hydrogen ions in {Paul} trap},
  journal = {Phys. Scr.},
  volume  = {T59},
  pages   = {423},
  year    = {1995},
  doi     = {10.1088/0031-8949/1995/T59/060}
}

@article{Leefer2017,
  author  = {Leefer, Nathan and
             Krimmel, Kai and
             Bertsche, William and
             Budker, Dmitry and
             Fajans, Joel and
             Folman, Ron and
             H{\"a}ffner, Hartmut and
             Schmidt-Kaler, Ferdinand},
  title   = {Investigation of two-frequency {Paul} traps for
             antihydrogen production},
  journal = {Hyperfine Interact.},
  volume  = {238},
  pages   = {12},
  year    = {2017},
  doi     = {10.1007/s10751-016-1388-0}
}

@article{Micke2019,
  author  = {P. Micke and J. Stark and S. A. King and T. Leopold and T. Pfeifer and L. Schm{\"o}ger and M. Schwarz and L. J. Spie{\ss} and P. O. Schmidt and J. R. Crespo L{\'o}pez-Urrutia},
  title   = {Closed-cycle, low-vibration 4 K cryostat for ion traps and other applications},
  journal = {Rev. Sci. Instrum.},
  volume  = {90},
  pages   = {065104},
  year    = {2019},
  doi     = {10.1063/1.5088593}
}

@article{Turchette2000,
  author  = {Q. A. Turchette and D. Kielpinski and B. E. King and D. Leibfried and D. M. Meekhof and C. J. Myatt and M. A. Rowe and C. A. Sackett and C. S. Wood and W. M. Itano and C. Monroe and D. J. Wineland},
  title   = {Heating of trapped ions from the quantum ground state},
  journal = {Phys. Rev. A},
  volume  = {61},
  pages   = {063418},
  year    = {2000},
  doi     = {10.1103/PhysRevA.61.063418}
}

@article{Brownnutt2015,
  author  = {M. Brownnutt and M. Kumph and P. Rabl and R. Blatt},
  title   = {Ion-trap measurements of electric-field noise near surfaces},
  journal = {Rev. Mod. Phys.},
  volume  = {87},
  number  = {4},
  pages   = {1419--1482},
  year    = {2015},
  doi     = {10.1103/RevModPhys.87.1419}
}

@article{Chan1999,
  author  = {H. W. Chan and J. C. Long and J. C. Price},
  title   = {Taber vibration isolator for vacuum and cryogenic applications},
  journal = {Rev. Sci. Instrum.},
  volume  = {70},
  number  = {6},
  pages   = {2742--2750},
  year    = {1999},
  doi     = {10.1063/1.1149839}
}

@misc{HerzanAEK2002,
  author       = {{Herzan LLC}},
  title        = {{AEK-2002} Acoustic Enclosure, Data Sheet},
  howpublished = {\url{https://www.herzan.com/wp-content/uploads/2018/02/AEK-2002-Data-Sheet.pdf}},
  note         = {Sound loss of 38~dB at 1~kHz}
}

@article{Dania2021,
  title = {Optical and electrical feedback cooling of a silica nanoparticle levitated in a Paul trap},
  author = {Dania, Lorenzo and Bykov, Dmitry S. and Knoll, Matthias and Mestres, Pau and Northup, Tracy E.},
  journal = {Phys. Rev. Res.},
  volume = {3},
  issue = {1},
  pages = {013018},
  numpages = {12},
  year = {2021},
  month = {Jan},
  publisher = {American Physical Society},
  doi = {10.1103/PhysRevResearch.3.013018},
  url = {https://link.aps.org/doi/10.1103/PhysRevResearch.3.013018}
}

@article{ZakkaBajjani2011,
  title = {Quantum superposition of a single microwave photon in two different `colour' states},
  author = {Zakka-Bajjani, Eva and Nguyen, Fran\c{c}ois and Lee, Minhyea and Vale, Leila R. and Simmonds, Raymond W. and Aumentado, Jos\'e},
  journal = {Nat. Phys.},
  volume = {7},
  issue = {8},
  pages = {599--603},
  year = {2011},
  month = {Aug},
  publisher = {Nature Publishing Group},
  doi = {10.1038/nphys2035},
  url = {https://doi.org/10.1038/nphys2035}
}

@article{McKay2016,
  title = {Universal Gate for Fixed-Frequency Qubits via a Tunable Bus},
  author = {McKay, David C. and Filipp, Stefan and Mezzacapo, Antonio and Magesan, Easwar and Chow, Jerry M. and Gambetta, Jay M.},
  journal = {Phys. Rev. Appl.},
  volume = {6},
  issue = {6},
  pages = {064007},
  year = {2016},
  month = {Dec},
  publisher = {American Physical Society},
  doi = {10.1103/PhysRevApplied.6.064007},
  url = {https://link.aps.org/doi/10.1103/PhysRevApplied.6.064007}
}

@article{Reagor2018,
  title = {Demonstration of universal parametric entangling gates on a multi-qubit lattice},
  author = {Reagor, M. and Osborn, C. B. and Tezak, N. and Staley, A. and Prawiroatmodjo, G. and Scheer, M. and Alidoust, N. and Sete, E. A. and Didier, N. and da Silva, M. P. and Acala, E. and Angeles, J. and Bestwick, A. and Block, M. and Bloom, B. and Bradley, A. and Bui, C. and Caldwell, S. and Capelluto, L. and Chilcott, R. and Cordova, J. and Crossman, G. and Curtis, M. and Deshpande, S. and El Bouayadi, T. and Girshovich, D. and Hong, S. and Hudson, A. and Karalekas, P. and Kuang, K. and Lenihan, M. and Manenti, R. and Manning, T. and Marshall, J. and Mohan, Y. and O'Brien, W. and Otterbach, J. and Papageorge, A. and Paquette, J.-P. and Pelstring, M. and Polloreno, A. and Rawat, V. and Ryan, C. A. and Renzas, R. and Rubin, N. and Russell, D. and Rust, M. and Scarabelli, D. and Selvanayagam, M. and Sinclair, R. and Smith, R. and Suska, M. and To, T.-W. and Vahidpour, M. and Vodrahalli, N. and Whyland, T. and Yadav, K. and Zeng, W. and Rigetti, C. T.},
  journal = {Sci. Adv.},
  volume = {4},
  issue = {2},
  pages = {eaao3603},
  year = {2018},
  month = {Feb},
  publisher = {American Association for the Advancement of Science},
  doi = {10.1126/sciadv.aao3603},
  url = {https://doi.org/10.1126/sciadv.aao3603}
}

@article{CastellanosBeltran2008,
  title = {Amplification and squeezing of quantum noise with a tunable {J}osephson metamaterial},
  author = {Castellanos-Beltran, M. A. and Irwin, K. D. and Hilton, G. C. and Vale, L. R. and Lehnert, K. W.},
  journal = {Nat. Phys.},
  volume = {4},
  issue = {12},
  pages = {929--931},
  year = {2008},
  month = {Dec},
  publisher = {Nature Publishing Group},
  doi = {10.1038/nphys1090},
  url = {https://doi.org/10.1038/nphys1090}
}

@article{Leghtas2015,
  title = {Confining the state of light to a quantum manifold by engineered two-photon loss},
  author = {Leghtas, Z. and Touzard, S. and Pop, I. M. and Kou, A. and Vlastakis, B. and Petrenko, A. and Sliwa, K. M. and Narla, A. and Shankar, S. and Hatridge, M. J. and Reagor, M. and Frunzio, L. and Schoelkopf, R. J. and Mirrahimi, M. and Devoret, M. H.},
  journal = {Science},
  volume = {347},
  issue = {6224},
  pages = {853--857},
  year = {2015},
  month = {Feb},
  publisher = {American Association for the Advancement of Science},
  doi = {10.1126/science.aaa2085},
  url = {https://doi.org/10.1126/science.aaa2085}
}

@article{Walker2019,
  title = {Measurement and feedback for cooling heavy levitated particles in low-frequency traps},
  author = {Walker, Lewis S. and Robb, Gordon R. M. and Daley, Andrew J.},
  journal = {Phys. Rev. A},
  volume = {100},
  issue = {6},
  pages = {063819},
  year = {2019},
  month = {Dec},
  publisher = {American Physical Society},
  doi = {10.1103/PhysRevA.100.063819},
  url = {https://link.aps.org/doi/10.1103/PhysRevA.100.063819}
}
 
\end{document}